\documentclass[traditabstract]{aa}   
\newcommand{\av}{A$_{\rm v}$}   
\usepackage{epsfig,graphicx,color}   
\usepackage{ctable}   
\usepackage{txfonts}   
\usepackage{ctable}   
   
\begin{document}

\title{Understanding star formation in molecular clouds}    
\subtitle{II. Signatures of gravitational collapse of IRDCs } 
 
  \author{N. Schneider \inst{1,2}   
  \and T. Csengeri     \inst{3}    
  \and R.S. Klessen    \inst{4,5,6}   
  \and P. Tremblin     \inst{7,8} 
  \and V. Ossenkopf    \inst{9}   
  \and N. Peretto      \inst{10} 
  \and R. Simon        \inst{9}  
  \and S. Bontemps     \inst{1,2}    
  \and C. Federrath    \inst{11,12}    
  }      %1 
 \institute{Universit\'e Bordeaux, LAB, UMR 5804, 33270 Floirac, France   
  \and   %2  
  CNRS, LAB, UMR 5804, 33270 Floirac, France   
  \and   %3 
  Max-Planck Institut f\"ur Radioastronomie, Auf dem H\"ugel, 53121 Bonn, Germany    
  \and   %4 
  Zentrum f\"ur Astronomie der Universit\"at Heidelberg,    
  Inst. f\"ur Theor. Astrophysik,    
  Albert-Ueberle Str. 2, 69120 Heidelberg, Germany   
  \and   %5 
  Department of Astronomy and Astrophysics, University of California, Santa Cruz, CA 95064, USA 
  \and   %6  
  Kavli Institute for Particle Astrophysics and Cosmology, Stanford University, SLAC National Accelerator  
  Laboratory, Menlo Park, CA 94025, USA 
  \and   %7 
  Astrophysics Group, University of Exeter, EX4 4QL Exeter, UK  
  \and   %8 
  Maison de la Simulation, CEA-CNRS-INRIA-UPS-UVSQ, USR 3441, CEA Saclay, 91191 Gif-sur-Yvette, France  
  \and   %9 
  I.\,Physikalisches Institut, Universit\"at zu K\"oln,   
  Z\"ulpicher Stra{\ss}e 77, 50937 K\"oln, Germany   
  \and   %10 
  School of Physics and Astronomy, Cardiff University, Queens Buildings, Cardiff CF24 3AA, UK17  
  \and   %11  
  Monash Centre for Astrophysics, School of Mathematical Sciences,  Monash University, VIC 3800, Australia    
  \and   %12 
  Research School of Astronomy\&Astrophysics, The Australian National University, Canberra,  ACT 2611, Australia 
}   
   
%\offprints{}   
   
\mail{nicola.schneider@obs.u-bordeaux1.fr}   
   
\titlerunning{IR-bright and IR-dark clouds}   
\authorrunning{N. Schneider}   
   
\date{\today}   
   
%\date{Received September 15, 1996; accepted March 16, 1997}   
   
\abstract {We analyse column density and temperature maps derived from
  {\sl Herschel} dust continuum observations of a sample of prominent,
  massive infrared dark clouds (IRDCs) i.e.  G11.11-0.12, G18.82-0.28,
  G28.37+0.07, and G28.53-0.25. We disentangle the velocity structure
  of the clouds using $^{13}$CO 1$\to$0 and $^{12}$CO 3$\to$2 data,
  showing that these IRDCs are the densest regions in massive giant
  molecular clouds (GMCs) and not isolated features.  The probability
  distribution function (PDF) of column densities for all clouds have
  a {\sl power-law distribution} over all (high) column densities,
  regardless of the evolutionary stage of the cloud: G11.11-0.12,
  G18.82-0.28, and G28.37+0.07 contain (proto)-stars, while
  G28.53-0.25 shows no signs of star formation.  This is in contrast
  to the purely log-normal PDFs reported for near and/or mid-IR
  extinction maps. We only find a log-normal distribution for lower
  column densities, if we perform PDFs of the column density maps of
  the {\sl whole} GMC in which the IRDCs are embedded.  By comparing
  the PDF slope and the radial column density profile of three of our
  clouds, we attribute the power law to the effect of large-scale
  gravitational collapse and to local free-fall collapse of pre- and
  protostellar cores for the highest column densities. A significant
  impact on the cloud properties from radiative feedback is unlikely
  because the clouds are mostly devoid of star formation.  Independent
  from the PDF analysis, we find infall signatures in the spectral
  profiles of $^{12}$CO for G28.37+0.07 and G11.11-0.12, supporting
  the scenario of gravitational collapse.  Our results are in line
  with earlier interpretations that see massive IRDCs as the densest
  regions within GMCs, which may be the progenitors of massive stars
  or clusters.  At least some of the IRDCs are probably the same
  features as {\sl ridges} (high column density regions with
  $N>$10$^{23}$ cm$^{-2}$ over small areas), which were defined for
  nearby IR-bright GMCs.  Because IRDCs are only confined to the
  densest (gravity dominated) cloud regions, the PDF constructed from
  this kind of a clipped image does not represent the (turbulence
  dominated) low column density regime of the cloud.}
   
   \maketitle   
%________________________________________________________________   
   
\section{Introduction} \label{intro}   
  
Infrared dark clouds (IRDCs) were detected as dark, cold ($<$25 K)
absorption features with high column density ($N>$10$^{22}$-10$^{23}$
cm$^{-2}$; e.g. Carey et al.  \cite{carey1998}, Peretto \& Fuller
\cite{peretto2010}) against the Galactic background at mid-IR
wavelengths (Egan et al.  \cite{egan1998}). Simon et al.
(\cite{simon2006a}), Teyssier et al.  (\cite{teyssier2002}), and
Kainulainen et al. (\cite{kai2011a}) showed that some IRDCs are
embedded in giant molecular clouds (GMCs), and thus are not isolated
features as originally suggested by Egan et al.  (1998).  Peretto \&
Fuller (2010) demonstrated on the basis of a large sample ($>$10000
clouds) that IRDCs span a large range of mass and size, with and
without star formation activity. The most massive and largest IRDCs
are proposed to represent the earliest stage of {\sl \textup{massive}}
star formation (e.g. Rathborne et al.  \cite{rathborne2006},
Nguyen-Luong et al. 2011, Beuther et al.  \cite{beuther2013}) because
they are thought to have not yet started to form stars or are at the
verge of star-formation.  On the other hand, an increasing number of
studies (e.g. Carey et al. \cite{carey2000}, Rathborne et al.
\cite{rathborne2011}, Sakai et al.  \cite{sakai2013}) now reveal that
many of the massive IRDCs show signatures of \textup{{\sl
    \textup{active}} and {\sl \textup{ongoing (massive) star
      formation}}} such as hot core and outflow emission and IR hot
spots.

\begin{table*}  \label{tab} 
\caption{Coordinates, distance $D$, and physical parameters of IRDCs, obtained from {\sl Herschel} FIR data,  
CO data, and near-IR/mid-IR extinction.}     
\begin{center}   
\begin{tabular}{lccccccccccccl}   
\hline  
\hline    
Region             & $l$    &  $b$   &  $D$ & $\langle N_{dust} \rangle$ & $\langle N_{bulk} \rangle$ & $\langle N_{cont} \rangle$ &  $M_{dust}$ & $M_{ext}$ & $M_{IRDC}$ & $M_{GMC}$ & $\Sigma_{dust}$ \\  
                   & [$^{\circ}$] & [$^{\circ}$] & [kpc] & {\tiny [10$^{21}$ cm$^{-2}$}] & {\tiny [10$^{21}$ cm$^{-2}$]} &  
{\tiny [10$^{21}$ cm$^{-2}$]} & {\tiny [10$^4$ M$_\odot$]} & {\tiny [10$^4$ M$_\odot$]} & {\tiny [10$^4$ M$_\odot$}] &  
{\tiny [10$^5$ M$_\odot$}] & {\tiny [M$_\odot$/pc$^2$}] \\   
%                   &        &        &      & {\tiny {\sl Herschel}} & {\tiny CO bulk}  & {\tiny CO cont.} &  
%{\tiny dust} & {\tiny extinc.} & {\tiny CO IRDC} & {\tiny CO GMC} & {\tiny dust} \\   
\                  &        &        &      &(1)   & (2)  & (3) & (4)  & (5)  & (6) & (7) & (8)  \\   
\hline       
{\sl G11.11-0.12} & 11.118 & -0.118 & 3.6  & 22.5 &  4.6 & 3.0 &  8.4 &  -  & 2.2 & 0.5 & 418 \\   
{\sl G18.82-0.28} & 18.822 & -0.285 & 4.8  & 24.1 &  9.9 & 7.1 &  3.4 & 1.9 & 2.3 & 1.3 & 448 \\   
{\sl G28.37+0.07} & 28.373 &  0.076 & 5.0  & 39.9 & 16.1 & 7.0 & 12.4 & 6.8 & 6.3 & 9.0 & 741 \\   
{\sl G28.53-0.25} & 28.531 & -0.251 & 5.7  & 25.6 & 11.0 & 14.0&  9.6 & 7.4 & 5.2 & 1.5 & 474 \\   
\end{tabular}   
\end{center}   
\vskip0.1cm 
\noindent (1) Average H$_2$-column  density ($N_{dust}$) from {\sl Herschel} FIR data within the ellipse  
defining the IRDC (Simon et al. 2006b). \\   
\noindent (2) Average H$_2$-column density from $^{13}$CO 1$\to$0 column density (assuming that $^{13}$CO is  
optically thin) with $N(H_2){\rm [cm^{-2}]}=4.92 \times 10^5\,  N(^{13}{\rm CO})$ in the same ellipse and assuming  
an excitation temperature of 15 K. The velocity range corresponds to the bulk  
emission of the IRDC. For G11.11-0.12, we used $^{12}$CO 3$\to$2 data with  
$N(H_2){\rm [cm^{-2}]}=2.3 \times N(^{12}{\rm CO})$  (Strong et al. \cite{strong1988}) with the $^{12}$CO column  
density $N(^{12}{\rm CO})$.\\   
\noindent (3) Lower limit for the 'contamination' column density from CO data, determined for all velocities lower and higher than  
that of the IRDC.\\   
\noindent (4) Mass $M=N_{dust} \, 2 \, m_H \, \mu \, A D^2 (\pi/180)^2 \,\,$ [M$_{\odot}$] 
with $m_H$=1.67$\times$10$^{-24}$ g, $\mu$=2.3 (mean atomic weight per molecule), within the area  
$A$ [deg$^2$] of the ellipse defining the IRDC. \\  
\noindent (5) Mass of the cloud from near-IR/mid-IR extinction, given in KT and BTK. \\  
\noindent (6,7) Lower limit of mass of the IRDC (6) and associated GMC (7) from CO bulk emission. \\ 
\noindent (8) Surface density $\Sigma_{dust}=M_{dust}/A$. The mass $M$ and area $A$ refer to the values  
within the ellipse defining the IRDC.   
\end{table*}  
 
How do these observations fit together?  Are IRDCs a special category
of molecular clouds/clumps that are dominated by turbulence during
their earliest stages of evolution ? Or are they only the densest
condensations in massive star-forming clouds and then dominated by
gravity as IR-bright clouds?  To address these questions, it is
crucial to start with studying the gas reservoir of IRDCs, i.e.  the
column density structure as the most convenient observable. In
particular the probability distribution function (PDF) of column
density serves as a key property in characterising various physical
processes that shape the structure of molecular clouds (see e.g.
Federrath \& Klessen \cite{fed2013} and references therein).
Observational studies, based on near-IR extinction or {\sl Herschel}
dust column density maps, showed that the PDF of star-forming clouds
has a log-normal distribution for low column densities and a power-law
tail for higher column densities, where the power-law tail is either
interpreted as due to external pressure (Kainulainen et al.
\cite{kai2011b}), self-gravity (Froebrich \& Rowles
\cite{froebrich2010}; Schneider et al.  2013, 2015), or a combination
of both (Tremblin et al. \cite{tremblin2014}).  Numerical simulations
(e.g. Padoan et al.  \cite{padoan1997}; Kritsuk et al.
\cite{kritsuk2007}; Federrath et al. \cite{fed2008a}) have shown that
supersonic turbulence in isothermal gas (without self-gravity) can
reproduce a log-normal distribution. When self-gravity is switched on
in the models, a power-law tail develops on the end of the PDF with
high column density(Klessen \cite{klessen2000}; Vazquez-Semadeni et
al.  \cite{vaz2008}; Kritsuk et al.  \cite{kritsuk2011}; Federrath \&
Klessen 2013; Girichidis et al.  \cite{giri2014}).  The only PDFs
reported for IRDCs were either constructed from ALMA dust continuum
(G0.253+0.016, Rathborne et al.  \cite{rathborne2014}) or from
extinction maps obtained from near- and mid-IR data (Kainulainen \&
Tan \cite{kai2013} and Butler, Tan, \& Kainulainen \cite{butler2014},
called KT and BTK in the following). The latter are best fitted by
log-normal distributions.
 
In this paper we investigate a sample of four well-known IRDCs from
the catalogue of Simon et al. (2006b).  The IRDC G11.11-0.12 (the
'snake') lies at a distance of 3.6 kpc and has a linear extent of
$\sim$30 pc.  G18.82-0.28, G28.37+0.07, and G28.53-0.25 are smaller
(size of a few pc) and further away (Table~\ref{tab}), and were
studied by KT and BTK, who named them Cloud A, C, and D, respectively.
The objective of this paper is to show that these IRDCs, which are
amongst the most massive in the Galaxy, are dominated by gravity and
show the same properties (e.g. average column density, PDF shape,
spectral line profiles) as the central cloud regions, called {\sl
  \textup{ridges}} (Schneider et al. \cite{schneider2010}, Hill et al.
\cite{hill2011}, Hennemann et al.  \cite{hennemann2012}), in IR-bright
GMCs.
 
\section{Observations and data analysis} \label{obs}   
 
\noindent {\bf Column density maps  from Herschel}\\   
\noindent   
We use {\sl Herschel}\footnote{Herschel is an ESA 
  space observatory with science instruments provided by European-led 
  Principal Investigator Consortia and with important participation 
  from NASA.} (Pilbratt et al.  \cite{pilbratt2010}) archive data from 
the {\it Herschel} Infrared GALactic plane survey (Hi-GAL), 
Molinari et al. \cite{molinari2010}) program.  The PACS (Poglitsch et 
al. \cite{poglitsch2010}) data were reduced using scanamorphos v23 
(Roussel \cite{roussel2013}). The SPIRE (Griffin et al. 
\cite{griffin2010}) data were reduced employing the HIPE10 pipeline 
and not the ROMAGAL software (Traficante et al. 
\cite{traficante2011}) that was developed for the Hi-Gal project. We 
preferred  to use the {\sl Herschel Interactive Processing Environment} 
(HIPE; Ott et al.  \cite{ott2011}) because it is the officially 
developed and well-documented data reduction software for the {\sl 
  Herschel} satellite.  The column density and temperature maps are 
determined from a pixel-to-pixel grey-body fit to the 160 $\mu$m 
(PACS), and 250, 350, 500 $\mu$m (SPIRE) wavelengths.  These maps have 
an angular resolution of $\sim$36$''$, corresponding to the nominal 
Gaussian FWHM value of 36.4$''$ for the 500 $\mu$m map given in the 
SPIRE Handbook (v2.5, 2014). To perform the SED fit, all lower 
wavelengths maps with high angular resolution were smoothed to 36$''$. 
All {\sl Herschel} maps have an absolute flux calibration, using  
Planck-corrections from the {\sc zeroPointCorrection} task in HIPE10 
for SPIRE and IRAS maps for PACS.  For the SED fit, we fix the 
specific dust opacity per unit mass (dust+gas) approximated by the 
power law $\kappa_\nu = 0.1 \, (\nu/1000 {\rm GHz})^\beta$ cm$^{2}$/g 
and $\beta$=2, and leave the dust temperature and column density as 
free parameters (see Schneider et al. (2015) for further details). We 
estimate the final uncertainties in the column density maps to be 
around $\sim$30--50\%, mainly arising from the uncertainty in the 
assumed form of the opacity law and possible temperature gradients 
along the line-of-sight (Russeil et al. \cite{russeil2013}, Roy et al. 
\cite{roy2013}).  \vspace{0.15cm} 
 
\noindent {\bf Molecular line data: $^{13}$CO 1$\to$0 and $^{12}$CO 3$\to$2} \\ 
\noindent 
Data cubes of $^{13}$CO 1$\to$0 (110.2 GHz) emission were taken from 
the Galactic Ring Survey archive (GRS), a survey covering the first 
Galactic quadrant (Jackson et al. \cite{jackson2006}), obtained with 
the Five Colleges Radio Astronomy Observatory (FCRAO). The angular 
resolution is $\sim$45$''$ at a velocity resolution of 0.21 km 
s$^{-1}$. We obtained $^{12}$CO 3$\to$2 (345.8 GHz) data (angular resolution of 
14$''$ and velocity resolution of 1 km s$^{-1}$)  from 
the JCMT/COHRS archive (Dempsey et al. \cite{dempsey2013}).  Both CO 
data sets are given on a main beam brightness temperature scale. 
\vspace{0.15cm} 
 
%%%%%%%%%%%%%%%%%%%%%%%%%%%%%%   
%%%  Cloud C  %%%%%%%%%%%%%%%   
\begin{figure*}[!htpb]   
\begin{centering}   
\includegraphics [width=7.5cm, angle={0}]{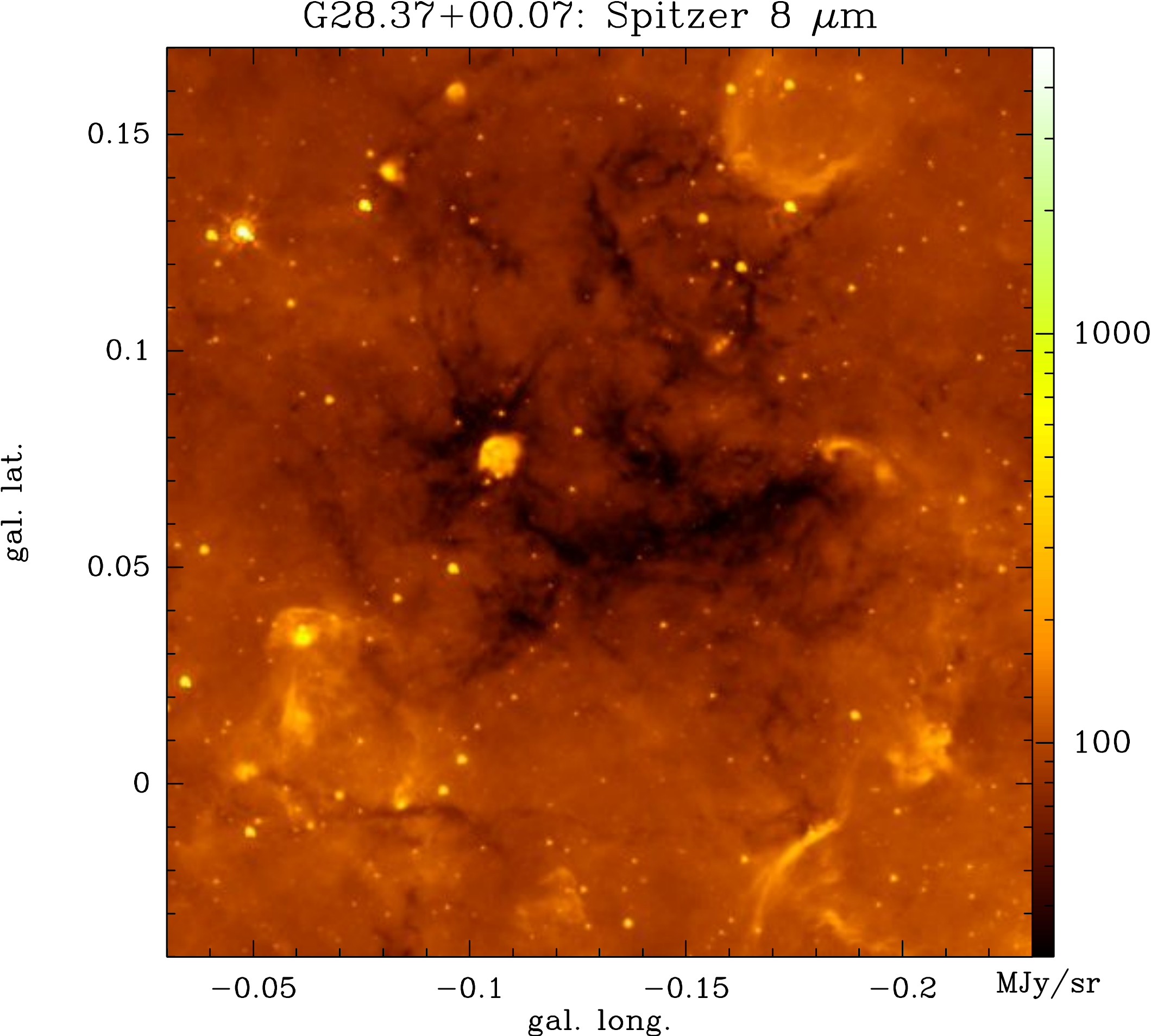}   
\includegraphics [width=7.5cm, angle={0}]{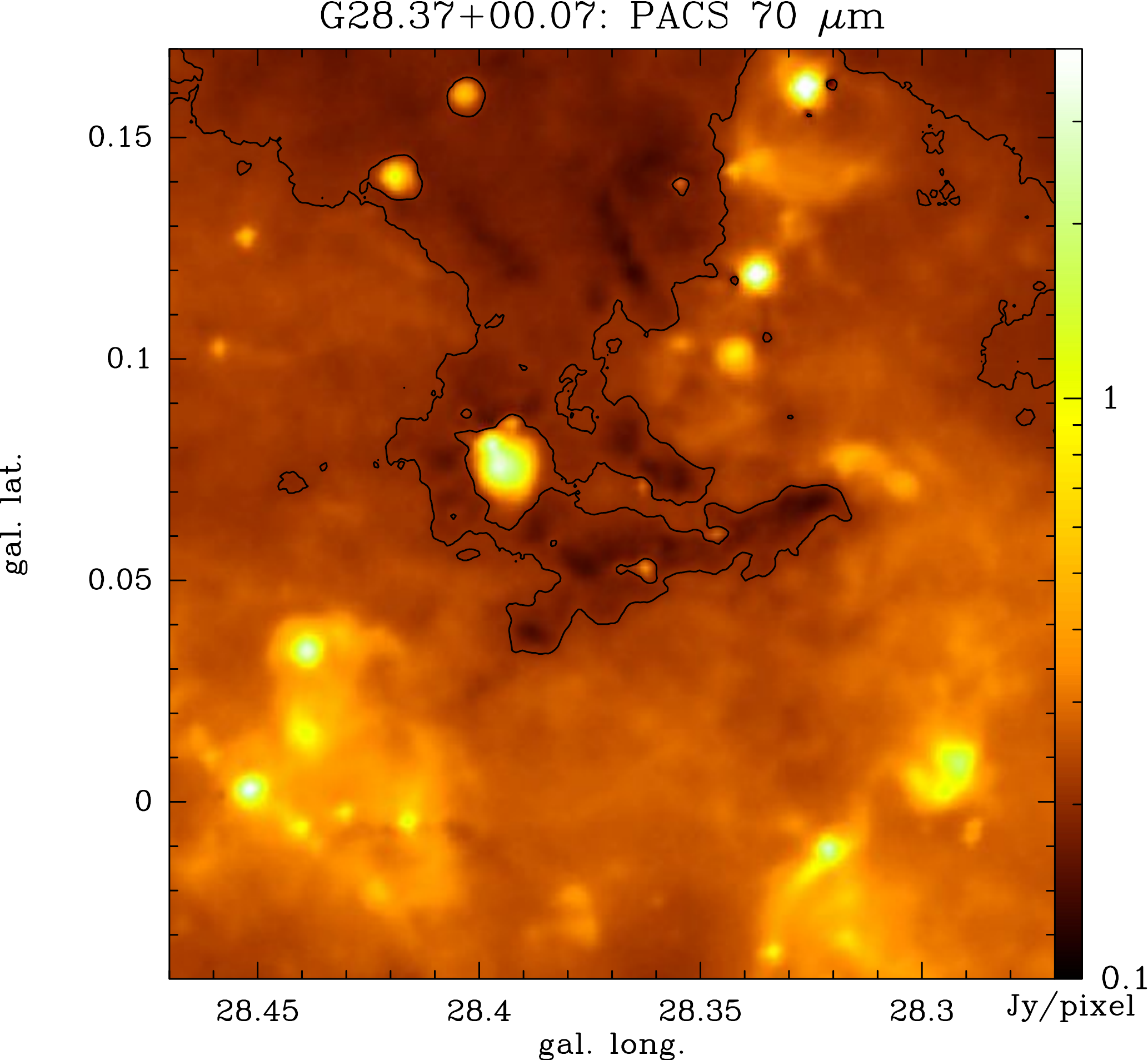}   
\includegraphics [width=7.5cm, angle={0}]{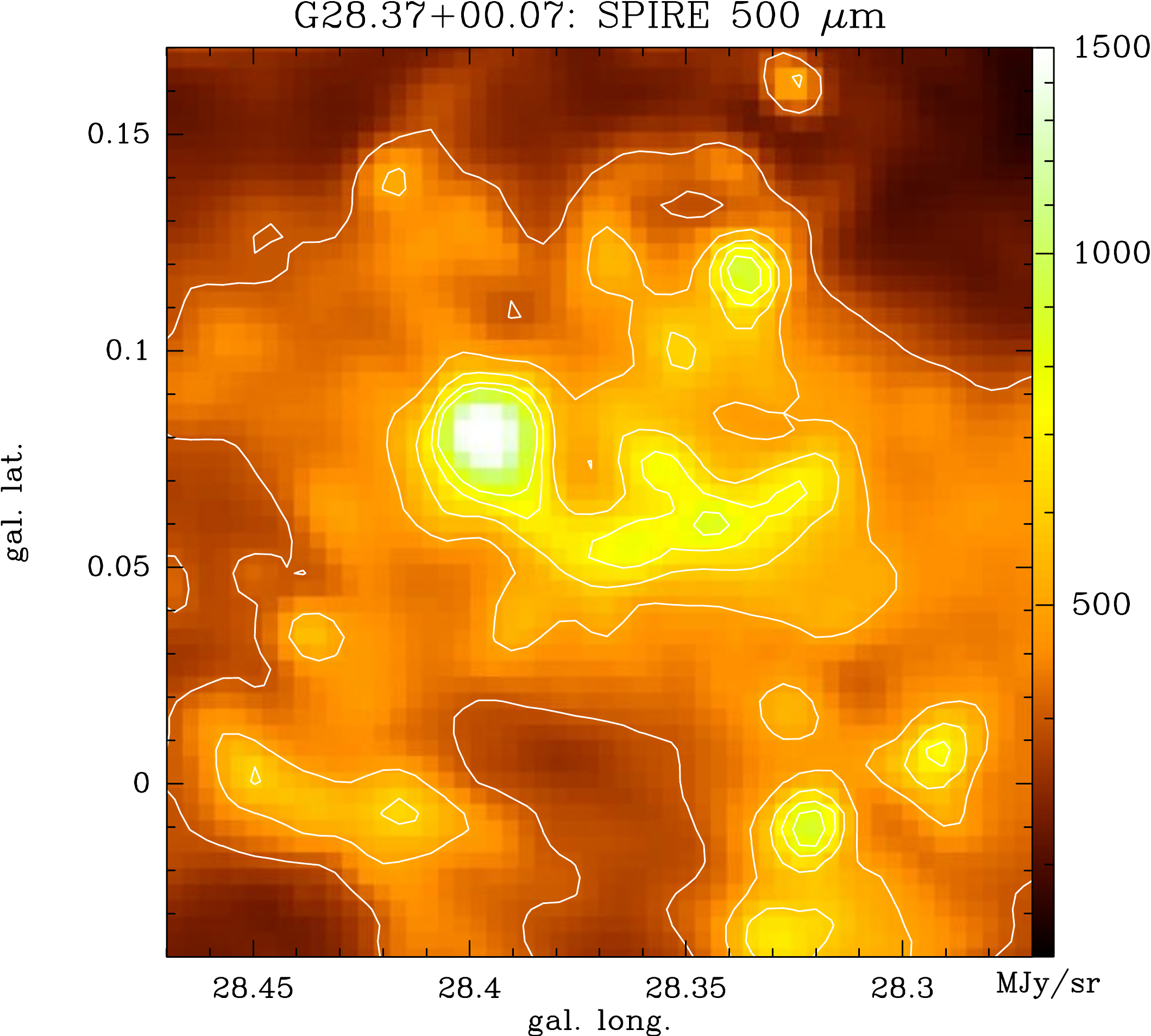} 
\includegraphics [width=7.5cm, angle={0}]{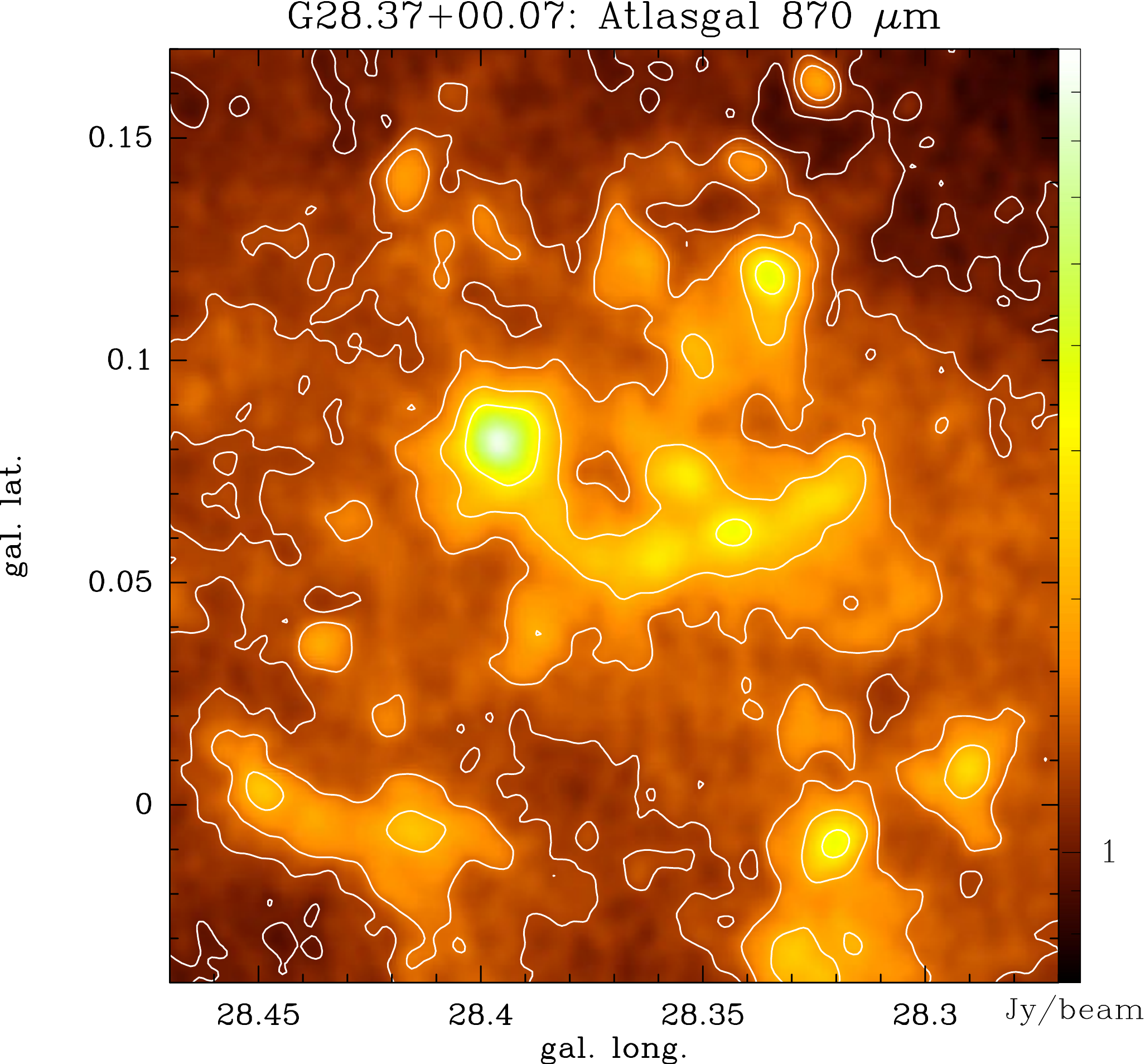} 
\caption[] {{\bf Top:} {\sl Spitzer} 8 $\mu$m ($\sim$2.5$''$ 
  resolution) and PACS 70 $\mu$m maps ($\sim$11$''$ resolution).  The 
  {\sl Spitzer/IRAC} data were taken from the public available GLIMPSE 
  archive.  {\bf Bottom:} SPIRE 500 $\mu$m map (36$''$ resolution) and 
  ATLASGAL 870 $\mu$m maps (23$''$ resolution) of G28.37+0.07.  The 
  central dark feature in the {\sl Spitzer} and PACS images is the 
  IRDC, appearing bright at longer wavelengths (SPIRE and ATLASGAL).} 
\label{herschel-cloud-c} 
\end{centering}  
\end{figure*}   
 
\begin{figure*}[htbp] \label{herschel2-cloud-c}   
\begin{centering}   
\includegraphics[width=7.5cm]{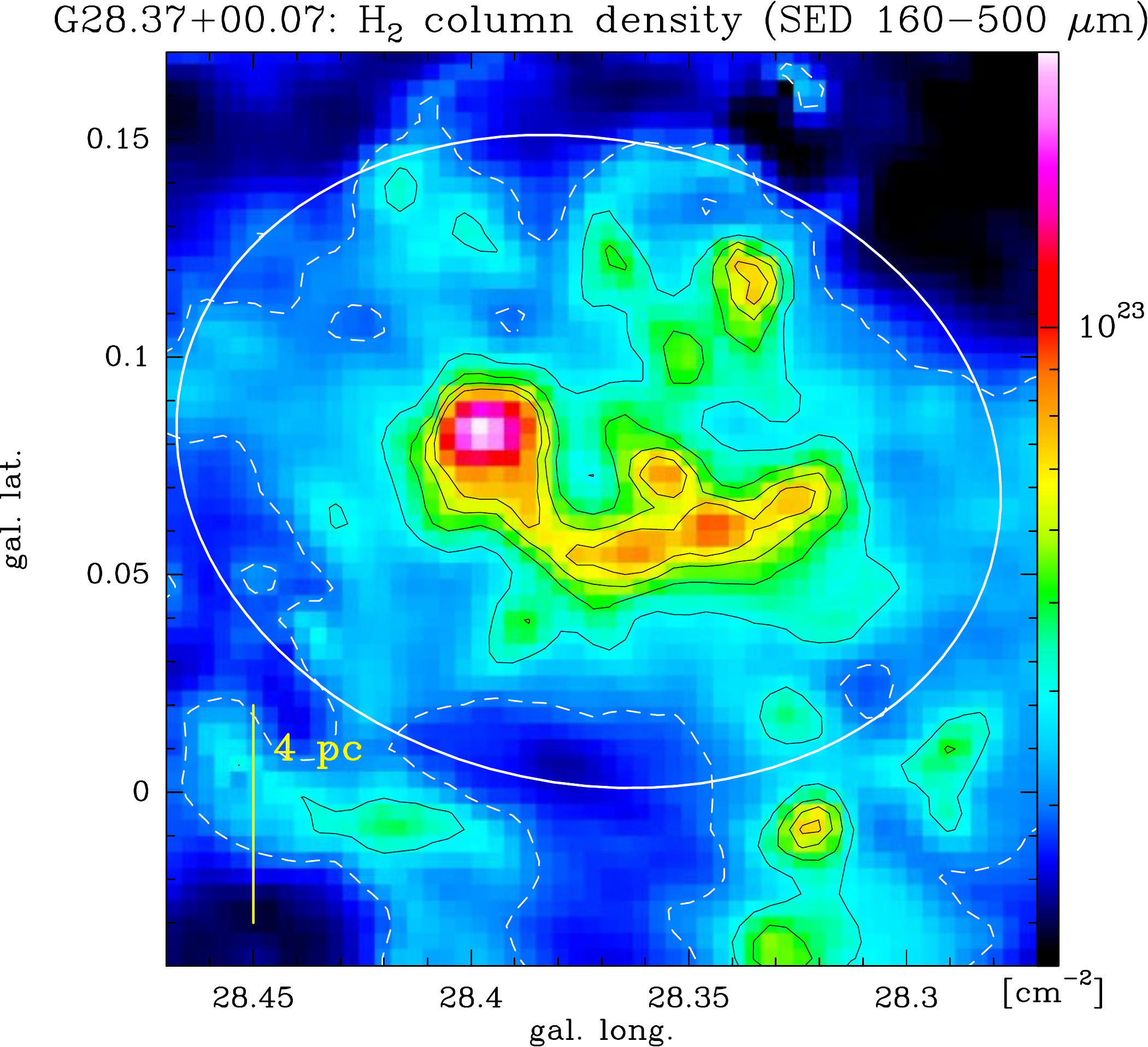} 
\includegraphics[width=7.5cm]{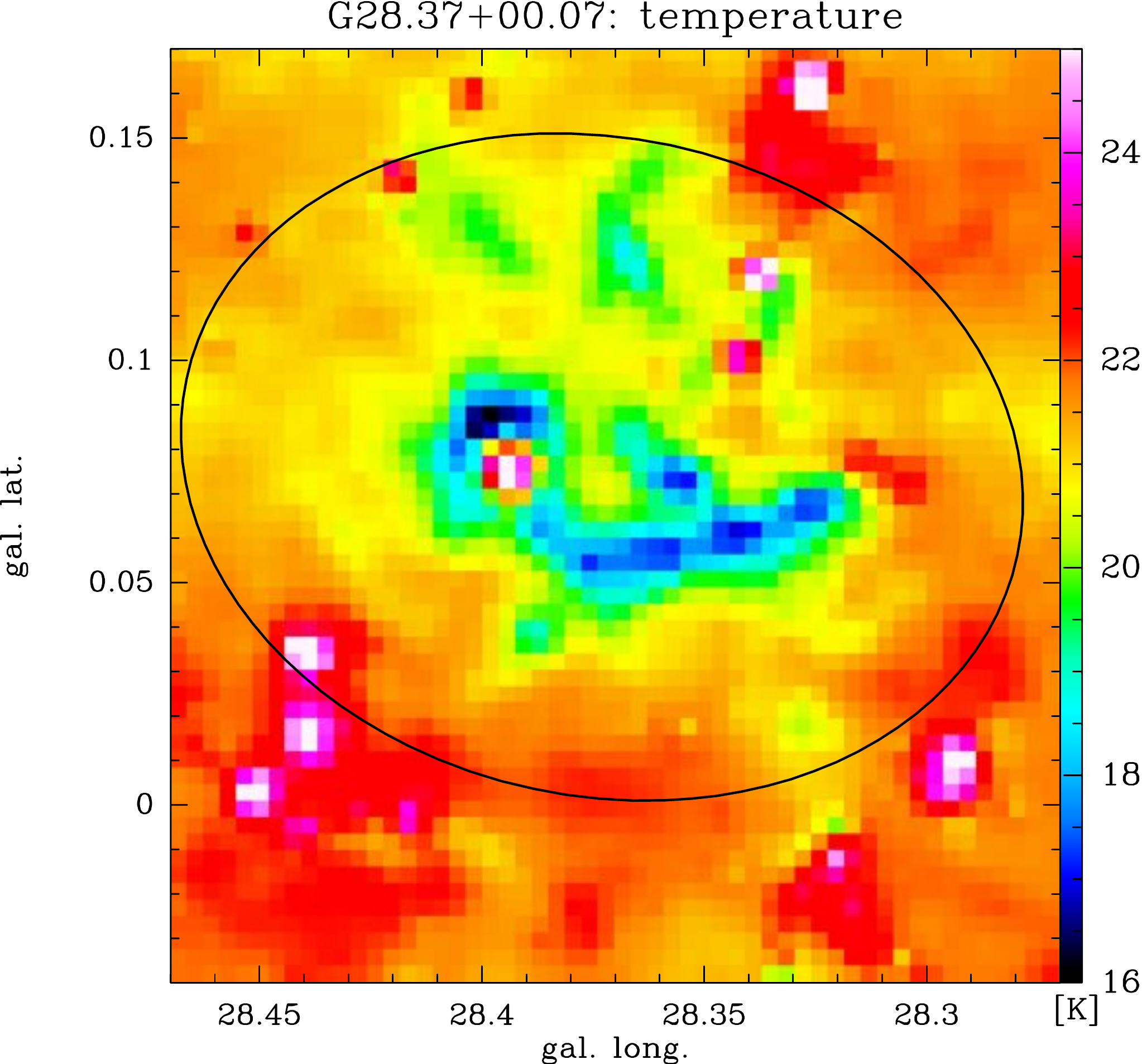} 
\caption[] {{\bf Left:} H$_2$-column density map from an SED fit to 
  {\sl Herschel} fluxes at 160-500 $\mu$m.  The black contours 
  indicate the levels $N_{dust}$=4 to 7$\times$10$^{22}$ cm$^{-2}$ in 
  steps of 10$^{22}$ cm$^{-2}$, the white-dashed contour outlines the 
  approximate completeness level ($\sim$3$\times$10$^{22}$ 
  cm$^{-2}$), and the white ellipse is taken from Simon et al. 
  (2006b), defining the IRDC. {\bf Right:} temperature map of 
  G28.37+0.07.} 
\end{centering}   
\end{figure*} 
 
\noindent {\bf ATLASGAL} \\ 
\noindent 
The ATLASGAL survey (Schuller et al. \cite{schuller2009}) imaged $420$
deg$^2$ of the Galactic plane ($l$=$\pm60^\circ$, $b$=$\pm1.5^\circ$)
at a 19.2$''$ spatial resolution at 870 $\mu$m with the LABOCA camera
Siringo et al. (\cite{siringo2009}) on the APEX Telescope.  The data,
together with the catalogue of compact sources, is presented in
Csengeri et al.  (\cite{csengeri2014}). Here, we make use of the
publicly available ATLASGAL
data\footnote{http://atlasgal.mpifr-bonn.mpg.de/}.  The sensitivity of
the data is dominated by the noise level of the ATLASGAL survey, which
in these regions is $\sim60$~mJy/beam.  Ground-based bolometer
observations intrinsically filter emission from larger scales when
removing the correlated noise during data reduction.  In the original
LABOCA observations the low-level emission is therefore filtered out
above $\sim$2.5$'$ angular scales, and only the densest regions are
visible. The more extended emission has been recovered using the high
sensitivity all-sky survey of the Planck/HFI instrument at 353~GHz,
following the method of Weiss et al.  (\cite{weiss2001}), and
presented in more detail in Csengeri et al.  (in prep).  As shown in
the figures presenting the ATLASGAL data, the combined data set is
sensitive to the cold dust at all spatial scales.  For display
reasons, we smoothed the data slightly to 23$''$ angular resolution.
We only display the maps to show the large scale distribution of dense
and cold gas.  We do not use the ATLASGAL data to determine the column
density. This will be done in a future work in which we produce column
density maps at higher angular resolution using ATLASGAL and selected
{\sl Herschel} bands.
 
\section{Results and discussion} \label{results}   
 
In this section we show as a representative case all continuum and 
molecular line data for G28.37+0.07 (Cloud C), which was studied in 
Simon et al.  (2006b), KT, and BTK.  Maps for the other three clouds 
are shown in Appendix A. 
 
\vspace{-0.2cm} 
\subsection{G28.37+0.07 with {\sl Herschel} and ATLASGAL} \label{cont}  
 
Figure~\ref{herschel-cloud-c} displays G28.37+0.07 as an extended {\sl
  \textup{dark feature}} in the 8 $\mu$m {\sl
  Spitzer/IRAC}\footnote{Data taken from the public available GLIMPSE
  archive http://irsa.ipac.caltech.edu/data/SPITZER/GLIMPSE/} and 70
$\mu$m PACS maps except with an {\sl \textup{emission peak}} at
$l$=28.396$^\circ$, $b$=0.081$^\circ$.  The cloud is overall bright at
500 $\mu$m and 870 $\mu$m, but also at 8 $\mu$m (Carey et al.  2000,
BTK). It corresponds to an IRAS source (IRAS18402-403), and contains
massive protostellar cores (Zhang et al. 2009), and is thus most
likely a protocluster.  Because this peak is also prominent at longer
wavelengths, the protocluster is still embedded in a dense, cold,
molecular envelope.  In particular, the 500 and 870 $\mu$m maps,
tracing the large-scale distribution of \textup{{\sl \textup{cold}},}
molecular gas, show that the IRDC is not an isolated feature.  This
also becomes obvious in the column density and temperature maps
(Fig.~2), where this central region is outlined by low temperatures of
approximately 16 to 19 K, except for the locally heated protostellar
environment, and column densities of 6$\times$10$^{22}$ cm$^{-2}$ to a
few 10$^{23}$ cm$^{-2}$.  Comparing our column density maps with those
of KT and BTK shows that the highest column densities above a column
density of few 10$^{-23}$ cm$^{-2}$ (BTK give a threshold of A$_v
\sim$200) are missing in the BTK studies because (i) there is
extinction saturation i.e. mid-IR foreground emission is as bright as
the emission towards the cloud, and (ii) it is a bright 8 $\mu$m
region: the densest part of G28.37+0.07 is emitting strongly at 8
$\mu$m, so it does not appear as an extinction feature. Therefore
recovering the column density structure from an extinction map at
these positions is impossible (Peretto \& Fuller 2010).
 
\begin{figure*}[!htpb]   
\begin{centering}   
\includegraphics [width=8cm, angle={0}]{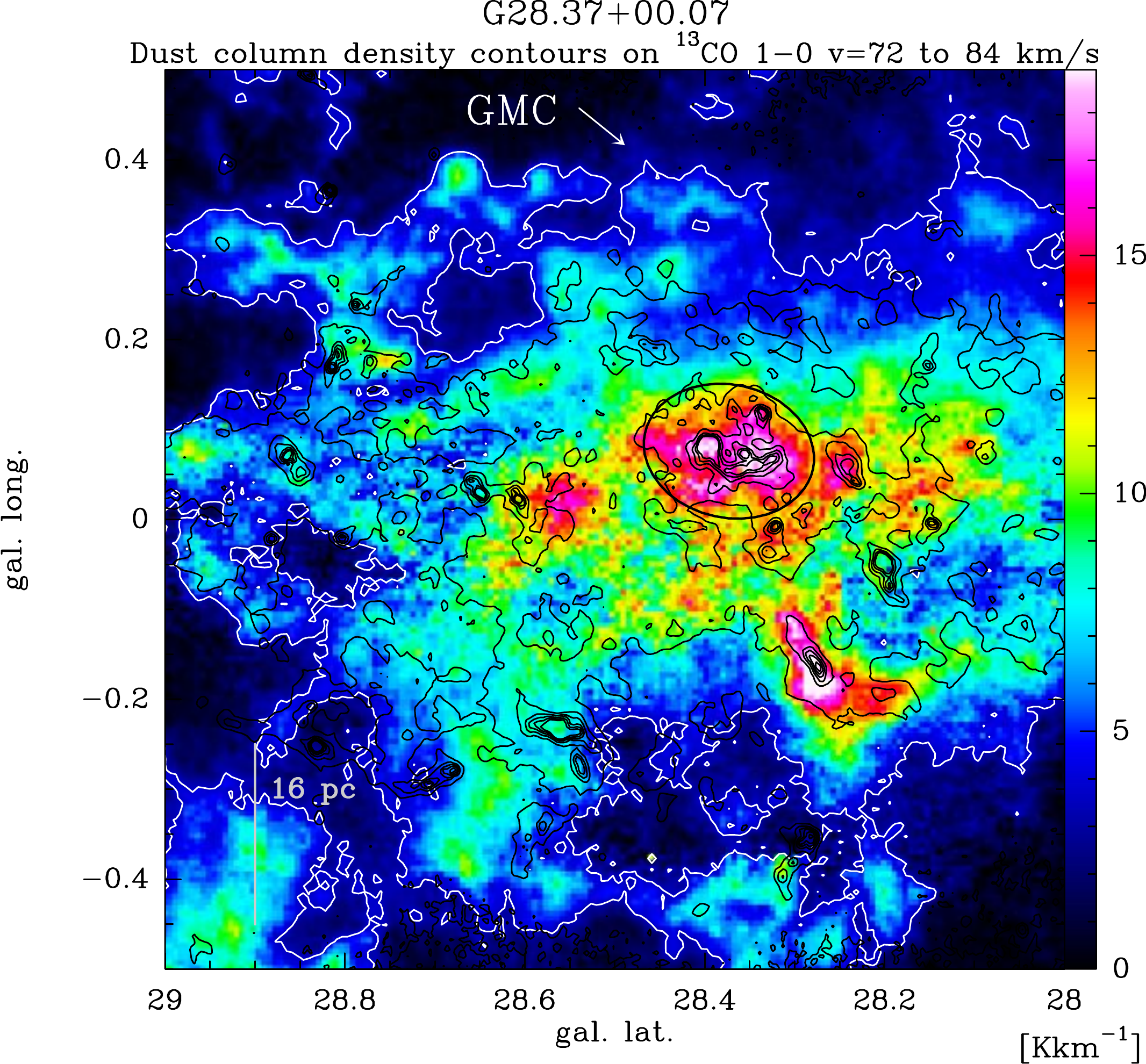}   
\includegraphics [width=8cm, angle={0}]{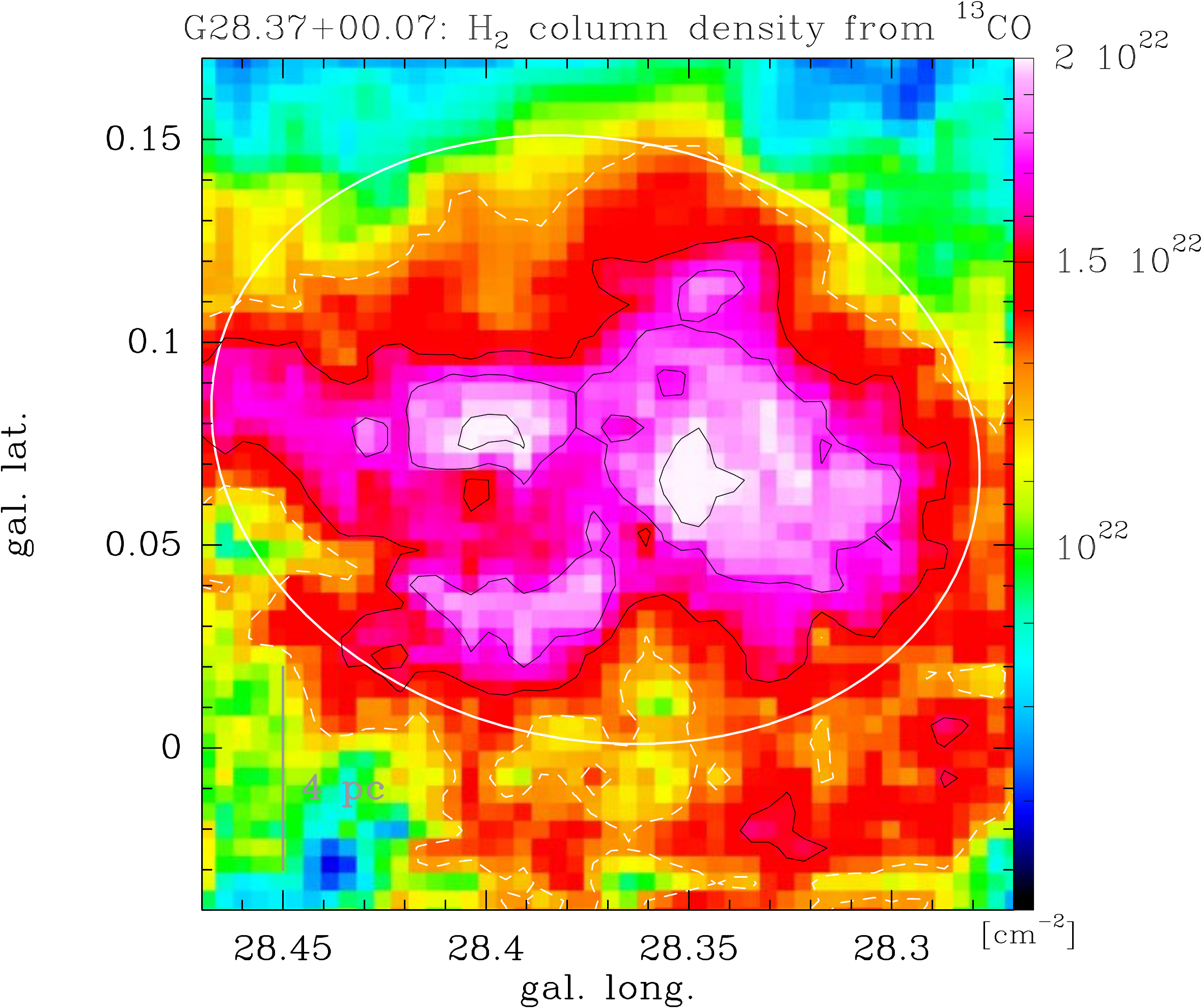}   
\caption[] {{\bf Left:} line integrated $^{13}$CO 1$\to$0 emission in
  colour scale between v=72 and 84 km s$^{-1}$, the velocity range of
  the bulk emission of the IRDC ,and the associated GMC. The white
  contour indicates the 3.5 K kms$^{-1}$ level, corresponding to
  \av=2, which defines the extent of the GMC and was used for mass
  estimation. The {\sl Herschel} column density is overlaid as black
  contours (levels 2 to 6$\times$10$^{22}$ cm$^{-2}$ in steps of
  10$^{22}$ cm$^{-2}$) and the IRDC is outlined by an ellipse (Simon
  et al.  2006b). {\bf Right:} H$_2$-column density map of the centre
  region of the GMC, obtained from $^{13}$CO 1$\to$0. The white-dashed
  contour indicates a column density of 1.25$\times$10$^{22}$
  cm$^{-2}$, the remaining black contours the levels 1.5, 1.75, and
  2$\times$10$^{22}$ cm$^{-2}$.}
\label{cloud-c}   
\end{centering}  
\end{figure*}   
 
\subsection{G28.37+0.07 in $^{13}$CO 1-0 emission} \label{co}  
\subsubsection{The IRDC as part of a GMC}  
To identify the exact velocity range of G28.37+0.07, we 
followed the method outlined in Simon et al. (2006b) using the GRS 
$^{13}$CO 1$\to$0 data to establish a morphological match between the 
{\sl Herschel} column density map and $^{13}$CO emission.  Simon et 
al.  (2006b) already listed this source with a velocity (v$_{lsr}$) of 
78.6 km s$^{-1}$ and a velocity dispersion of 8.3 km s$^{-1}$. 
Figure~\ref{cloud-c} (left) shows the large-scale velocity integrated 
(72 to 84 km s$^{-1}$) $^{13}$CO map with the dust column density 
overlaid as contours. We define the 'border' of the GMC by the \av=2 
level (see Sec.~\ref{mass}). The right panel of Fig.~\ref{cloud-c} 
shows a zoom into one of the cloud peak emission regions that include 
the IRDC. This map is given in H$_2$-column density, determined from 
the $^{13}$CO 1$\to$0 map, and can be compared to the {\sl Herschel} 
H$_2$-map calculated from dust emission 
(Fig.~\ref{herschel2-cloud-c}). 
 
The overall correspondence on the GMC scale between $^{13}$CO and dust
emission is good.  The two most prominent peaks in $^{13}$CO emission
correlate well with contours of dust column density, indicating that
most of the dust seen with {\sl Herschel} indeed comes from gas of the
molecular cloud at that velocity range i.e. the bulk emission of the
cloud. All $^{13}$CO emission outside this velocity range is due to
diffuse emission and/or clouds along the line-of-sight. We estimate
this contribution to sum up to a column density of
$\sim$7$\times$10$^{21}$ cm$^{-2}$ (Sec.~\ref{los}). On the IRDC
scale, a one-to-one correspondence between H$_2$ from $^{13}$CO and
dust is not expected because $^{13}$CO can become optically thick and
no longer traces the coldest and highest density regions.  However,
the maps agree well and the prominent peak, which is also seen at 500
and 870 $\mu$m emission and column density
(Fig.~\ref{herschel-cloud-c}), shows up as a peak in $^{13}$CO (and
H$_2$) in Fig.~\ref{cloud-c}.  This correlation, and the fact that
this peak has the same velocity as seen in N$_2$H$^+$ 1$\to$0
(Tackenberg et al.  \cite{tack2014}), provides evidence that this high
column density clump is clearly an intrinsic part of the IRDC.
 
A more quantitative comparison between the H$_2$-maps from $^{13}$CO
and dust shows that both maps display a gradient of decreasing columm
density, down to \av$\sim$15 for the map obtained from $^{13}$CO and
\av$\sim$30 for the {\sl Herschel} map.  Considering a line-of-sight
confusion for the {\sl Herschel} map of around \av =7
(Sec.~\ref{los}), and the large uncertainties for both methods in
determining H$_2$ column density maps, the agreement is rather good.
However, this finding is in contrast with the extinction map of BTK,
which shows a much steeper gradient in extinction going down to values
of \av=3. In our map (H$_2$ from $^{13}$CO), this low value is only
found at the borders of the whole GMC.
 
In summary, we have shown, using the $^{13}$CO map, that G28.37+0.07
is a part of a much larger molecular cloud complex.  It constitutes
the cloud centre region with highest (column) densities.  We obtain
the same results for the other clouds in this study.  Appendix A shows
a $^{12}$CO 3$\to$2 map of G11.11-0.12 and $^{13}$CO 1$\to$0 maps of
G18.82-00.28 and G28.53-00.25. Because of the spatial limit of the
COHRS $^{12}$CO survey, the GMC associated with G11.11-0.12 (Fig. A.2)
is not fully covered towards lower galactic latitudes. However,
G11.11-0.12 is clearly not an isolated feature but forms a dense
region within this GMC that has a mass of at least 0.5$\times$10$^5$
M$_\odot$ (Table 1).  The IRDCs G18.82-00.28 and G28.53-00.25 (Figs.
A.4 and A.6) are part of more massive clouds ($\sim$1.5$\times$10$^5$
M$_\odot$) that were fully included in the GRS $^{13}$CO survey.  Our
results are thus in line with earlier findings (e.g. Carey et al.
\cite{carey2000}, Simon et al.  \cite{simon2006a}, Teyssier et al.
\cite{teyssier2002}), which see many IRDCs as an intrinsic part of a
larger molecular cloud.
 
\subsubsection{Mass determination and link to IR-bright GMCs} \label{mass}  
 
For the mass determination from $^{13}$CO, we define the extent of the
GMC above an \av\ level of 2 mag. This threshold is commonly used
(e.g. Lada et al. \cite{lada2010}, Heiderman et al.
\cite{heiderman2010}). To be consistent with Simon et al. (2001,
2006a), we employ an N(H$_2$)/N($^{13}$CO) conversion factor of
4.92$\times$10$^5$ (see Simon et al. \cite{simon2001} for further
details).  For the excitation temperature, we take a value of 15 K.
This is a compromise between the average dust temperature of 21 K,
determined from the map shown in Fig.~\ref{herschel2-cloud-c}, and the
excitation temperature obtained from the $^{12}$CO 3$\to$2
data\footnote{$T_{\rm ex} = 16.6 \times [\ln(16.6/T_{\rm mb}({\rm
    CO32})+0.036)]^{-1}[K]$}. From $^{12}$CO, we dervive $T_{\rm
  ex}$=11 (16, 22) K for an observed main beam temperature of $T_{\rm
  mb}$= 4 (6, 8) K (see Fig.~\ref{spectra-cloud-c}). The dust
temperature is most likely overestimated by a few degrees (Peretto et
al.  \cite{peretto2013}), and it is not clear how well the dust is
mixed with the gas. Taking T$_{ex}$=15 K seem to be reasonable choice
and leads to a value of $\sim$3.5 K kms$^{-1}$ for the line integrated
$^{13}$CO emission for \av=2.  The total mass of the GMC above this
level (indicated by a white contour in Fig.~\ref{cloud-c}) is then
9.0$\times$10$^6$ M$_\odot$, and the equivalent radius $\sim$40 pc.
 
The IRDC G28.37+0.07 occupies only a small spatial fraction within the
GMC. From the $^{13}$CO measurements, we are consistent with Simon et
al. (2006a) and find a peak H$_2$ column density of
16(21)$\times$10$^{21}$ cm$^{-2}$ for T$_{ex}$=15(20) K, and a mass of
6.3$\times$10$^4$ M$_\odot$. From the dust continuum, we obtain higher
values of average and peak column density i.e. $\langle N_{dust}
\rangle \sim$4$\times$10$^{22}$ cm$^{-2}$ and
$N_{peak}\sim$3$\times$10$^{23}$ cm$^{-2}$, respectively, and a high
surface density ($\Sigma_{dust} \sim$740 M$_\odot$ pc$^{-2}$).  The
surface density ($\Sigma_{dust}=M_{dust}/A$) refers to the area $A$
within the ellipse defining the IRDC.
 
The values of column densities and surface densities for the other 
IRDCs of our sample are also high i.e. $\langle N_{dust} 
\rangle\sim$2.5$\times$10$^{22}$ cm$^{-2}$ and $\Sigma_{dust} >$400 
M$_\odot$ pc$^{-2}$. These massive IRDCs may thus correspond to {\sl 
  \textup{ridges}} that were defined as the central regions within a GMC with 
column densities above $\sim$10$^{22-23}$ cm$^{-2}$ at temperatures 
below 20 K for the IR-bright cloud Vela C (Hill et al. 
\cite{hill2011}) and the DR21 ridge\footnote{Note that the DR21 ridge 
  itself is an IRDC (Marston et al.  \cite{marston2004}), and that the 
  IRDC G035.39-0.33 in W48 was already qualified as a {\sl ridge} 
  (Nguyen-Luong et al.  \cite{quang2011}) with the same physical 
  properties.}  (Schneider et al.  2010, Hennemann et al.  2012). 
 
%%%%%%%%%%%%%%%%%%%%%%%%%%%%%%   
%%%  PDF IRDC cloud_c %%%%%%%%%%%%%%%   
\begin{figure}[!htpb]   
\begin{centering}   
\includegraphics [width=8cm, angle={90}]{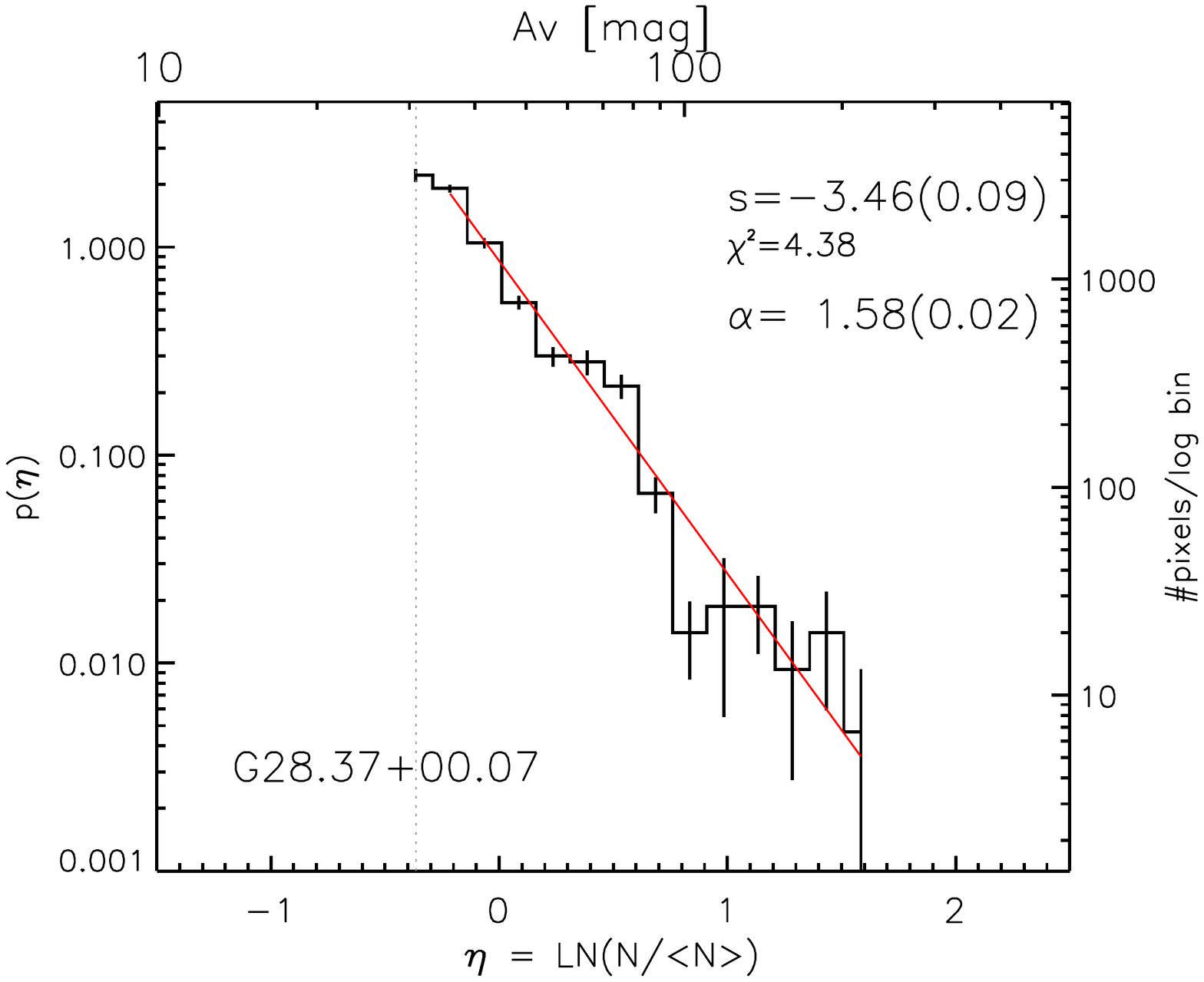}   
 
\vspace{0.6cm} 
\includegraphics [width=8cm, angle={90}]{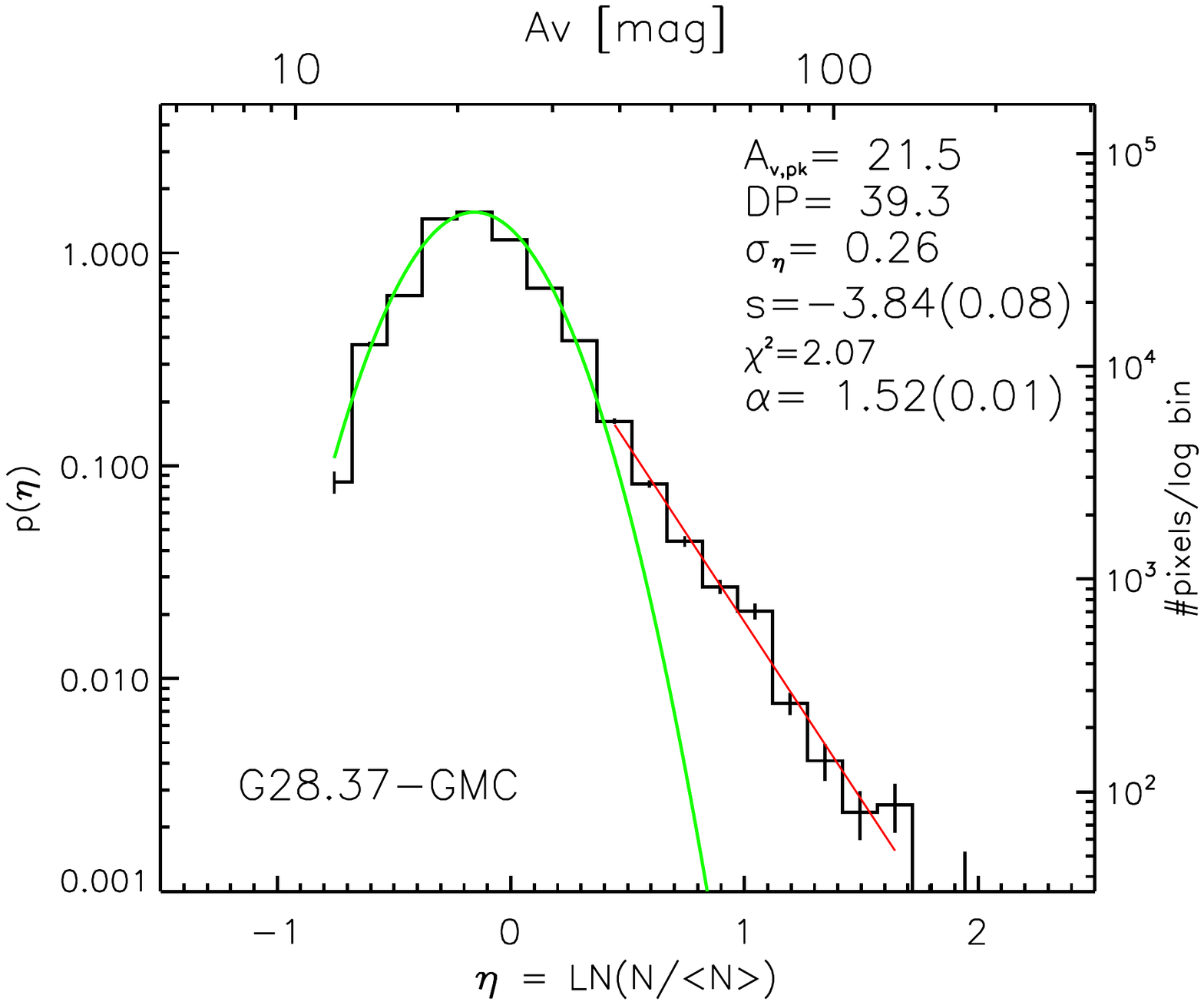}   
\caption[] {PDFs of G28.37+0.07 and its associated GMC, derived from 
  the {\sl Herschel} column density map. The upper panel shows the PDF 
  of the IRDC, only from the pixels inside the ellipse shown in 
  Fig.~2. The lower panel shows the PDF obtained for the whole GMC (all 
  pixels inside the contour indicated in Fig.~3, right).  The 
  error bars were calculated using Poisson statistics.  The left 
  y-axis gives the normalised probability $p(\eta)$ and the right y-axis indicates 
  the number of pixels per log bin. The upper x-axis is the visual 
  extinction and the lower x-axis is the logarithm of the normalised 
  column density.  The green curve shows a log-normal fit to the PDF 
  from the GMC. The red line indicates a power-law fit to the high 
  column density tail with the slope $s$ together with its error and 
  the reduced $X^2$ goodness-of-fit. The exponent $\alpha$ of an 
  equivalent spherical density distribution $ \rho(r) \propto 
  r^{-\alpha}$ is also indicated in the panel.  The dashed line in the 
  upper panel indicates the completeness level; the PDF left of this 
  line is incomplete.} 
\label{pdf-cloud-c}   
\end{centering}  
\end{figure}   
 
\subsubsection{Line-of-sight confusion towards G28.37+0.07} \label{los}  
   
We estimate the $^{13}$CO emission along the line-of-sight to assess
how much of H$_2$ column density may arise from clouds and diffuse
emission not related to our target cloud.  The average H$_2$-column
density, using the same conversion factor N(H$_2$)/N($^{13}$CO) of
4.92$\times$10$^5$ as for the determination of the masses of the GMC,
in the velocity ranges {\sl \textup{outside}} the bulk emission of the
IRDC and within the ellipse is $\langle N _{cont}
\rangle$=7$\times$10$^{21}$ cm$^{-2}$. The H$_2$ column density
$\langle N_{dust} \rangle$ determined from the {\sl Herschel} map is
$\sim$40$\times$10$^{21}$ cm$^{-2}$ and that from $^{13}$CO is
$\langle N_{bulk} \rangle \sim$16$\times$10$^{21}$ cm$^{-2}$. However,
if $^{13}$CO becomes optically thick and/or freezes out in the highest
density regions, it underestimates the H$_2$ column density. On the
other hand, the {\sl Herschel} column density map is affected by
line-of-sight contamination and thus overestimates the H$_2$ column
density.  Considering these effects and the uncertainties of the
methods, a factor of 2 disagreement between the column density
determined from CO and dust is acceptable.
 
\section{Probability distribution functions  of IRDCs} \label{pdf}    
 
\subsection{Do IRDCs have a log-normal or power-law PDF ?} \label{pdf-detail} 
 
\subsubsection{PDFs from dust continuum} \label{pdf-dust} 
 
The PDFs\footnote{We define $\eta\equiv \ln(N/ \langle N \rangle)$ as
  the natural logarithm of the column density $N$, divided by the mean
  column density $\langle N \rangle$, and the quantity $p_\eta(\eta)$
  then corresponds to the PDF of $\eta$ with the normalisation
  $\int_{-\infty}^{+\infty} p_\eta d\eta=\int_0^{+\infty} p_N\, dN=1$
  (see Schneider et al. 2015 for details).} of dust column density for
the IRDC G28.37+0.07 and its associated GMC are shown in
Fig.~\ref{pdf-cloud-c}, those for G11.11-0.12, G18.82-0.28, and
G28.53-0.25 are listed in Appendix B. To be consistent with other
studies, we express the PDF in visual extinction \av\ using the
conversion $N(H_2)$/\av=0.94$\times$10$^{21}$ cm$^{-2}$ mag$^{-1}$
(Bohlin et al.  1978).
 
All PDFs are sampled with a bin size of 0.15 (in $\eta$) using maps of
a grid of 14$''$. In Schneider et al. (2015), we investigated the
effect of different bin sizes on the PDF and found that the PDF
properties, and in particular the slope, do not change significantly
using different bins. As shown in their Fig. A.1, the best compromise
between high sampling and resolution lies for bin sizes of 0.1 to 0.2.
We thus decided to adopt a bin size of 0.15. The PDFs are constructed
from pixels above the approximate completeness limit within the
ellipses defining the IRDC.  This level was obtained from the column
density maps (e.g., Fig.~\ref{herschel2-cloud-c}) where we determined
the lowest contour level (indicated by a white dashed line) that is
still continuous.  That implies that above this contour, the majority
of pixels is still found within the ellipse defining the IRDC, and
below this contour, the map is incomplete.  Accordingly, for lower
column densities, the pixel distribution is also not complete and can
create an artificial fall-off of the PDF, which should not be confused
with a log-normal PDF.  The PDF of the IRDC alone obtained in this
way, as well as the PDFs for the other IRDC clouds (see Fig. B.1), is
consistent with a pure {\sl \textup{power-law distribution}}. In
contrast, the PDF of the whole GMC (Fig.~\ref{pdf-cloud-c}, bottom) in
which the IRDC is embedded indeed shows a log-normal part plus a
power-law tail, similar to what was found for star-forming clouds
(e.g. Kainulainen et al. 2011b, Schneider et al. 2013).
 
To exclude a possible log-normal distribution for our observed PDF of
the IRDC, we performed a two-sample Kolmogorow-Smirnov (KS)-test,
comparing our distribution with a set of pure log-normal distributions
with different widths ($\sigma$=0.2 to 1.7) and normalisations (see
Appendix C for details). The probabilities $p$ in all cases of broad
PDFS are very low so we exclude the possibility that our observed PDF
is purely log-normal.  We also made a KS-test using our PDF of the
IRDC (Fig.~\ref{pdf-cloud-c}, top), but including the data points left
of the completeness level and ignoring the data points above
\av\,$\sim$100.  The two-sample KS-test with the synthetic PDFs also 
shows small values for $p$ (see Table C.1).  The p-value for a
PDF with $\sigma$=1.7 (fitted by BTK) is 2.2$\times$10$^{-6}$ (or
4.7$\times$10$^{-4}$, depending on normalisation). The best-fitting
log-normal has a width of $\sigma$=0.4 (p-value = 0.25 (0.28)), which
implies that the best-fit might be consistent at the 2-sigma level.
However, the PDF is too narrow to be consistent with what was found in
BTK.
  
We performed a linear regression fitting to the power-law part of the 
PDF. The starting point was chosen as the break point of the PDF 
between log-normal and power-law seen in the PDF of the GMC.  For 
example, for G28.37+0.07 (Fig.~\ref{pdf-cloud-c}), the break occurs at 
\av $\sim$40 and we thus started the fitting for the IRDC at this 
value. The reduced $\chi^2$ values of the power-law fits scatter, but 
are all above one and thus indicate that the general assumption of a 
power-law dependence is justified. 
 
In Appendix B, we show the PDFs of the other IRDCs (alone and 
including the embedding GMC). The result is the same as for 
G28.37+0.07 i.e. a log-normal plus power-law tail for the GMC and a 
power-law distribution for the IRDC. For G28.53-0.25, there is an 
indication for a log-normal distribution below \av\ $\sim$30. However,  the PDFs of the GMC are less reliable because effects of 
line-of-sight contamination become more important and can lead to a 
change in slope for the power-law tail (Schneider et al. 2015). 
 
%%%  PDF 13CO %%%%%%%%%%%%%%%   
\begin{figure}[!htpb]   
\begin{centering}   
\includegraphics [width=6cm, angle={0}]{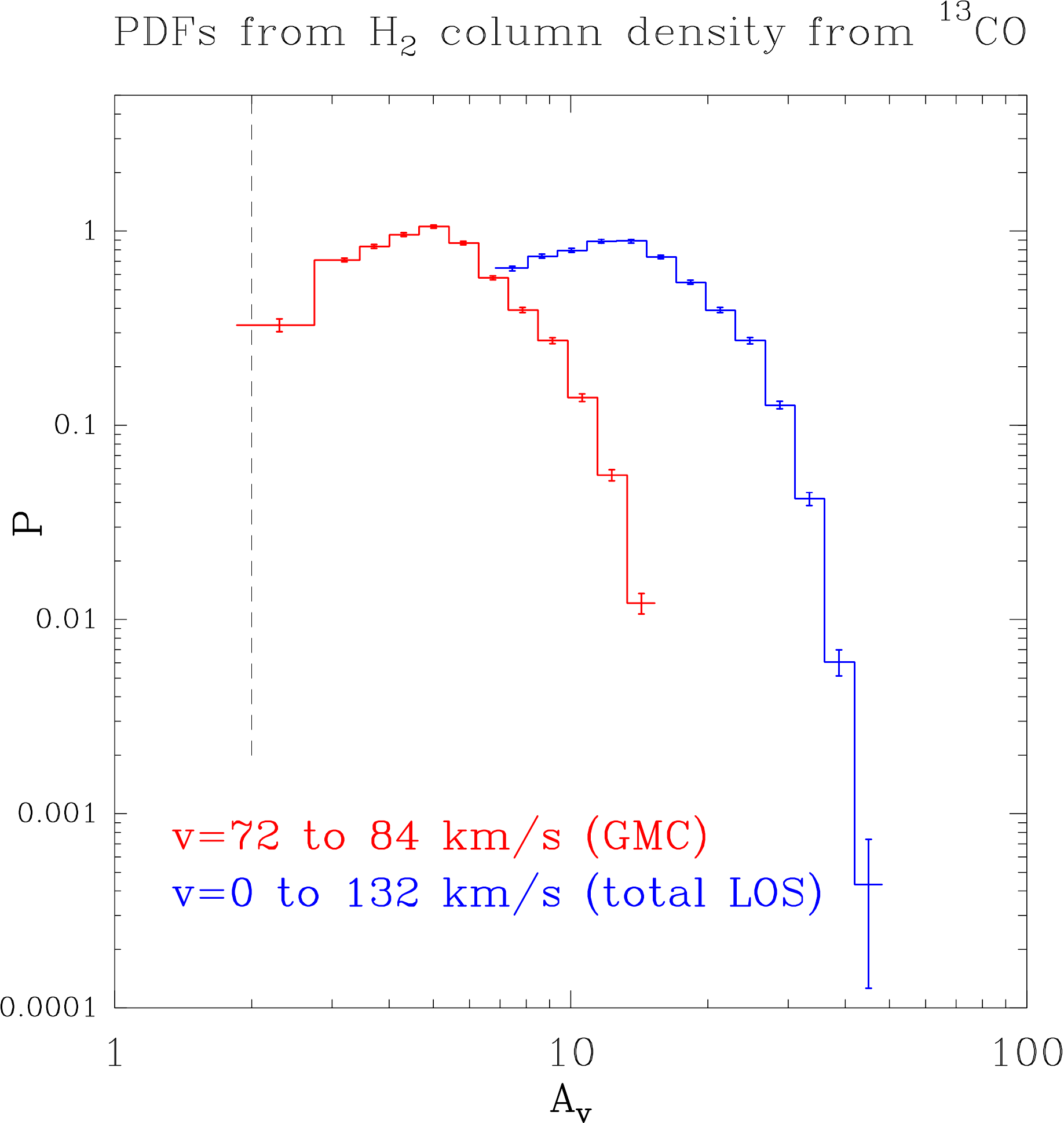}   
\caption[] {PDFs of the whole GMC including G28.37+0.07, obtained from 
  H$_2$ column density maps derived from $^{13}$CO emission.  The blue 
  PDF contains all emission along the line-of-sight (LOS) towards the 
  GMC (v=0 to 132 km s$^{-1}$); the red PDF contains  only emission coming from 
  the bulk emission of the cloud (v=72 to 84 km s$^{-1}$). The long 
  dashed line is the \av = 2 level outlining the GMC.} 
\label{pdf13co}   
\end{centering}  
\end{figure}   
 
\subsubsection{Probability distribution functions from $^{13}$CO} \label{pdf-co} 
 
Figure~\ref{pdf13co} shows PDFs obtained from the $^{13}$CO data.  The
red PDF is constructed from the map displayed in Fig.~\ref{cloud-c}
and contains only emission from the GMC in the velocity range 72 to 84
km s$^{-1}$. The PDF is similar to a log-normal but because it is
cut-off at higher column densities (around \av = 10) most likely
because of an increasing optical depth, we refrain from trying to fit
the data.  In addition, the highest column density range can not be
traced by the $^{13}$CO line because its critical density is below the
densities of the clumps or cores (around 10$^{4-6}$ cm$^{-3}$) that
populate the high \av-range.  The blue PDF includes all emission along
the line-of sight towards the GMC between velocities of 0 to 132 km
s$^{-1}$. This PDF is shifted by \av $\sim$8--10 towards higher column
density values compared to the red PDF. Though the $^{13}$CO line
still saturates in each individual cloud along the sightline above \av
$\sim$10, the total H$_2$-PDF can still go up to higher values. These
findings confirm our direct calculation of the contaminating column
density (Sec.~\ref{los}). The blue PDF corresponds very well over the
whole \av-range (10--40) with the PDF obtained from the dust column
density (Fig.~\ref{pdf-cloud-c}), indicating that the conversions
H$_2$ from $^{13}$CO and H$_2$ from dust are rather consistent.
 
%%%  PDF IRDC - scales %%%%%%%%%%%%%%%   
\begin{figure}[!htpb]   
\begin{centering}   
\includegraphics [width=6cm, angle={0}]{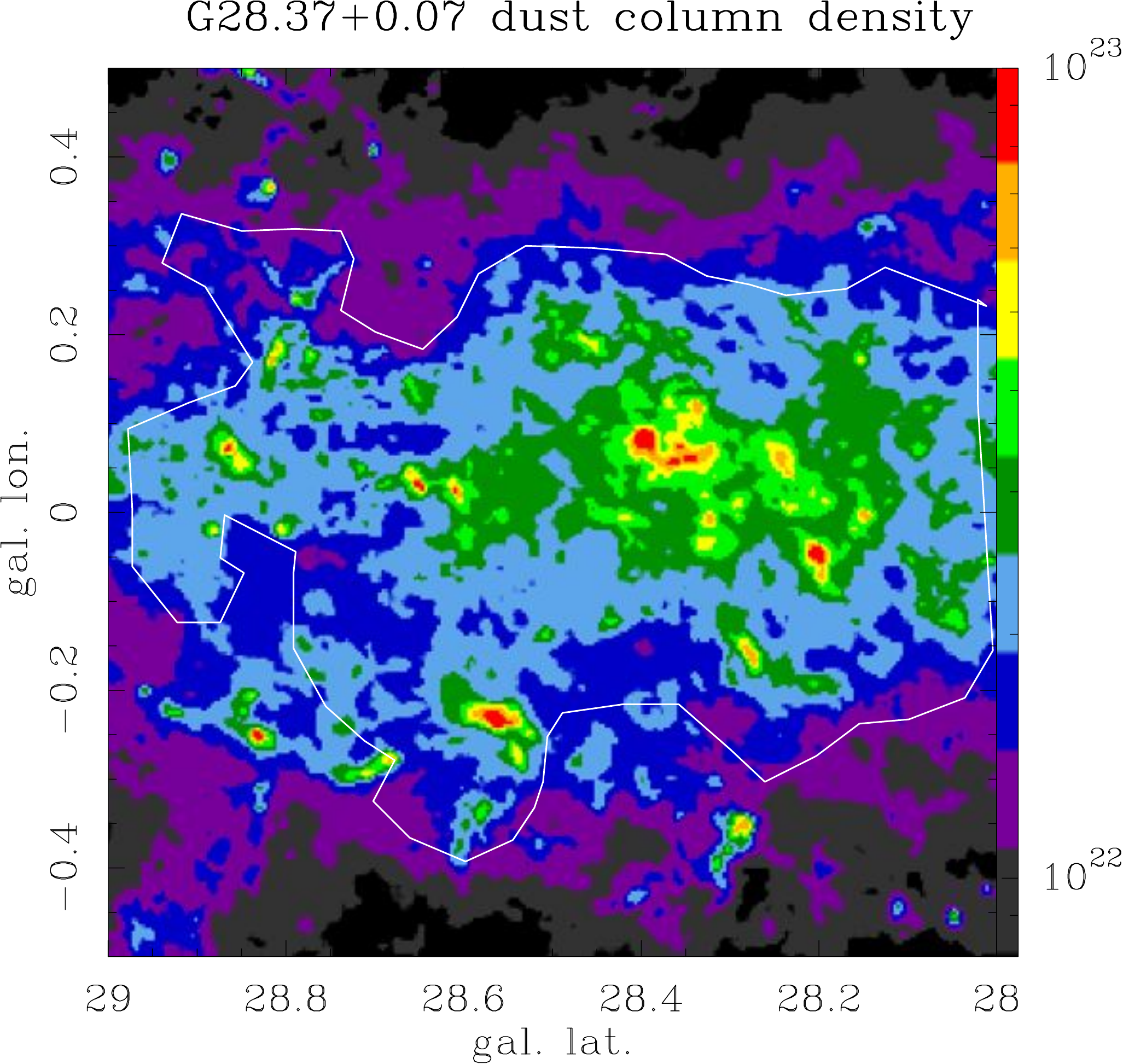}   
\includegraphics [width=5.5cm, angle={0}]{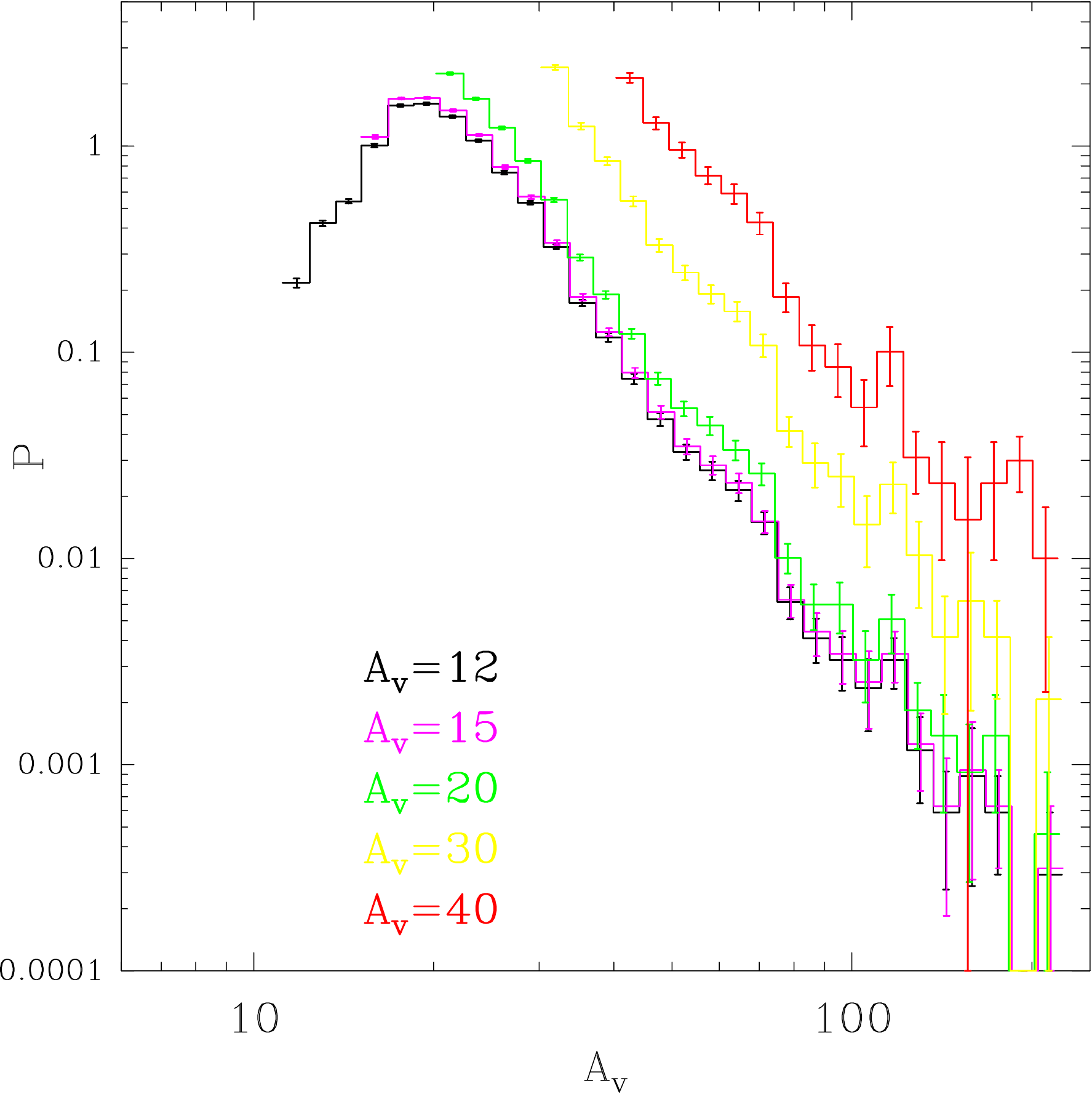}   
\caption[] {{\bf Top:} H$_2$ column density from dust continuum 
    with the white contour, roughly outlining the GMC in 
    which G28.37+0.07 is embedded.  {\bf Bottom:} PDFs of the GMC, 
    derived from the dust column density map from the left above 
    different \av\,-thresholds (indicated with different colours and 
    given in the panel).} 
\label{allpdfs}   
\end{centering}  
\end{figure}   
 
\section{Clipping effects and completeness level} \label{pdf-more} 
 
Comparing the PDFs shown in Fig.~\ref{pdf-cloud-c} reveals that the
PDF constructed only from the pixels defining the IRDC has no
log-normal part in contrast to that obtained for the whole GMC.  The
reason is that the PDF of the IRDC alone is not sampled down to the
lowest \av-ranges and a PDF from an image where the low column density
pixels are ignored is just composed of a power-law distribution
(Schneider et al. 2015).
 
To illustrate this cropping effect, we made PDFs from pixels above
different \av\,-thresholds for the whole GMC (Fig.~\ref{cloud-c})
using the Herschel dust column density map (see right panel of
Fig.~\ref{cloud-c}).  Figure~\ref{allpdfs} shows how the PDF changes
from a distribution that can be described by a log-normal plus
power-law distribution (for pixels above \av\,=12) into a purely
power-law distribution (for pixels above \av\,=20).  Note that the
curves 'shift' due to different normalisations (the area/number of
pixels decreases with increasing \av\,-level). We emphasise that the
contour outlining the GMC still comprises higher column density gas
(\av\, $>$10), which is well above the threshold of typically \av\,=1
or 2 (e.g. Lada et al.  \cite{lada2010}) that is commonly used to
'define' the extent of a molecular cloud. The dust continuum map,
however, suffers from line-of-sight contamination on a high level of
at least \av\,$\sim$7--10 because (i) this is the value at the map
borders outside of the GMC (Fig.~\ref{allpdfs}), and (ii) it also
corresponds to the H$_2$ column density derived from $^{13}$CO adding
all emission outside of the bulk emission of the GMC (Sec.~\ref{co}).
 
In any case, the main point of this excercise is to demonstrate that
the PDF of the IRDC (Fig.~\ref{pdf-cloud-c}, top) can be fitted by a
pure power law\footnote{Note that fitting to binned data is not always
  statistically robust as discussed in Virkar \& Clauset
  (\cite{vikar2014}). However, we make PDFs of $\eta$=ln($N/\langle N
  \rangle$) in linear binning with bin size d$\eta$=0.15, while Virkar
  \& Clauset use log-binning (of $N/\langle N \rangle$ when applied to
  our case).}  above \av\,-values of $\sim$30.  This PDF can be
directly compared to that shown in BTK (their Fig.~3, left panel),
which shows a broad PDF fitted by a log-normal distribution with a
width of $\sigma$=1.7.  The two PDFs differ in the lowest and highest
column density ranges. The authors BTK state a completeness limit of
\av = 3, implying that their extinction map of the IRDC goes down to
that value at the borders\footnote{In Kainulainen \& Tan
  (\cite{kai2013}) the same map of G28.37+0.07 was produced with a
  slightly different method and a value of \av = 7 was given for the
  same completeness level, better fitting with our findings.}. As
outlined in Sec.~\ref{co}, we show that the IRDC is the centre region
of a GMC and the column density values remain high (at least on a
level of \av $\sim$10 to 20) at the borders. Though the higher angular
resolution extinction map may resolve smaller spatial structures in
which high and low column density regions are mixed, the border
regions look homogeneous at a low \av\ value.  In addition, the
highest column density pixels (above \av\, $\sim$100-200) are missing
in BTK because they were masked out.  However, if we compare only the
\av\ range between $\sim$30 to $\sim$100 (which we find to be the
column density range of the IRDC), the PDF of BTK could be well fit by
a power-law distribution and our results would be consistent.
 
\section{Global and local gravitational collapse of IRDCs} \label{gravity}  
 
The power-law distribution we find for all PDFs of the IRDCs in our
sample is in accordance with the power-law tails found for IR-bright
clouds (Lombardi et al.  \cite{lombardi2008}; Kainulainen et al.
\cite{kai2009}; Hill et al. \cite{hill2011}; Schneider et al. 2012,
2013, 2015; Russeil et al. 2013; Tremblin et al.  \cite{tremblin2014};
Alves de Oliveira et al.  \cite{catarina2014}).  In these studies, a
clear turnover from a log-normal distribution for low column densities
(typically from \av\,$<$1 up to \av\ of a few magnitudes) into a
power-law tail was observed. Self-gravity as the dominating (over
pressure and magnetic fields) process to form this power-law tail was
advocated by numerical models (Klessen \cite{klessen2000}, Federrath
et al.  \cite{fed2008b}, Kritsuk et al.  \cite{kritsuk2011}) and
observationally supported by Froebrich \& Rowles
(\cite{froebrich2010}) and Schneider et al.  (\cite{schneider2013}).
Recently, Rathborne et al. (2014) obtained a PDF from ALMA continuum
data for the IRDC G0.253+0.016 in the central molecular zone (CMZ),
which has a log-normal shape and some excess at highest column
densities they attribute to self-gravity within the most massive
cores.  In the following, we will present additional evidence from
radial column density profiles and molecular line observations that
massive IRDCs can be in gravitational collapse, which is reflected in
the power-law distribution of the column density.
     
%%%%%%%%%%%%%%%%%%%%%%%%%%%%%%   
%%% Radial profile of Cloud D %%%%%%%%%%%%%%%   
\begin{figure}[!htpb]   
\begin{centering}   
\includegraphics [width=7cm, angle={00}]{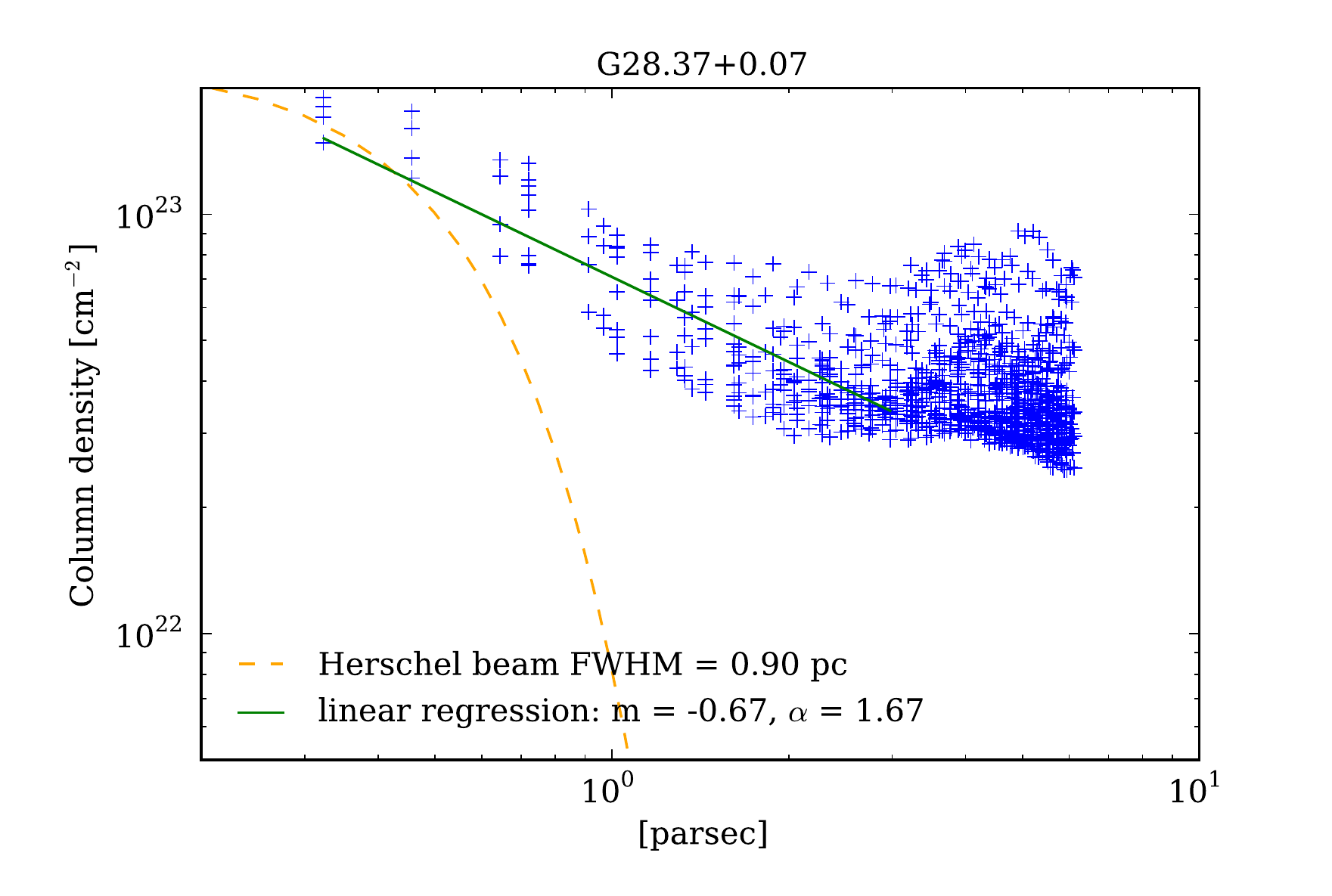}   
\includegraphics [width=7cm, angle={00}]{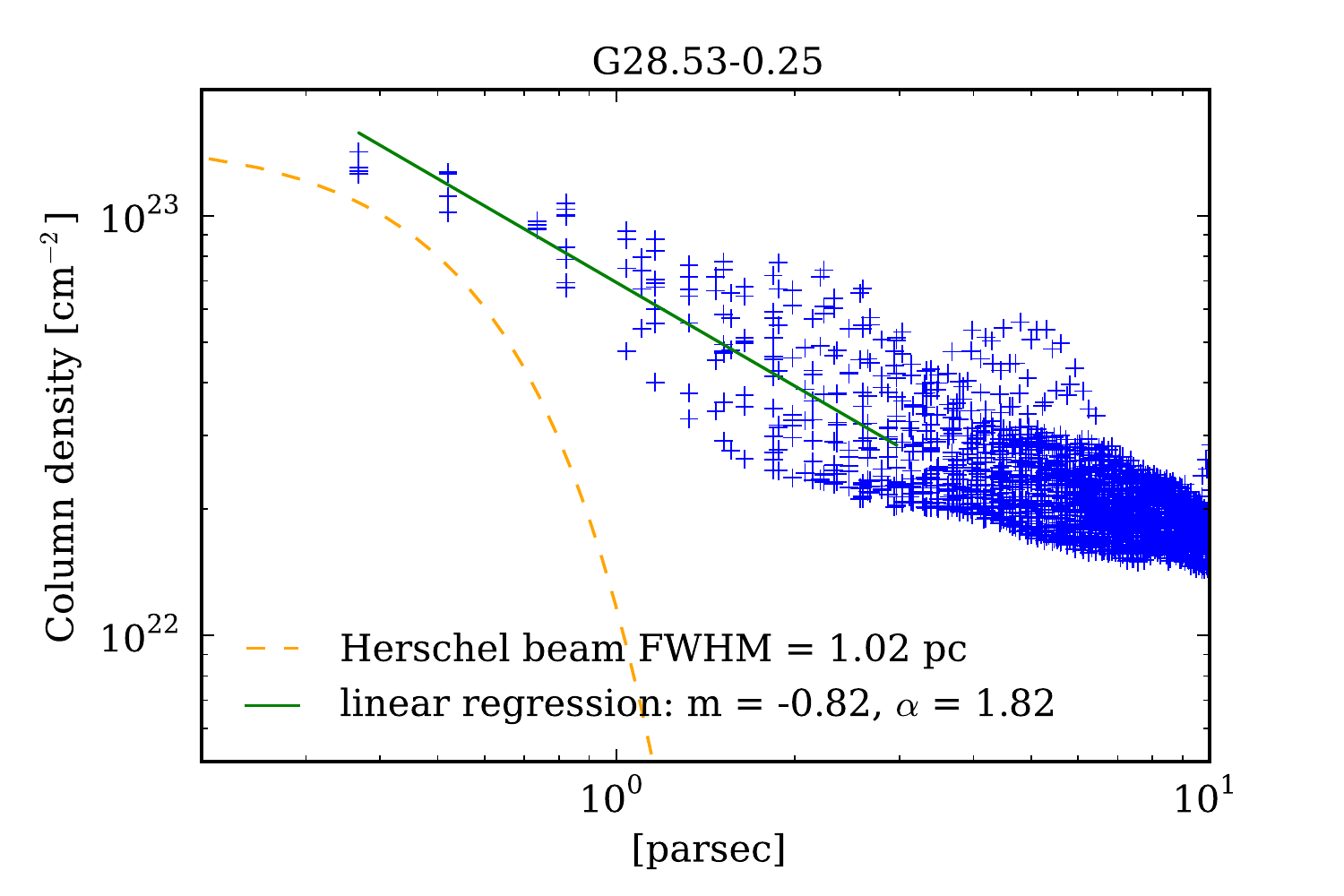}   
\includegraphics [width=7cm, angle={00}]{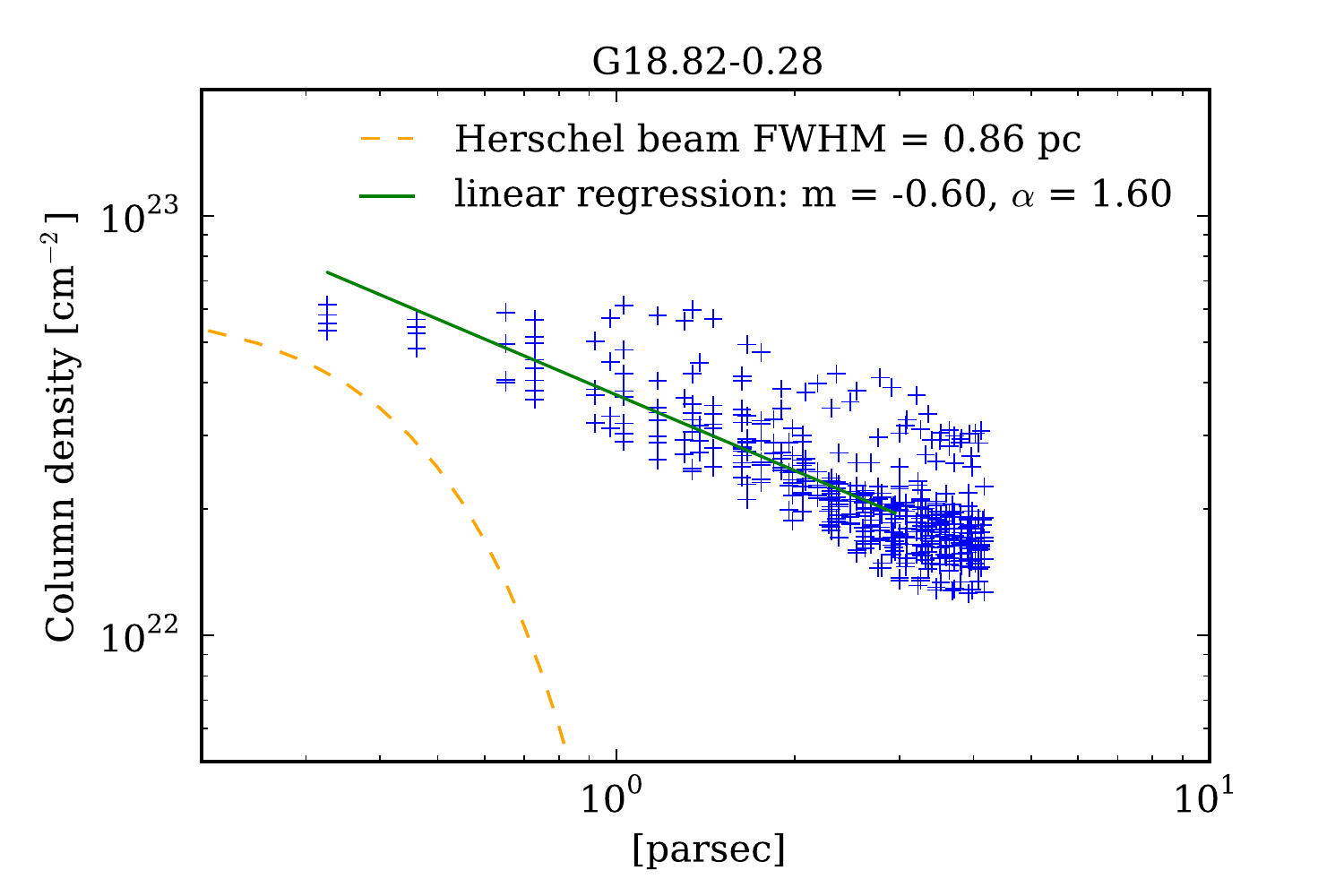}   
\caption[] {Radial column density profile for G28.37+0.07, 
  G28.53-0.25, and G18.82-0.28. Each cross in the panels represents 
  one pixel value in the map. The x-axis gives the distance to the 
  centre of the ellipse (x-axis).  The dashed orange line indicates 
  the Herschel beam (Gaussian profile with a FWHM of 36$''$ displayed 
  in log-log format). The exponent $\alpha$ was deduced from fitting 
  the column density profile ($N \propto r^{1-\alpha}$).  For the fit, 
  we included only the pixels that were inside a radius of 3 pc. } 
\label{profile-cloud-d}   
\end{centering}  
\end{figure}   
 
\begin{table}  \label{slopes} 
  \caption{Exponents $\alpha$ and their errors $\Delta\alpha$, assuming a spherical density distribution  
    $ \rho(r) \propto r^{-\alpha}$, derived from the slope $s$ of the power-law distribution from the  
    PDF (colum 2) and from the column density profiles (column 3).}     
\begin{center}   
\begin{tabular}{lccccc}   
\hline  
\hline    
 &    &  \\ 
Region            & $\alpha$(PDF)  &  $\Delta\alpha$(PDF) & $\alpha$(profile) & $\Delta\alpha$(profile)\\  
\hline    
{\sl G28.37+0.07} & 1.58  & 0.02 & 1.67 & 0.02 \\ 
{\sl G28.53-0.25} & 1.97  & 0.08 & 1.82 & 0.06 \\ 
{\sl G18.82-0.28} & 1.54  & 0.02 & 1.60 & 0.03 \\ 
\end{tabular} 
\end{center}   
\end{table}    
 
\subsection{The link between the PDF power-law tail and the column density profile}  
 
Provided that the power-law distribution of the PDF is only due to
gravity, and if we assume spherical symmetry, the power-law slope $s$
of the PDF is related to the exponent $\alpha$ of a radial density
profile according to $ \rho(r) \propto r^{-\alpha}$ and $\alpha=-2/s
+1$ (Federrath \& Klessen \cite{fed2013}). For the PDFs of the IRDCs
in our sample (Fig.~\ref{pdf-cloud-c} and \ref{pdf-irdcs}), we obtain
values between $\sim$1.5 and 2 for $\alpha$\footnote{Note that these
  are lower limits for $\alpha$ because the power-law slope steepens
  because of line-of-sight contamination (Schneider et al.  2015).}
This is consistent with a structure dominated by {\sl
  \textup{self-gravity}}, i.e.  local free-fall of individual cores
and clumps and global collapse (Girichidis et al. 2014, Schneider et
al. 2015).  The extent to which the variations in the slope can be
attributed to the evolutionary state of the cloud is not clear. As was
shown in simulations (Ballesteros-Paredes et al.  2011, Federrath \&
Klessen 2013), the power-law flattens with time/increasing star
formation efficiency.  In our small sample of IRDCs, however, the low
value of $\alpha$ points towards an early state in cloud evolution.
 
To derive a PDF-independent value for the exponent $\alpha$, we fitted
the radial column density profiles of G28.37+0.07, G28.53-0.25, and
G18.82-0.28 (Fig.~\ref{profile-cloud-d}).  From our sample, only these
sources can roughly be approximated as ``spherical'', G11.11-0.12 is
too filamentary. Assuming again spherical geometry, the column density
$N$ can be expressed as $N \propto \rho(r) \times r \propto
r^{1-\alpha}$, thus the exponents $\alpha$ from the PDF and the column
density profile should correspond.  Within the error bars, this is
approximately the case (see Table~\ref{slopes}), we obtain from the
column density profile $\alpha$=1.67, 1.82, 1.60 for G28.37+0.07,
G28.53-0.25, and G18.82-0.28, respectively, compared to $\alpha$=1.58,
1.97, 1.54, deduced from the PDF. Figure~\ref{profile-cloud-d}
displays the column density of the pixels around the maximum column
density as a function of their distance to this maximum. We only
fitted the points within a circle of 3 pc around the maximum column
density. The values of $\alpha$ between 1.6 and 1.8 determined from
the column density profiles are consistent with {\sl
  \textup{gravitational collapse on large scales}}.  In the next
section, we demonstrate that complementary molecular line data indeed
point towards this kind of a scenario for our IRDCs.
 
%%%%%%%%%%%%%%%%%%%%%%%%%%%%%%   
%%% Spectra of Cloud C %%%%%%%%%%%%%%%   
\begin{figure*}[!htpb]   
\begin{centering}   
\includegraphics [width=14cm, angle={00}]{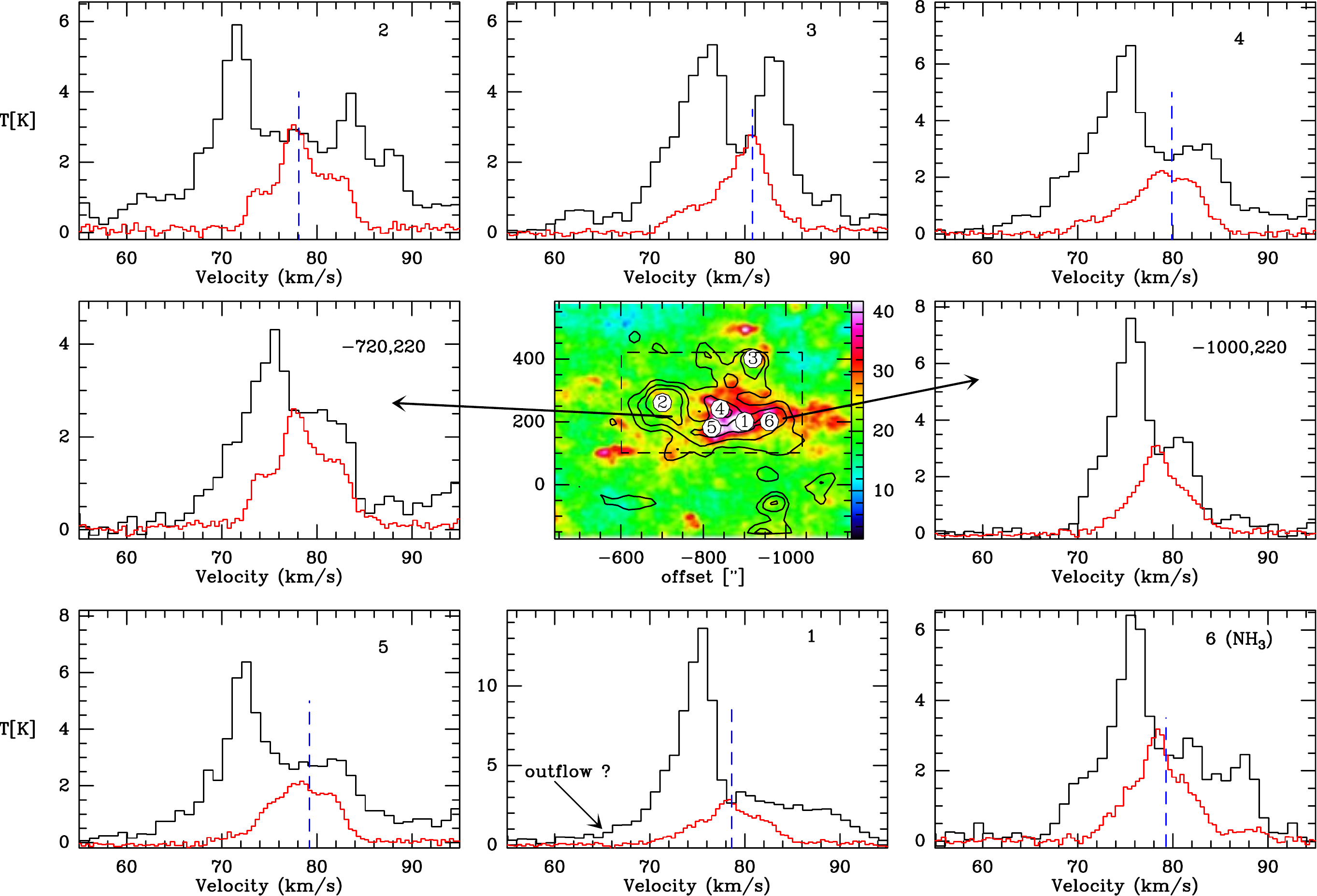}   
\caption[] {Middle panel shows in colour the line integrated $^{12}$CO
  3$\to$2 emission in K kms$^{-1}$ on a main beam temperature scale
  (between 72 and 84 km s$^{-1}$) of G28.37+0.07.  The dust column
  density from {\sl Herschel} is overlaid as black contours (levels 4,
  5, 7 10$^{22}$ cm$^{-2}$).  The dashed lines outline the area
  displayed in Fig.~\ref{allspectra}, and the numbering from 1 to 6
  indicates the position of submm-continuum sources seen with ATLASGAL
  and subsequently observed in N$_2$H$^+$ (Tackenberg et al. 2014).
  The $^{12}$CO 3$\to$2 (black) and $^{13}$CO 1$\to$0 (red) spectra at
  these positions are displayed in the panels around. Additionally, we
  show spectra from two positions (indicated by arrows) off the
  sources but still within the molecular cloud.  The blue-dashed line
  denotes the centre velocity of the N$_2$H$^+$ line.}
\label{spectra-cloud-c}   
\end{centering}  
\end{figure*}   
 
%%%%%%%%%%%%%%%%%%%%%%%%%%%%%%   
%%% global infall spectra of Cloud C %%%%%%%%%%%%%%%   
\begin{figure*}[!htpb]   
\begin{centering}   
\includegraphics [width=13cm, angle={00}]{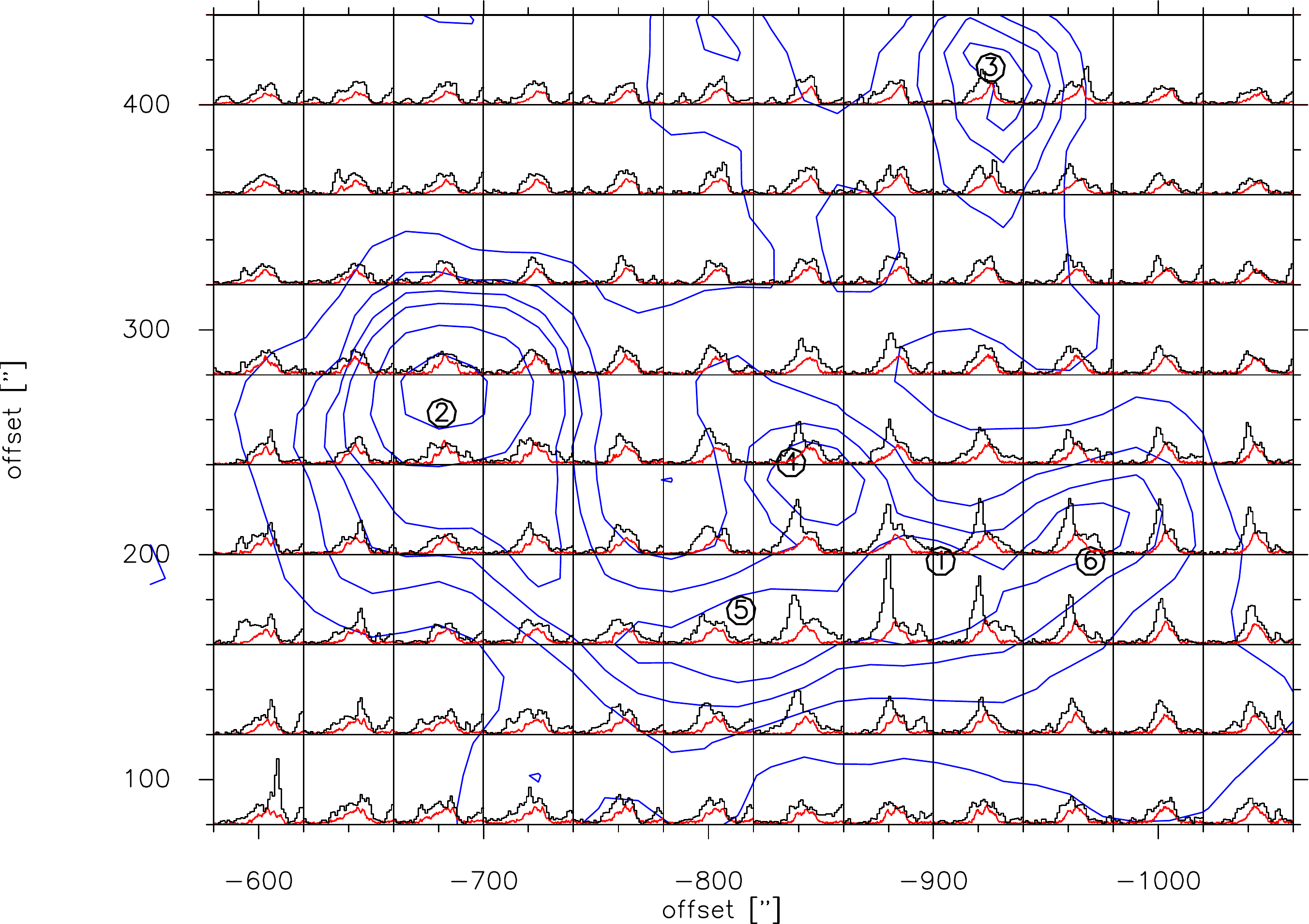}   
\caption[] {Main beam brightness temperature spectra of $^{12}$CO
  3$\to$2 (black) and $^{13}$CO 1$\to$0 (red) emission in the velocity
  range 55 to 95 km s$^{-1}$ and temperature range -0.2 to 12 K. The
  area corresponds to the centre region of the IRDC G28.37+0.07,
  outlined in Fig.~\ref{spectra-cloud-c}.  The dust column density
  from {\sl Herschel} is overlaid as blue contours (levels 4, 5, 7,
  10, 16 10$^{22}$ cm$^{-2}$), and the numbering from 1 to 6 indicates
  the position of submm-continuum sources labelled using ATLASGAL and
  subsequently observed in N$_2$H$^+$ (Tackenberg et al. 2014).}
\label{allspectra}   
\end{centering}  
\end{figure*}   
 
\subsection{Collapse signatures in molecular line profiles}  
 
The shape of molecular line profiles allows us to disentangle various
dynamic processes, such as outflows, rotation, and infall, within a
molecular cloud/clump/core.  The inwards motion of gas caused by
gravity in regions of star formation leads to a self-absorbed emission
line profile ('double-peak' or 'P-Cygni profile') of an optically
thick line.  An optically thin line must then peak in the self-aborbed
dip to exclude the possibility of several line components and
rotation. Simple early models (see e.g. Myers et al.  \cite{myers1996}
and references therein) are based on a gravitationally collapsing
isothermal sphere and predict a 'blue asymmetry' i.e. with increasing
infall speed, the blue peak in the double-peak profile becomes
brighter than the red peak.  However, recently Smith et al.
(\cite{smith2012}, \cite{smith2013}) showed that collapsing {\sl
  embedded} cores show more complex line profiles because of the
contribution of the enclosing filament.
  
In any case, there are a number of examples where infall signatures in
spectral profiles have been detected and successfully modelled, mostly
for isolated low-mass cores (e.g. Walker et al.  \cite{walker1994},
Tafalla et al.  \cite{tafalla2002}) and high-mass cores (e.g.
Csengeri et al. \cite{csengeri2011}).  Gravitational collapse of a
whole filament on a scale of a few pc was seen for the DR21 ridge
(Schneider et al.  \cite{schneider2010}) and Serpens-South (Kirk et
al.  \cite{kirk2013}). Global collapse of IRDCs was suggested for
SDC335 (Peretto et al. \cite{peretto2013}), G79.3+0.3 (Carey et al.
2000), and G32.03+0.05 (Battersby et al., priv. comm.).
 
For the IRDCs in our sample, we use the $^{12}$CO 3$\to$2 line from
the COHRS archive as an optically thick tracer, and $^{13}$CO 1$\to$0
from the GRS as the optically thin line.  For better comparison, we
smoothed the $^{12}$CO data to the resolution of the $^{13}$CO data
(45$''$). Both CO data sets are only available for G28.37+0.07. For
this source, we show in Fig.~\ref{spectra-cloud-c} the velocity
integrated $^{12}$CO map and spectra from selected positions, i.e. the
strongest submm continuum sources from ATLASGAL (Contreras et al.
\cite{contreras2013}, Csengeri et al. 2014) and two positions
off-source but within the cloud.  The continuum sources were also
observed in the optically thin N$_2$H$^+$ 1$\to$0 line at 93.173 GHz
with a velocity resolution of 0.2 km s$^{-1}$ (Tackenberg et al.
2014), and the centre velocities reported in their Table~3 are
indicated as a dashed blue line in Fig.~\ref{spectra-cloud-c}. The
same velocity of sources 2 and 6 are also derived by Shipman et al.
(\cite{shipman2014}) based on pointed observations of optically thin
lines, such as N$_2$H$^+$ and C$^{17}$O, using {\sl Herschel} and
APEX.
 
The $^{13}$CO 1$\to$0 line consists of several components, but there
is no 'double-peak' feature and the centre velocity of the main
Gaussian corresponds very well to the velocities determined with
N$_2$H$^+$, so that $^{13}$CO emission is optically thin or only
moderately optically thick. Most importantly, the main $^{13}$CO line
peaks in the gap of a double-peaked $^{12}$CO line, best visible for
sources 1 and 4. This sort of $^{12}$CO profile seen for basically all
spectra in the figure shows the classical infall signature profile
with a brighter blue peak than the red peak. The optically thin
$^{13}$CO line is blue shifted with respect to the self-absorption
gap, clearly visible for positions 1 and 6 and those off the sources
(offsets -720$''$,220$''$ and -1000$''$,220$''$), but less clearly for
the remaining positions.  Source 2 (most likely a protocluster, see
Sec.~\ref{cont}) is more complex with several line components,
associated with CH$_3$OH and H$_2$O masers (Pillai et al.
\cite{pillai2006}; Wang et al.  \cite{wang2008}, \cite{wang2014}) and
harbours a hot-core (Zhang et al.  \cite{zhang2009}). Some sources, in
particular source 1 for velocities lower than $\sim$70 km s$^{-1}$ ,
show broad wings that possibly indicate outflow emission from the
protostellar object.  This becomes more evident when the on-source
$^{12}$CO spectra are compared to two spectra (indicated by offset
--720$''$,220$''$ and --1000$''$,220$''$ in
Fig.~\ref{spectra-cloud-c}) that are located off source but still
within the IRDC. These also show the self-absorption dip but no
prominent wings. Though it is out of our scope to go into more detail
for outflows, it is important to recognise those as indicators for
star formation.
 
Though $^{12}$CO is not the best tracer for infall signatures, HCO$^+$
or HCN are better suited because they show a clearer profile
(Schneider et al. 2010), the general behaviour is the same. Since we
can rather safely exclude the possibility that the double-peak profile
of $^{12}$CO is due to several line components using the $^{13}$CO
spectra and the N$_2$H$^+$ velocity information, the most likely
explanation is that we observe inwards motion of gas.
 
There are several arguments as to why we think that the $^{12}$CO
profile is not simply caused by a low-density, subthermally excited
foreground cloud. First, we would not observe the blue asymmetry
profile systematically, but the two apparant peaks of the $^{12}$CO
line would be equally strong across the map. Outflows and rotation
would also give rise to both a red and a blue asymmetric line. Second,
the $^{13}$CO line may become optically thick at the positions with
highest (column) density and should then also show self-absorption
features (which is not the case). Third, the Herschel column density
maps indicate that the IRDCs are cold, with an increase in excitation
temperature into the cloud at least at the positions of protostellar
sources.  This contrasts with the requirement of a colder outer layer
causing self-absorption.
  
{\sl \textup{Local infall}} on the pre-and protostellar sources is
seen in H$_2$O observations (Shipman et al. 2014) of source 2 and
source 6 (labelled G28-MM and G28-NH$_3$ in Shipman et al. 2014),
where the water line appears in absorption and is systematically
redshifted relative to the systemic velocity of the clump.  However,
comparing $^{12}$CO and $^{13}$CO spectra across the whole IRDC (the
two off-source spectra in Fig.~\ref{spectra-cloud-c}, and
Fig.~\ref{allspectra}) shows that the double-peak profiles extend well
beyond the source positions and are found basically everywhere inside
the blue contours, outlining the IRDC.  We thus propose that the whole
IRDC is in {\sl \textup{global collapse}}.
 
In Appendix D, we show that infall $^{12}$CO profiles are also
observed for G11.11-0.12.  Because we have no complementary $^{13}$CO
data, the global collapse scenario there is more tentative. However,
single pointings in H$_2$O in G11.11.-0.12 (Shipman et al. 2014)
already show infall signatures similar to G28.37+0.07, so that we
conclude that global collapse could probably be a feature shared by
all {\sl \textup{massive}} IRDCs that are embedded in GMCs. Higher
angular resolution molecular line observations using, e.g.  HCO$^+$ or
HCN and the optically thin isotopologues H$^{13}$CO$^+$ or H$^{13}$CN
as line tracers, are required to investigate this scenario in more
detail.
 
\section{Summary and conclusions}  
 
Our study using {\sl Herschel}, ATLASGAL, and $^{12}$CO and $^{13}$CO
molecular data shows that massive IRDCs are embedded in more extended
molecular clouds and have similar physical properties (peak and
average column density, surface density) as {\sl \textup{ridges}},
i.e., the densest, central regions of GMCs.  The PDFs for all four
IRDCs obtained from {\sl Herschel} dust column density maps show a
power-law distribution, which we interpret as arising from
gravitational contraction. There is no log-normal part in the PDF
because the PDF is constructed from a cropped image, focussing on the
densest region within the GMC. On the other hand, the PDF of the
associated GMC shows a log-normal form for lower column densities and
a power law for high densities.
 
We give two additional and independent arguments for the dominance of
self-gravity for the IRDCs we present.  First, fitting the radial
column density profiles leads to similar exponents
$\alpha$=1.70$\pm$0.07 as derived from the slope of the PDF power-law
distribution ($\alpha$=1.66$\pm$0.18), and is consistent with
self-gravity ($\alpha$=1.5 to 2 assuming a spherical density
distribution).  Second, molecular line profiles of the optically thick
$^{12}$CO 3$\to$2 line show self-absorption at the velocity of the
bulk emission of the cloud, indicating locally infalling gas on proto-
and prestellar cores and on the global IRDC scale.
 
The self-gravitating scenario on all scales is consistent with what is
found in numerical models. Just to list a few, Ballesteros-Paredes et
al. (2011), Kritsuk et al. (2011), Federrath \& Klessen (2013),
Girichidis et al. (2014), Ward et al. (2014) obtain a
log-normal+power-law tail PDF (density or column density) for their
simulations including self-gravity and turbulence, but without
external pressure (such as radiative feedback). First, turbulence
creates a self-similar structure of the gas that is well represented
by a purely log-normal PDF.  As soon as self-gravity is switched on
and collapse proceeds for some time, the hierarchical structure in
clouds is affected and the (column) density PDF departs from a
log-normal and forms a power-law tail.  Studies focussing on the
\textup{{\sl \textup{spatial}} }cloud structure, using power spectra
or the $\Delta$-variance (see Schneider et al.  \cite{schneider2011}
and references therein) also show that molecular clouds lose their
hierarchical structure and show characteristic scales as soon as star
formation sets in.  Federrath \& Klessen (2013, Fig.~4) link the
change of the slope of the PDF to the star formation efficiency (SFE).
Their simulations are consistent with our observations.
 
The PDFs of the IRDCs we present all show a power-law distribution
with similar slopes, regardless of their evolutionary states:
G18.82-0.28 and G28.37+0.07 show bright peaks at 70 $\mu$m, and in
G11.11-0.12 the presence of IR-bright protostars was demonstrated
(Henning et al. 2010).  The IRDC\ G28.53-0.25 is in a very early stage
because it contains no (F)IR-bright peaks.  Despite its youth, this
cloud shows a power-law distribution over all column densities.  Given
the uncertainties in the column-density PDFs, the slope differences
are marginal and it is difficult to distinguish an SFE of 5\% from
20\% for most parameter sets of the simulations (note that the PDFs
from simulations differ from those of observations, because they are
averages over all three lines of sight from different directions).
Given that the IRDCs of our sample are all in a collapsing state, and
most of them have already started forming stars, they might well have
local SFE of $\sim$5\% in agreement with the models shown in Federrath
\& Klessen (2013).
 
\begin{acknowledgements}   
  We thank P. Girichidis for discussions on collapse scenarios.  N.S. and S.B.  
  acknowledge support by the ANR-11-BS56-010 ``STARFICH''. N.S., 
  V.O., and R.S.K. acknowledge funding from the DFG priority program 
  SPP 1573 ``Physics of the ISM'' (project number OS177/2-1 and KL 
  1358/19-1).  R.S.K.  acknowledges subsidies from the collaborative 
  research project SFB 881 (``The Milky Way System'', subprojects B1, 
  B2, and B5), and support from the ERC Framework Programme 
  FP7/2007-2013 via the ERC Advanced Grant STARLIGHT (project number 
  339177). R.S.K. thanks for the warm hospitality at the Department of 
  Astronomy and Astrophysics at the University of California at Santa 
  Cruz and the Kavli Institute for Particle Astrophysics and Cosmology 
  at Stanford University  during a sabbatical stay in 2014 and 
  2015.  T.Cs.  acknowledges financial support for the ERC Advanced 
  Grant GLOSTAR under contract no.  247078. 
\end{acknowledgements}

\appendix      
 
\section{Herschel maps of infrared dark clouds} 
 
In this section, we show the whole sample of FIR-data from {\sl
  Herschel} (70 $\mu$m and 500 $\mu$m data) and ATLASGAL (870 $\mu$m),
and column density and temperature maps obtained from SED fits for the
infrared dark clouds G11.11-0.12 ('snake'), G18.82-0.28 (Cloud A), and
G28.53-0.25 (Cloud D). In addition, molecular line data maps
($^{13}$CO 1$\to$0 for G18.82-0.28 and G28.53-0.25, and $^{12}$CO
3$\to$2 for G11.11-0.12) of the associated GMC are shown.

%%%%%%%%%%%%%%%%%%%%%%%%%%%%%%   
%%%  Total G11 IRDCs %%%%%%%%%%%%%%%   
\begin{figure*}[!htpb]   
\begin{centering}   
\includegraphics [width=8cm, angle={0}]{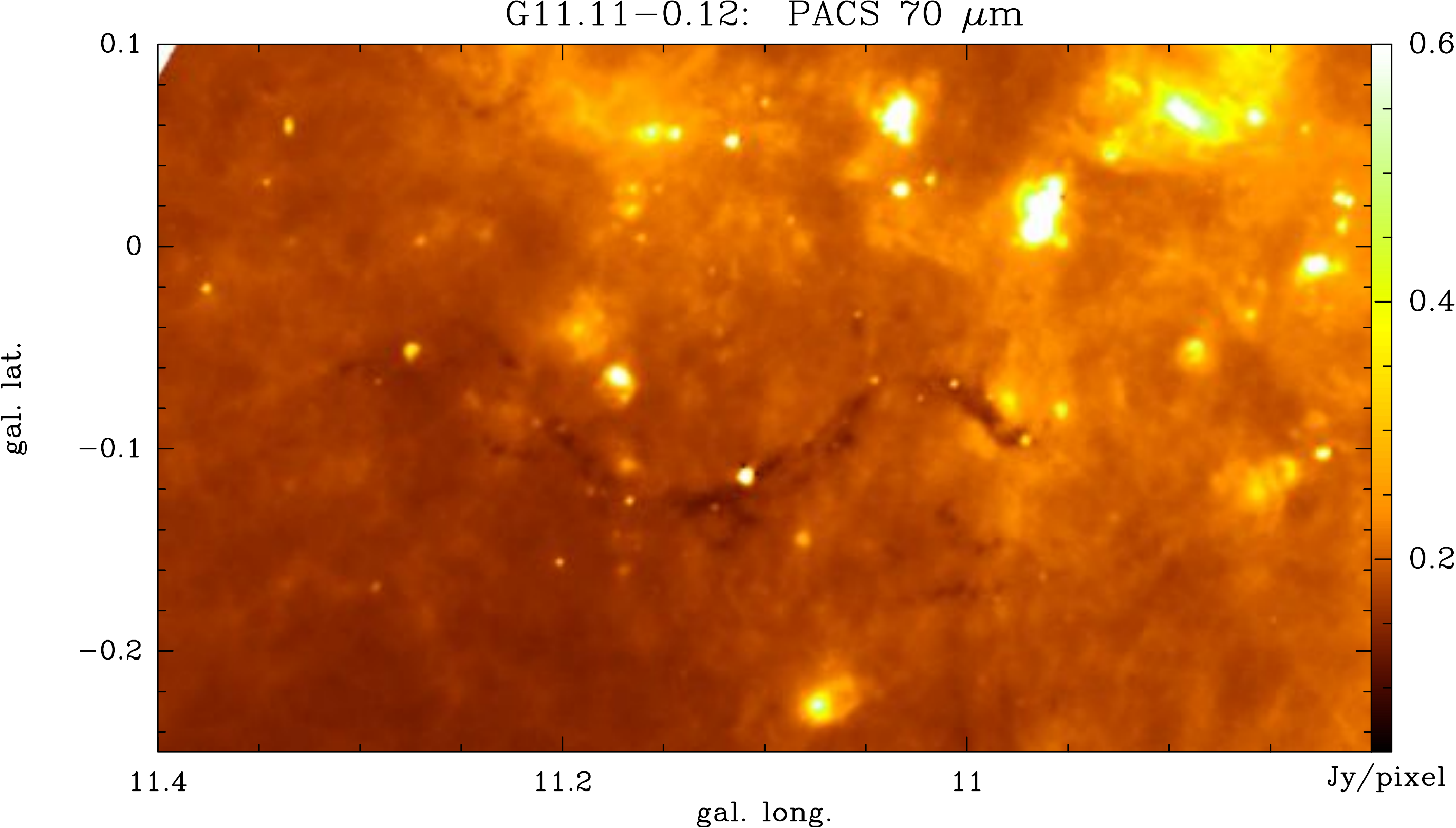}   
\hspace{0.3cm} 
\includegraphics [width=8cm, angle={0}]{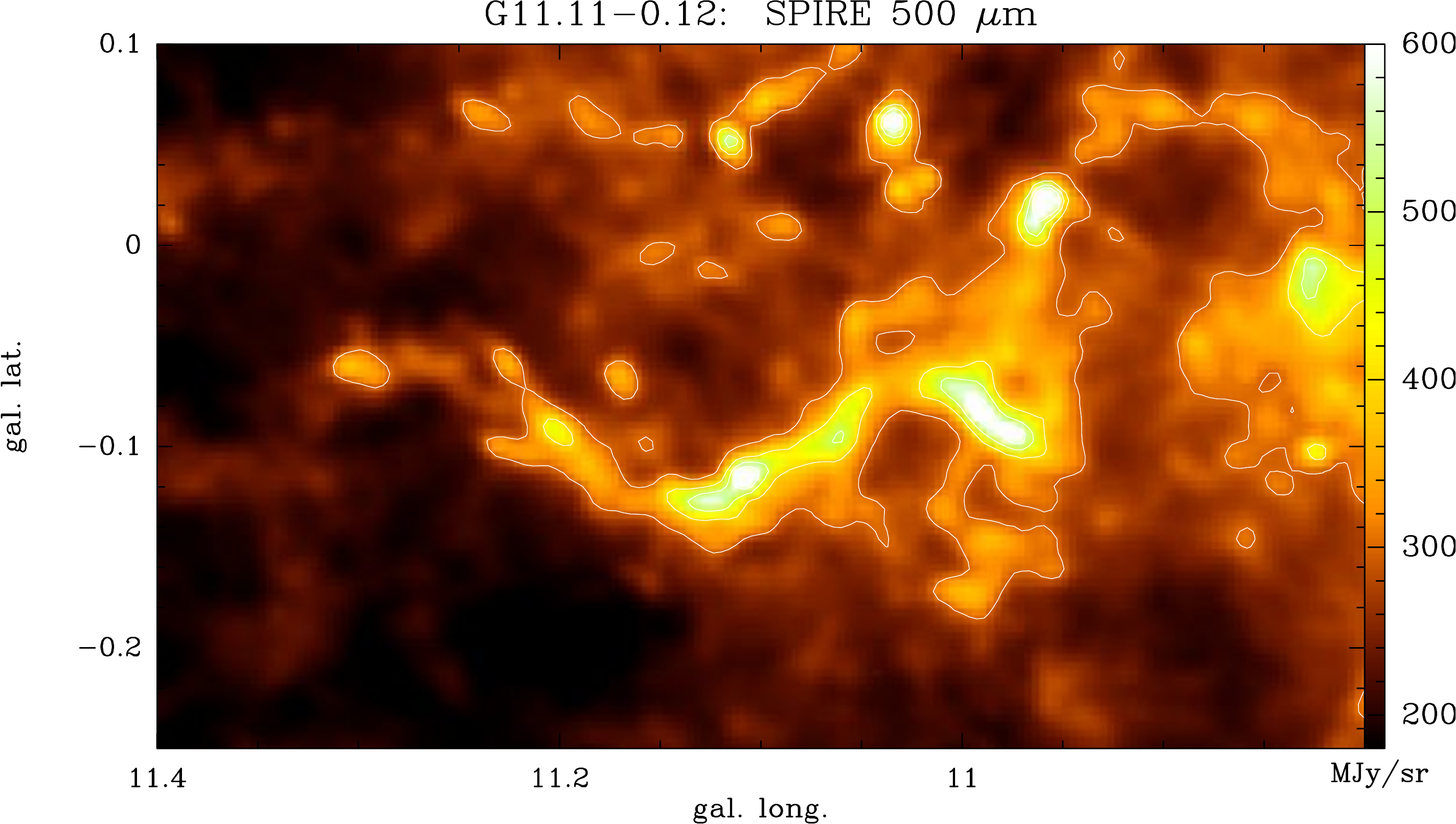} 
 
\vspace{0.5cm} 
\includegraphics [width=8cm, angle={0}]{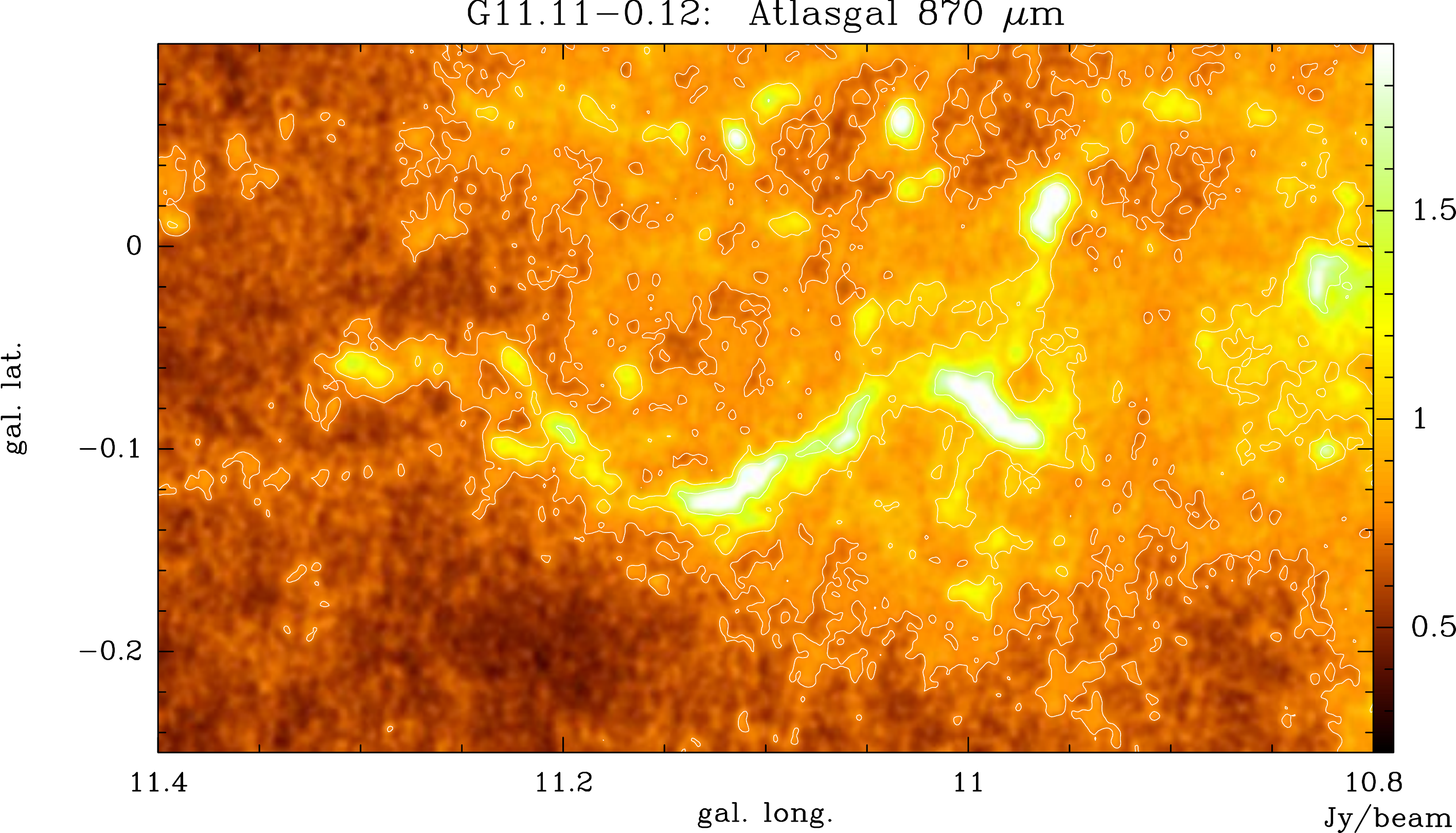} 
\hspace{0.3cm} 
\includegraphics [width=8cm, angle={0}]{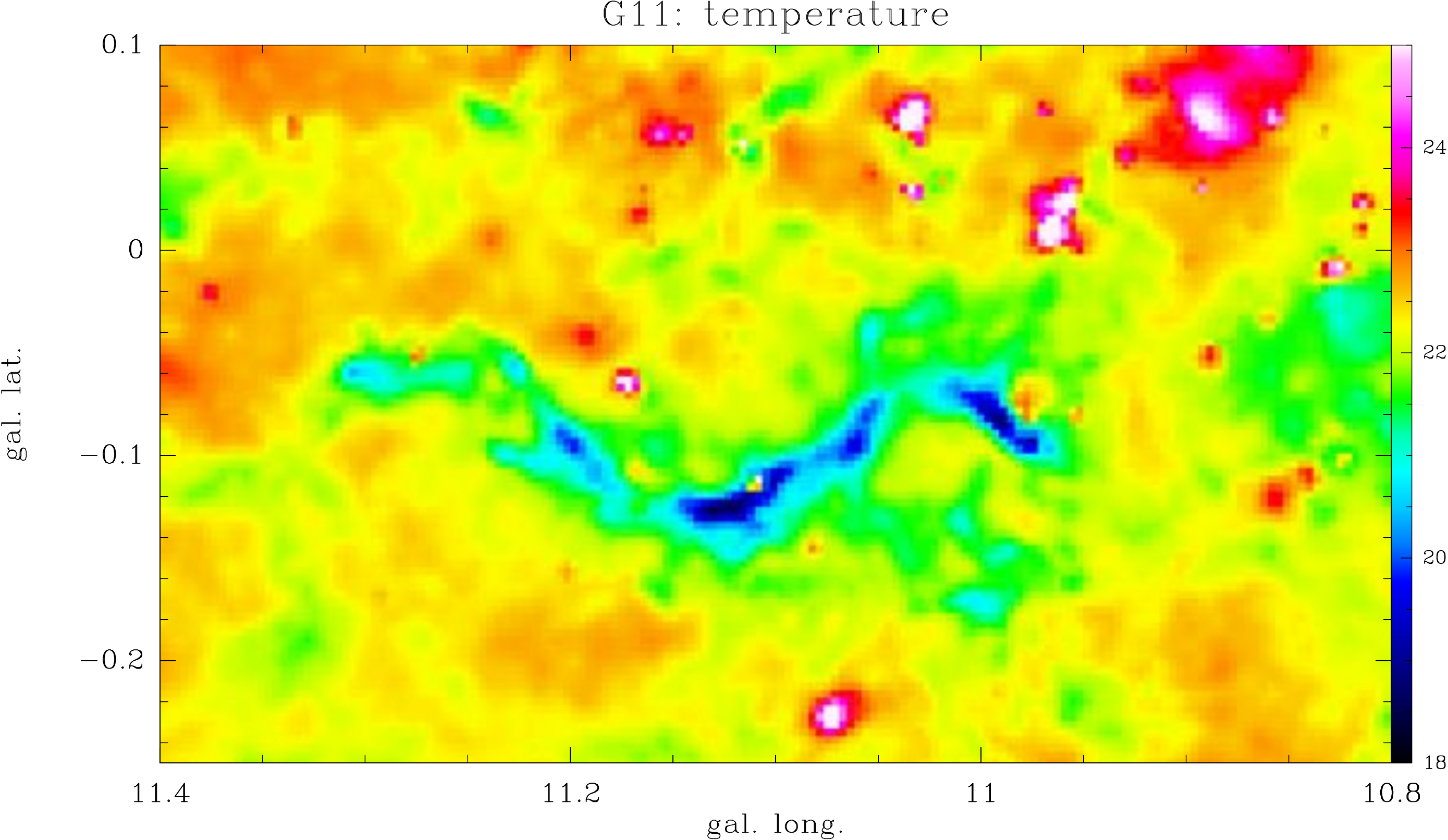} 
\caption[] {PACS 70 $\mu$m map, SPIRE 500 $\mu$m, and ATLASGAL 870
  $\mu$m maps of G11.11-0.12 (the 'snake'). The long filamentary
  structure is well visible as a dark (bright) feature in the 70
  $\mu$m (500, 870 $\mu$m) maps. Bottom right: temperature map from
  SED fit 160-500 $\mu$m.}
\label{g11-1}   
\end{centering}  
\end{figure*}   
 
\begin{figure*}[!htpb]   
\begin{centering}   
\includegraphics [width=8cm, angle={0}]{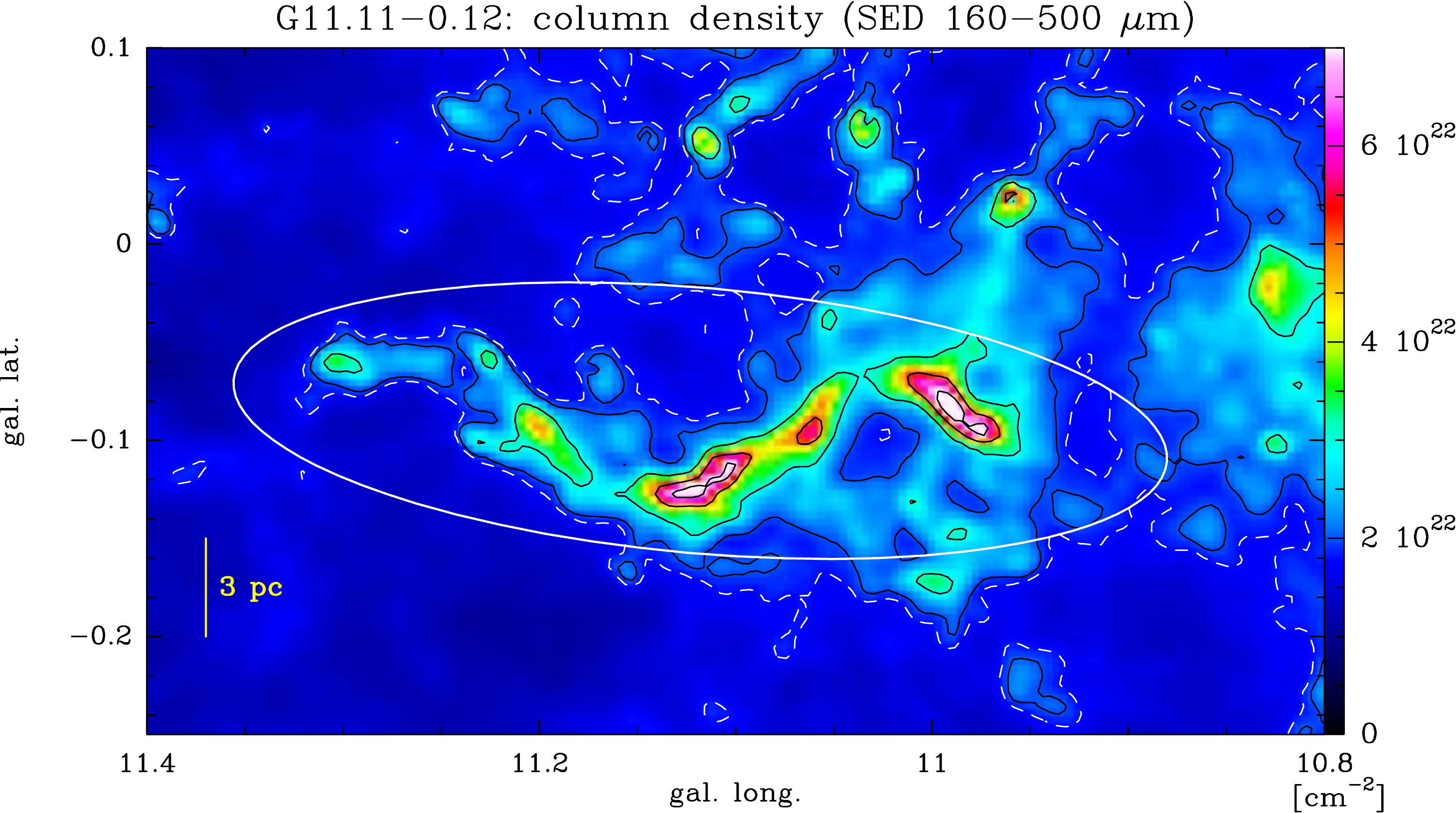}   
\hspace{0.8cm} 
\includegraphics [width=8cm, angle={0}]{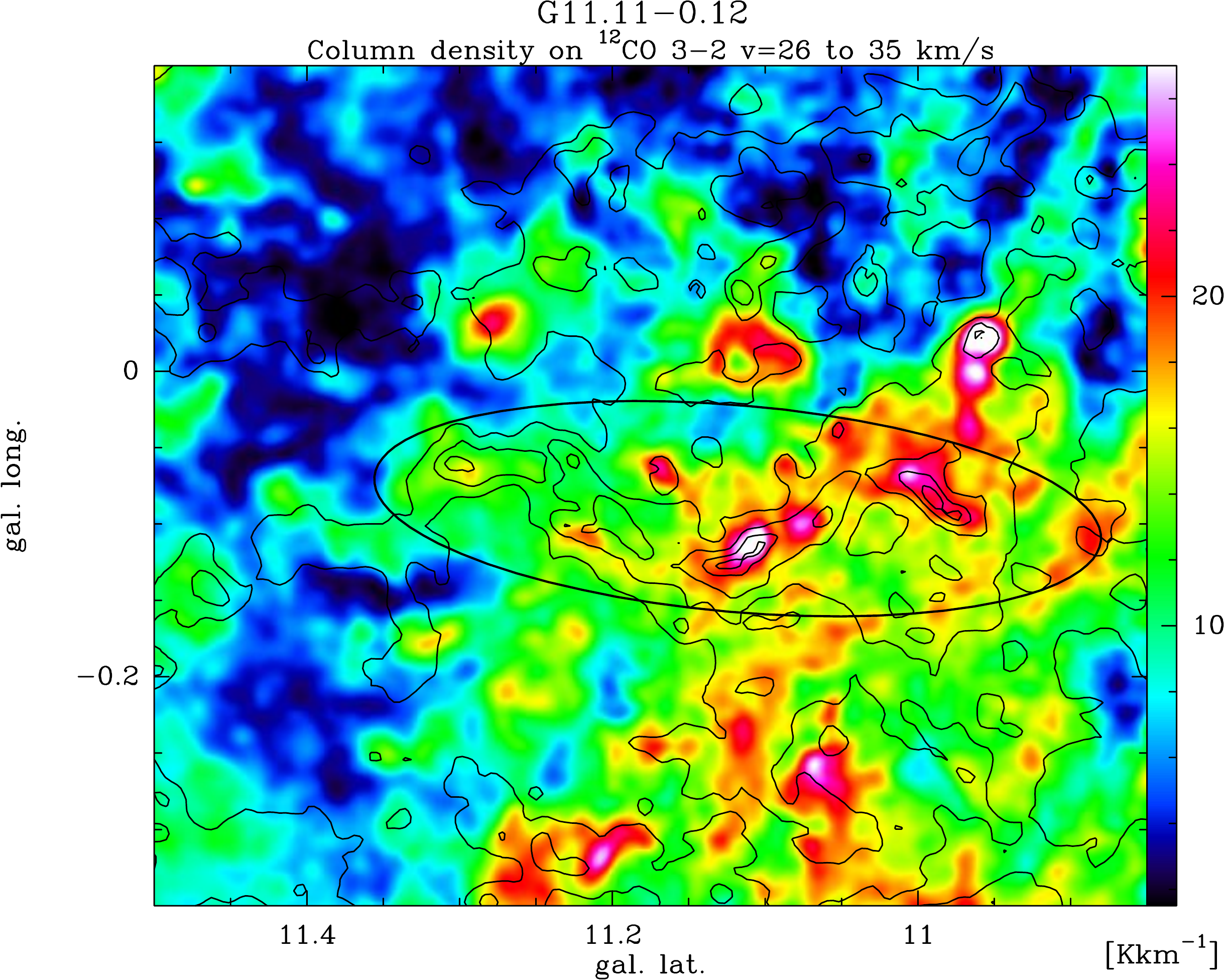} 
\caption[] {{\bf Left:} column density map of G11.11-0.12, obtained
  from SED fit to {\sl Herschel} 160, 250, 350, 500 $\mu$m data.  The
  black contours indicate the levels
  $N_{dust}$=2,3,5,7$\times$10$^{22}$ cm$^{-2}$, the white-dashed
  contour outlines the approximate completeness level, and ellipse is
  taken from Simon et al. (2006b), defining the IRDC.  {\bf Right:}
  line integrated $^{12}$CO 3$\to$2 emission in colour scale (in [K
  km/s]) between v=26 and 35 km s$^{-1}$, the velocity range of the
  bulk emission of G11.11-0.12 and the associated GMC.  The {\sl
    Herschel} H$_2$ column density is overlaid as black contours
  (levels 1.5, 2, 3, 5, 7 10$^{22}$ cm$^{-2}$) and the IRDC is
  outlined by an ellipse (Simon et al. 2006b).}
\label{g11-2}   
\end{centering}  
\end{figure*}   
 
\begin{figure*}[!htpb]   
\begin{centering}   
\includegraphics [width=8.5cm, angle={0}]{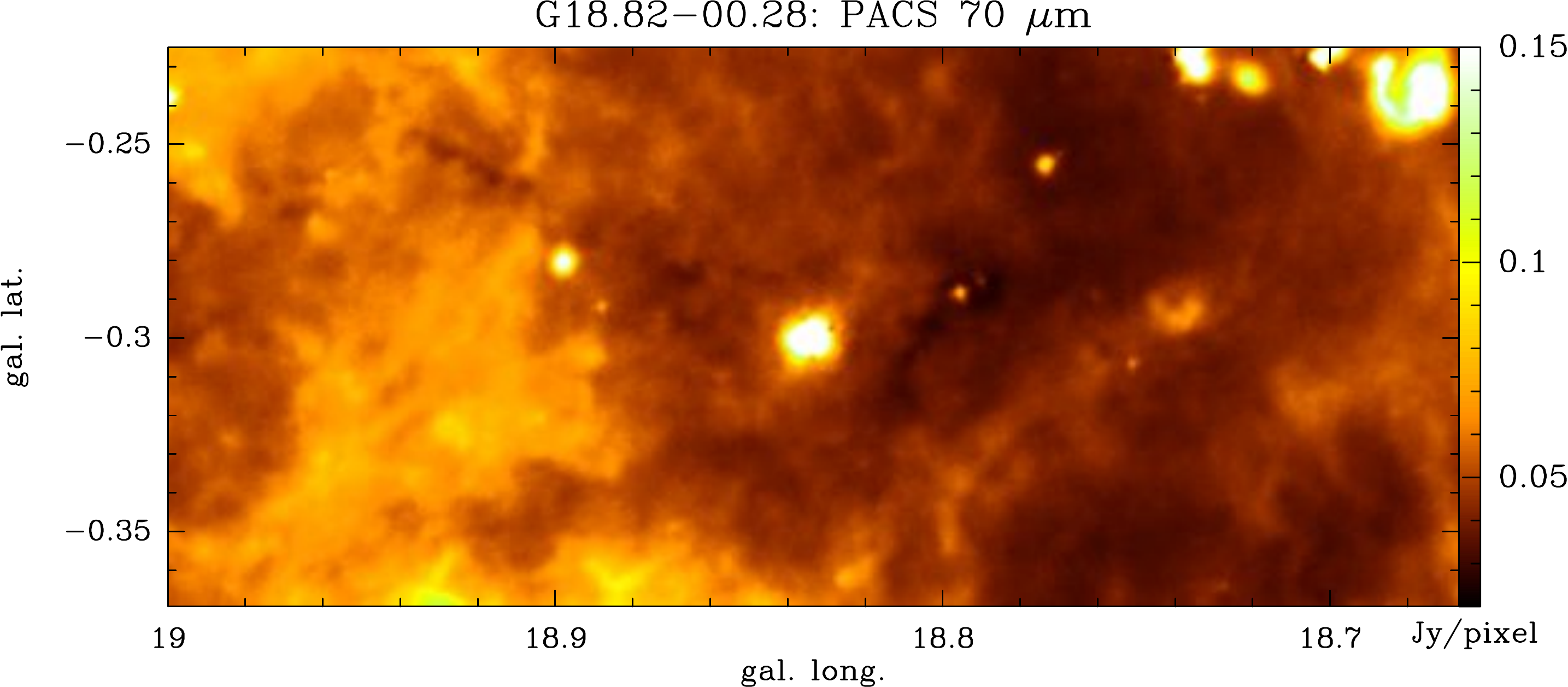}   
\hspace{0.3cm} 
\includegraphics [width=8.5cm, angle={0}]{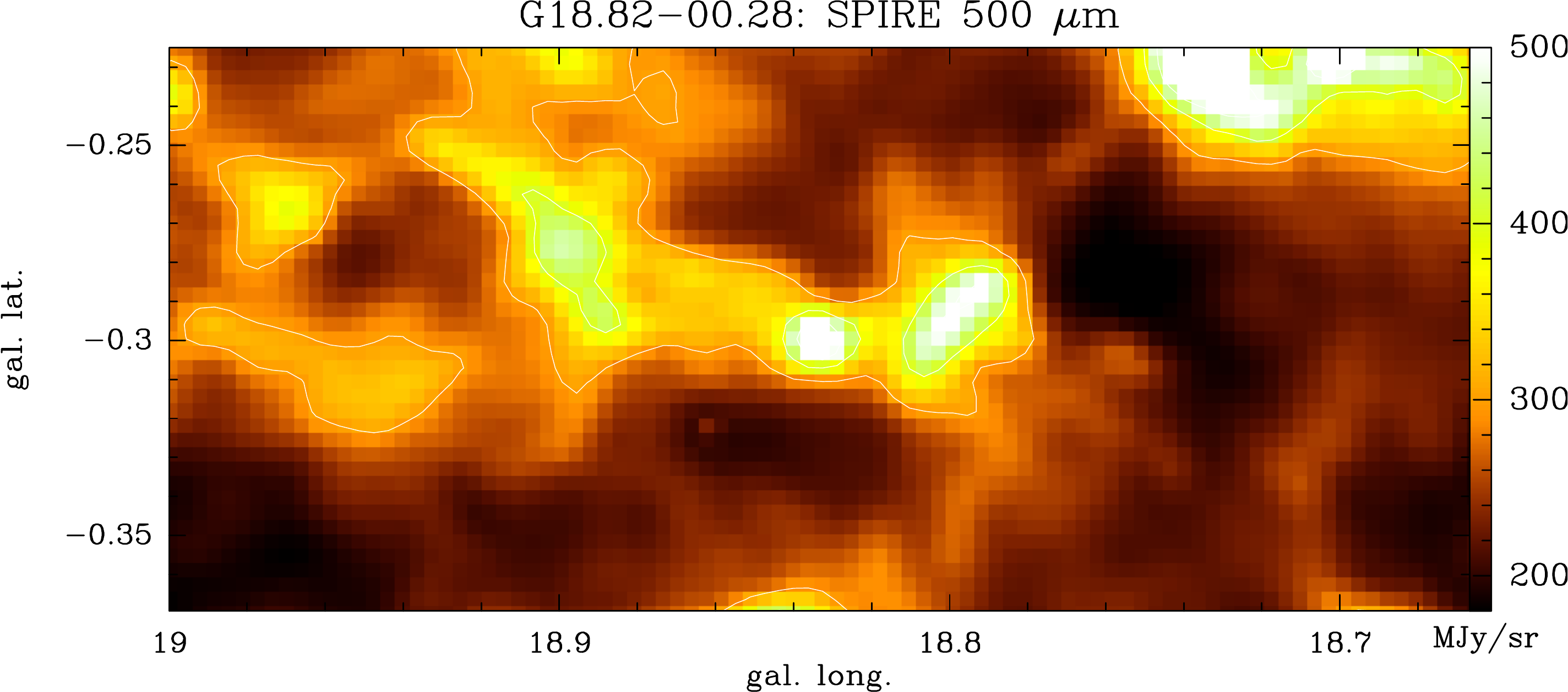} 
 
\vspace{0.5cm} 
\includegraphics [width=8.5cm, angle={0}]{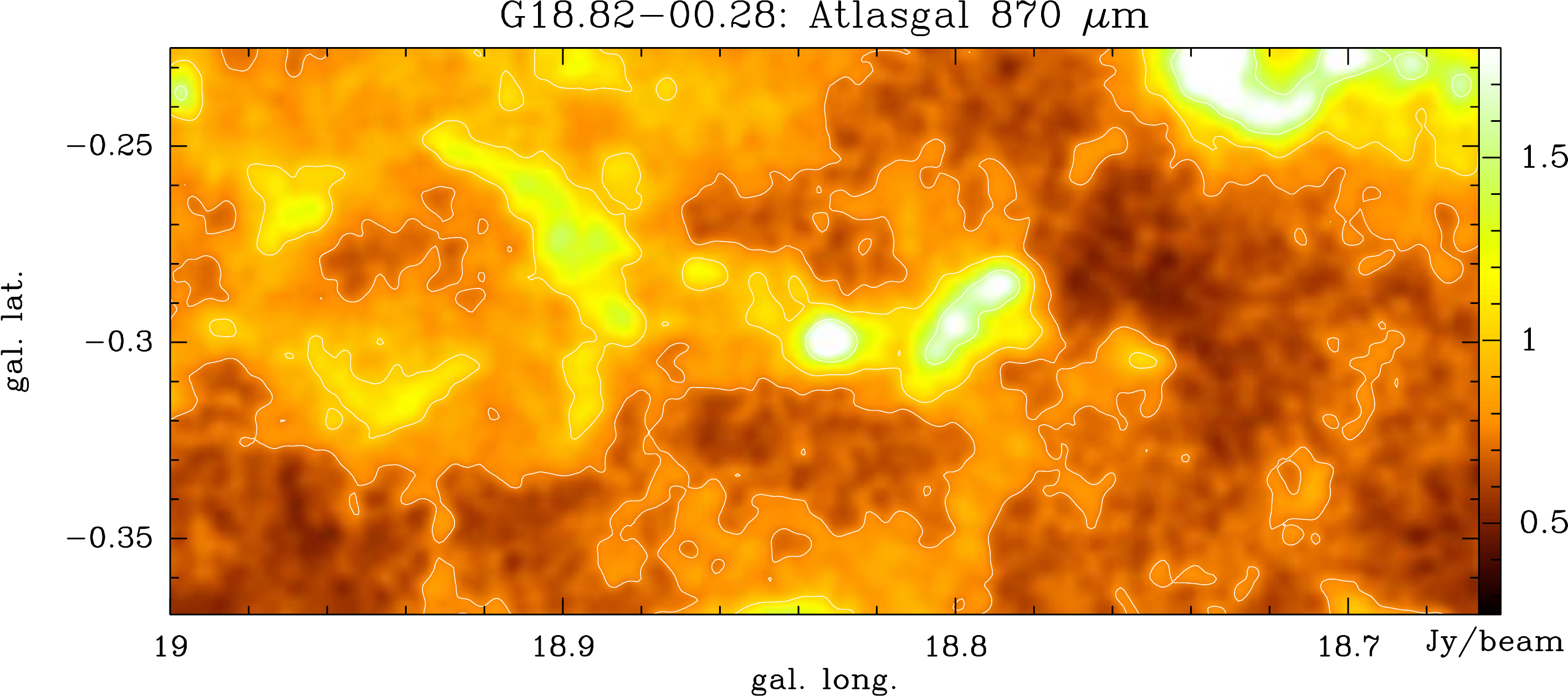} 
\hspace{0.3cm} 
\includegraphics [width=8.5cm, angle={0}]{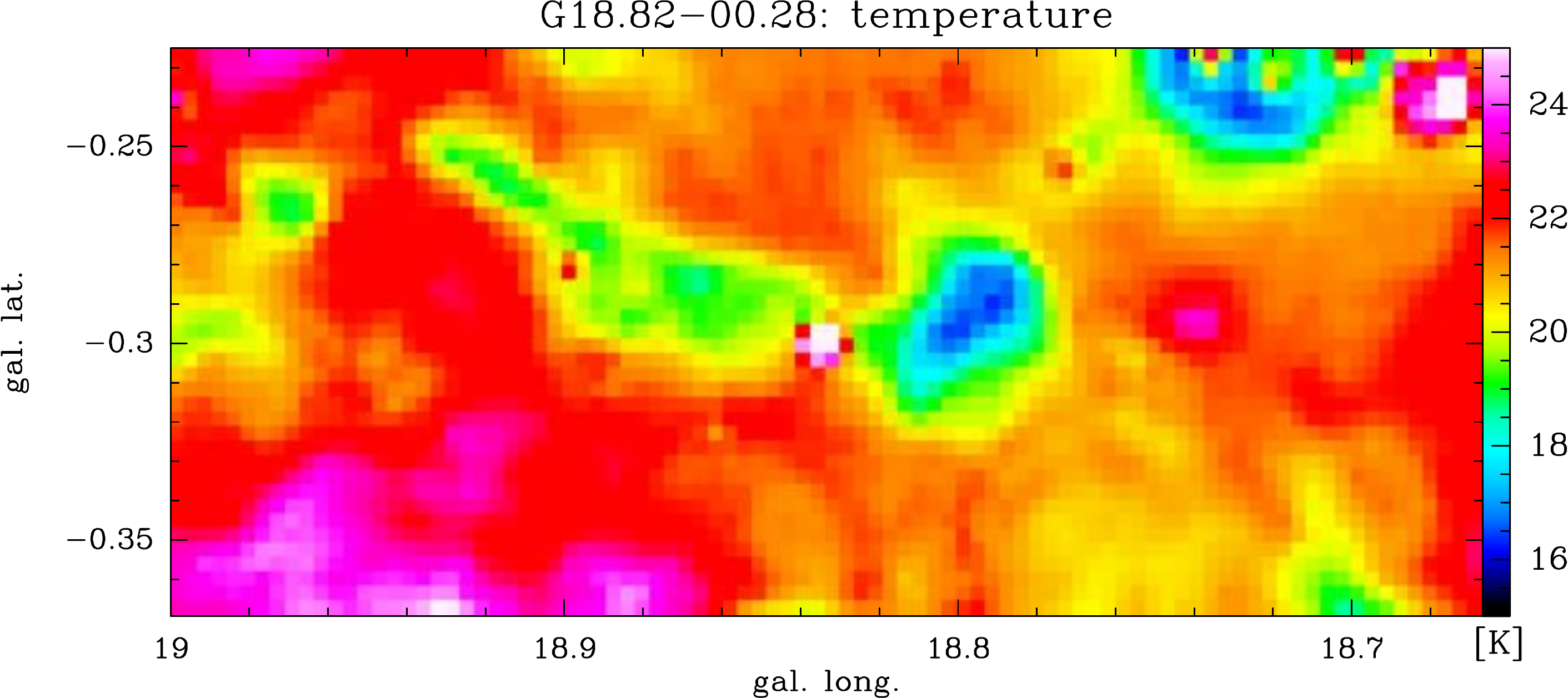} 
\caption[] {PACS 70 $\mu$m map, SPIRE 500 $\mu$m, and ATLASGAL 870
  $\mu$m maps of IRDC G18.82-0.28 (Cloud A).  The long filamentary
  structure is well visible as a dark (bright) feature in the 70
  $\mu$m (500, 870 $\mu$m) maps. Bottom right: temperature map from
  SED fit 160-500 $\mu$m.}
\label{cloud-a-1}   
\end{centering}  
\end{figure*}   
 
\begin{figure*}[!htpb]   
\centering   
\includegraphics [width=8.5cm, angle={0}]{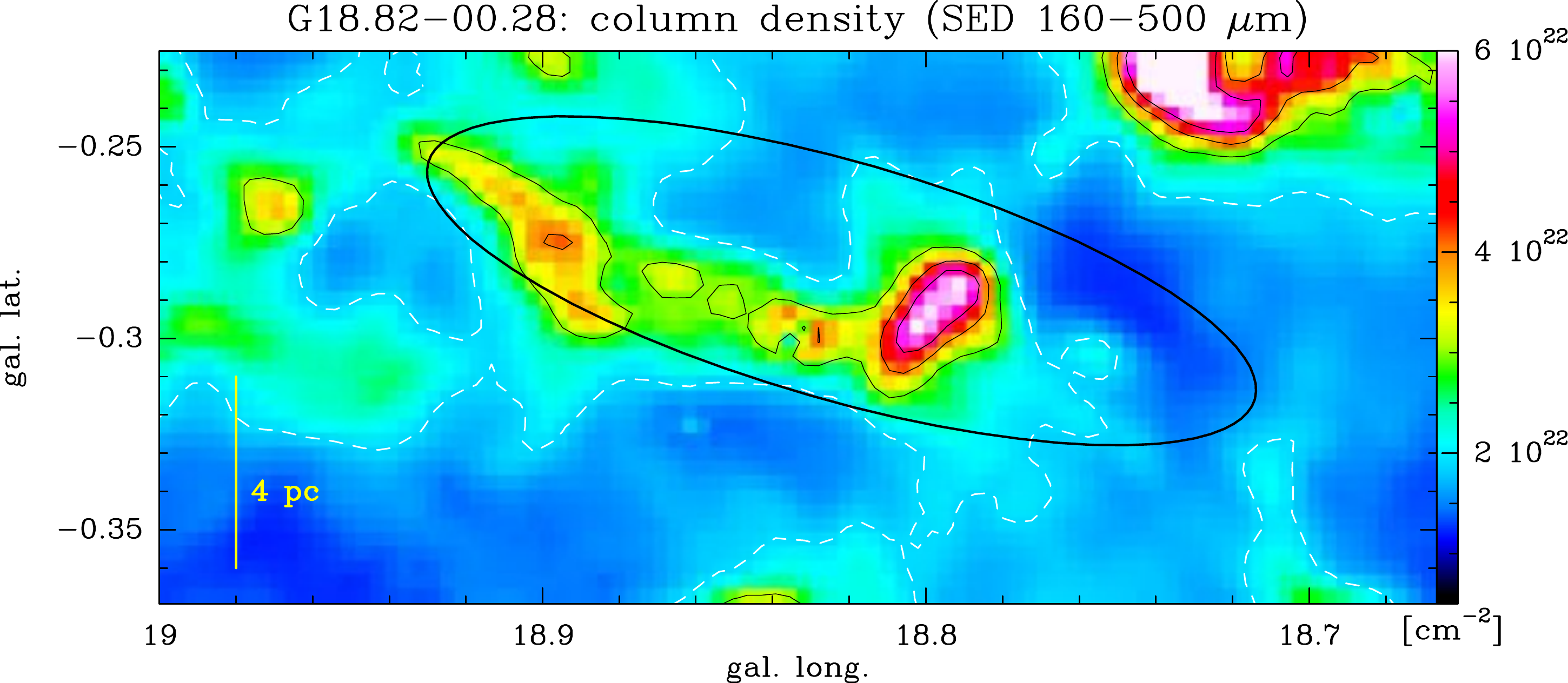}   
\hspace{0.8cm} 
\includegraphics [width=8.2cm, angle={0}]{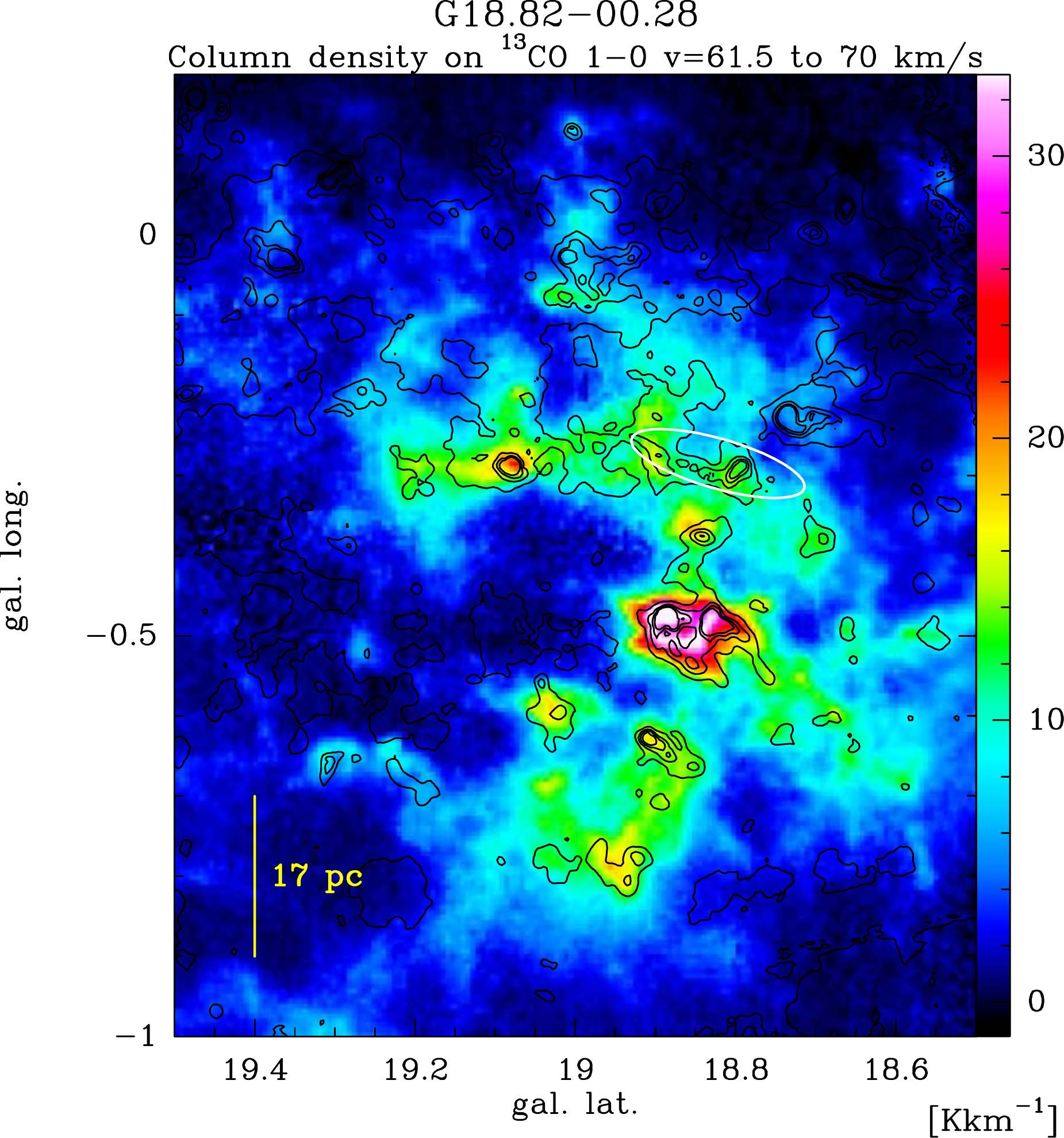}   
\caption[] {{\bf Left:} column density map of G18.82-0.28, obtained
  from SED-fit to {\sl Herschel} 160, 250, 350, 500 $\mu$m data.
  Black contours indicate the levels $N_{dust}$=2 to
  5$\times$10$^{22}$ cm$^{-2}$ in steps of 10$^{22}$ cm$^{-2}$,
  white-dashed contour outlines the approximate completeness level,
  and ellipse is taken from Simon et al. (2006b), defining the IRDC.
  {\bf Right:} line integrated $^{13}$CO 1$\to$0 emission in colour
  scale (in [K km/s]) between v=61 and 70 km s$^{-1}$, the velocity
  range of the bulk emission of G18.82-0.28 and the associated GMC.
  Note that this GMC was already identified as a coherent cloud
  complex in Schuller et al. \cite{schuller2009}. The {\sl Herschel}
  H$_2$ column density is overlaid as black contours (levels 1,2,3,4,5
  10$^{22}$ cm$^{-2}$) and the IRDC is outlined by an ellipse (Simon
  et al. 2006b).}
\label{cloud-a-2}   
\end{figure*}   
 
\begin{figure*}[!htpb]   
\begin{centering}   
\includegraphics [width=6.5cm, angle={0}]{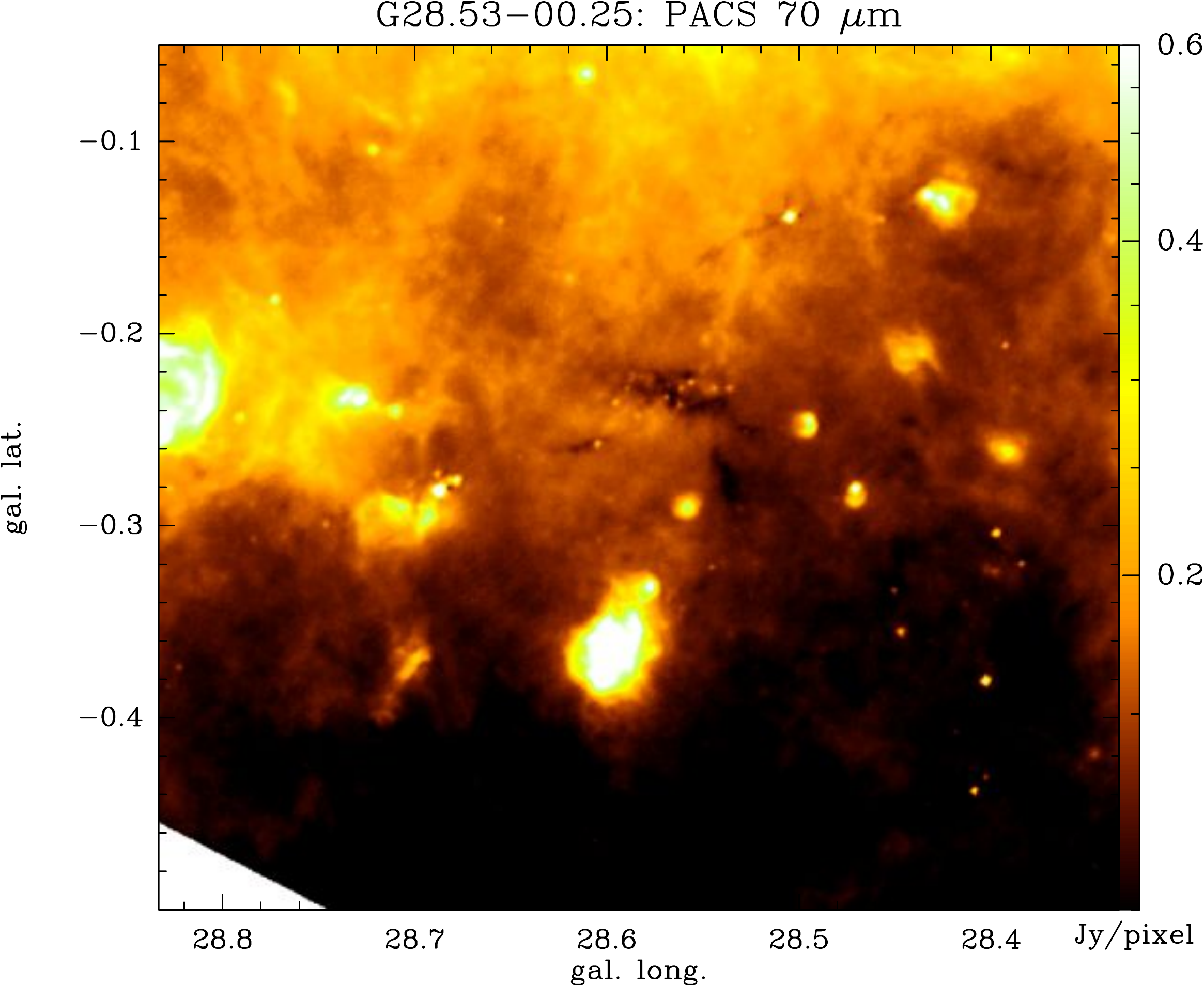}   
\hspace{0.3cm} 
\includegraphics [width=6.5cm, angle={0}]{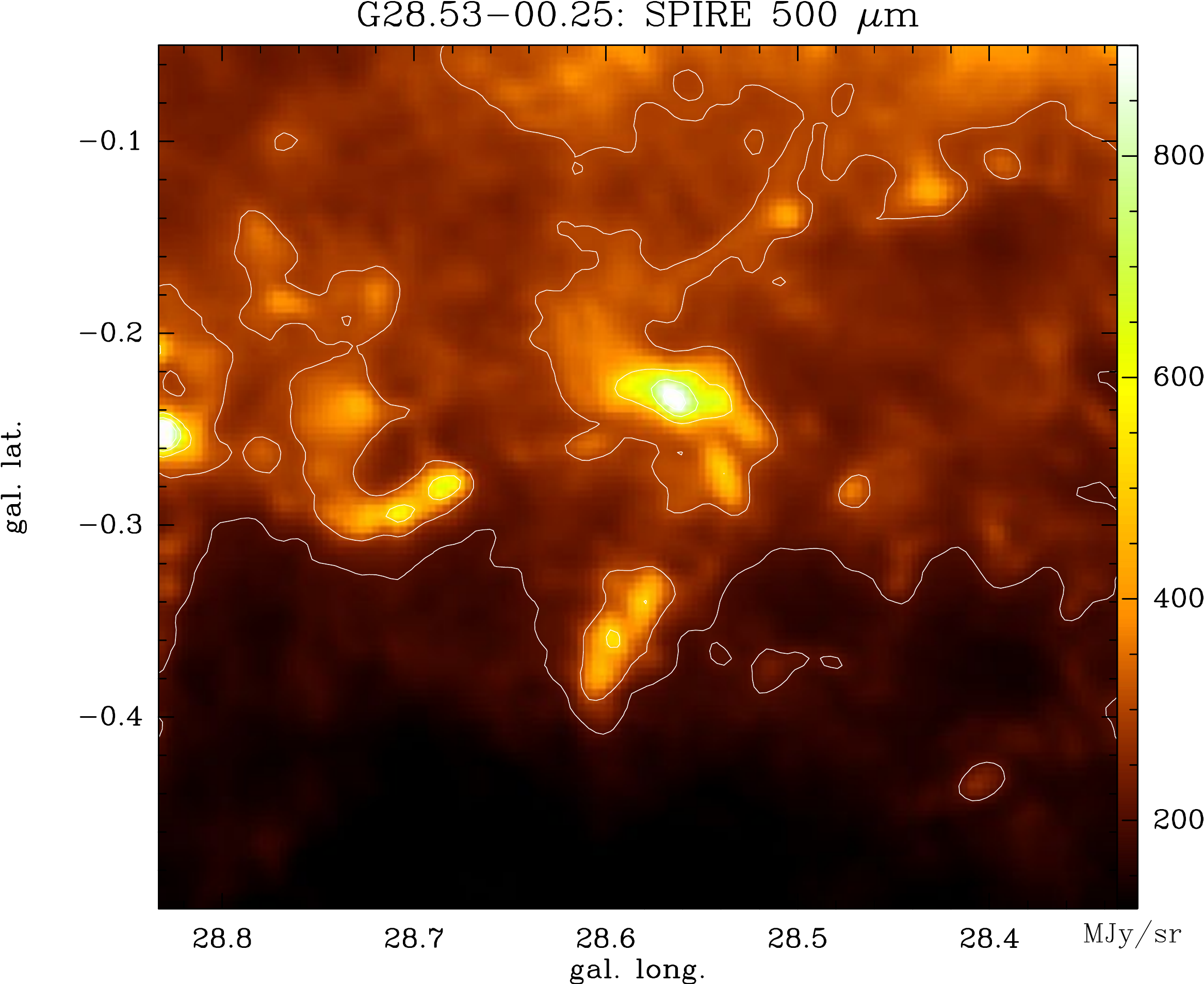} 
 
\vspace{0.5cm} 
\includegraphics [width=6.5cm, angle={0}]{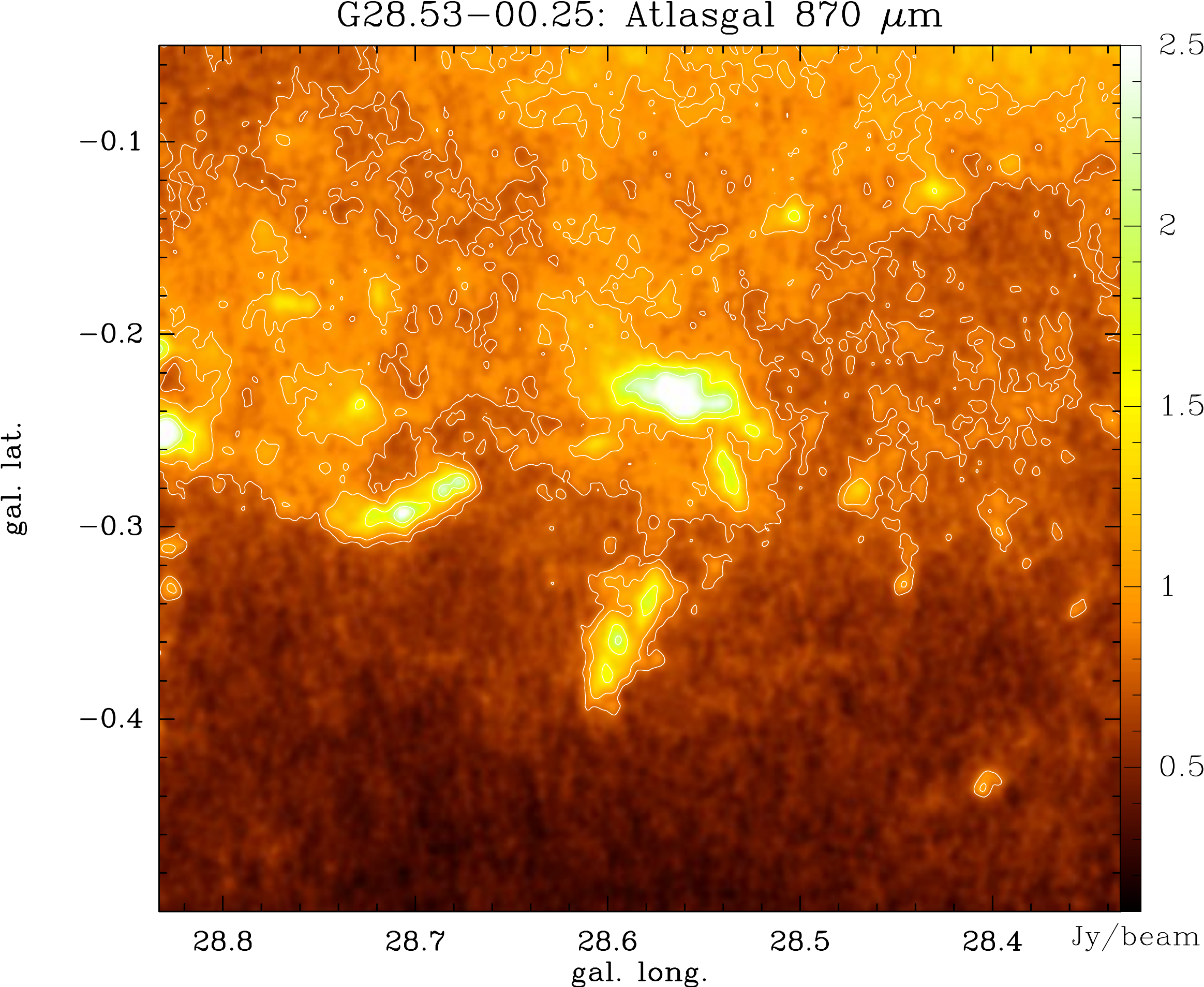} 
\hspace{0.3cm} 
\includegraphics [width=6.5cm, angle={0}]{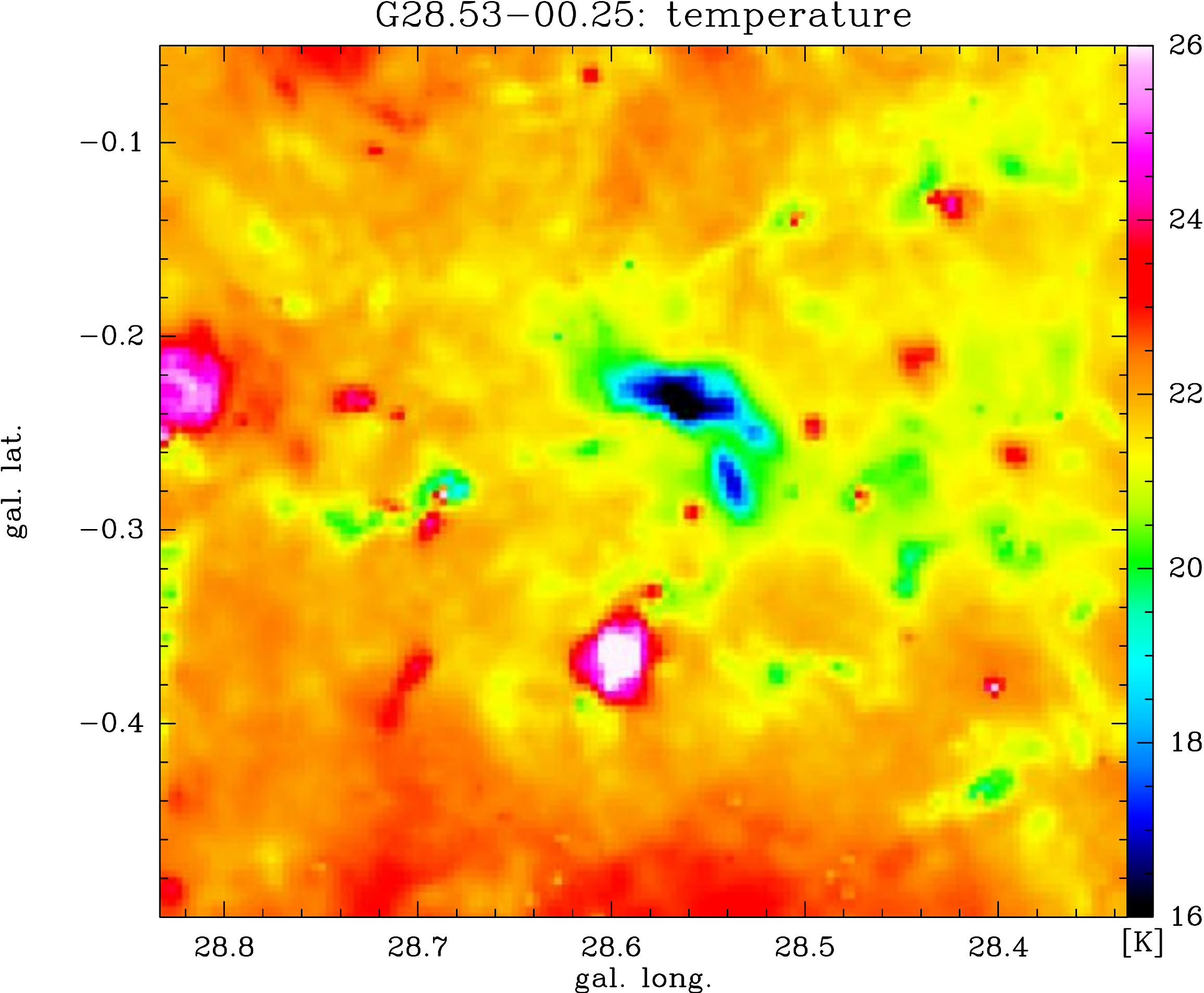} 
\caption[] {PACS 70 $\mu$m map (left), SPIRE 500 $\mu$m (middle) and
  ATLASGAL 870 $\mu$m maps of G28.53-0.25 (Cloud D).  The long
  filamentary structure is well visible as a dark (bright) feature in
  the 70 $\mu$m (500, 870 $\mu$m) maps.  Bottom right: temperature map
  from SED fit 160-500 $\mu$m}
\label{cloud-d-1}   
\end{centering}  
\end{figure*}   
 
\begin{figure*}[!htpb]   
\centering   
\includegraphics [width=6cm, angle={0}]{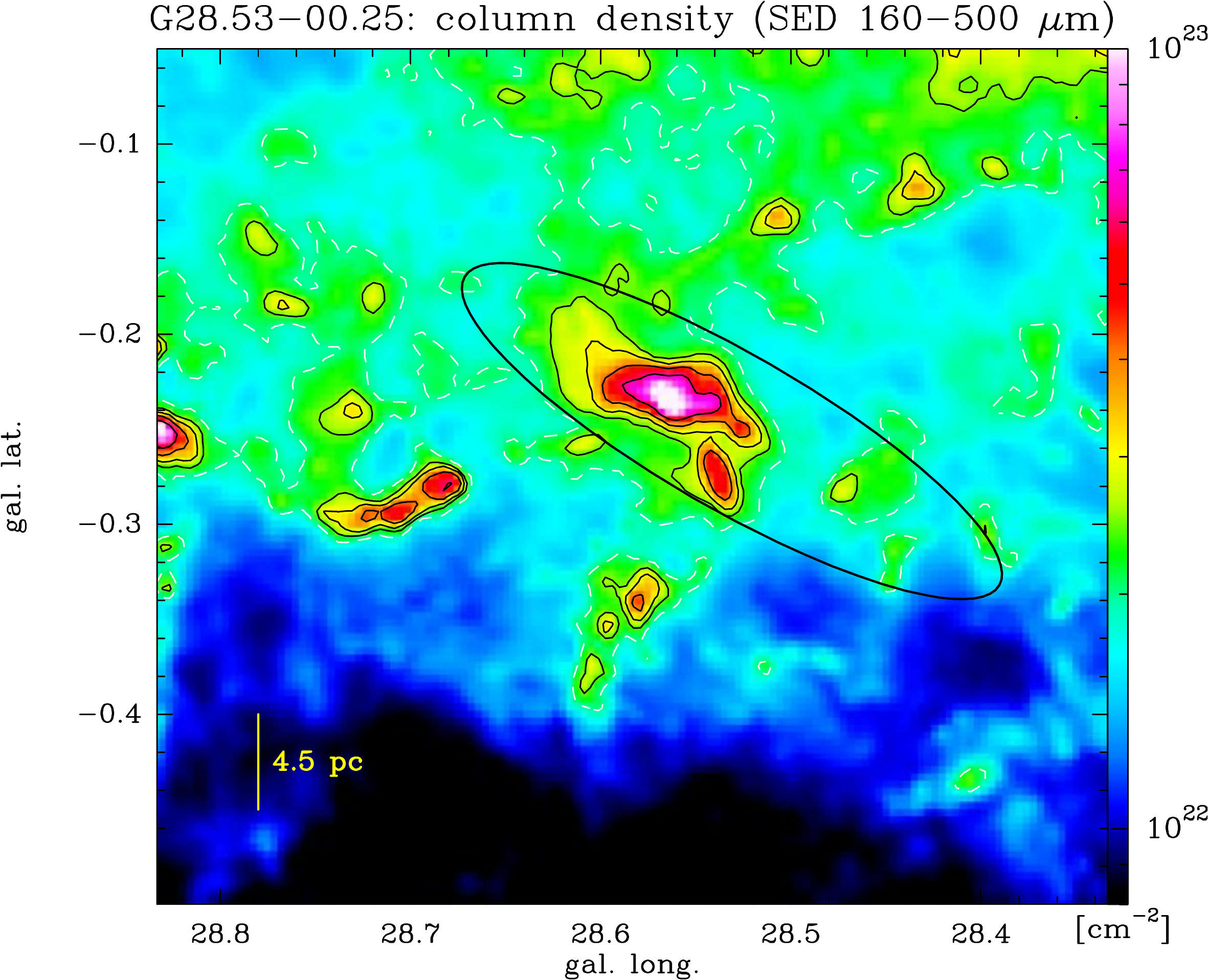}   
\hspace{0.8cm} 
\includegraphics [width=8cm, angle={0}]{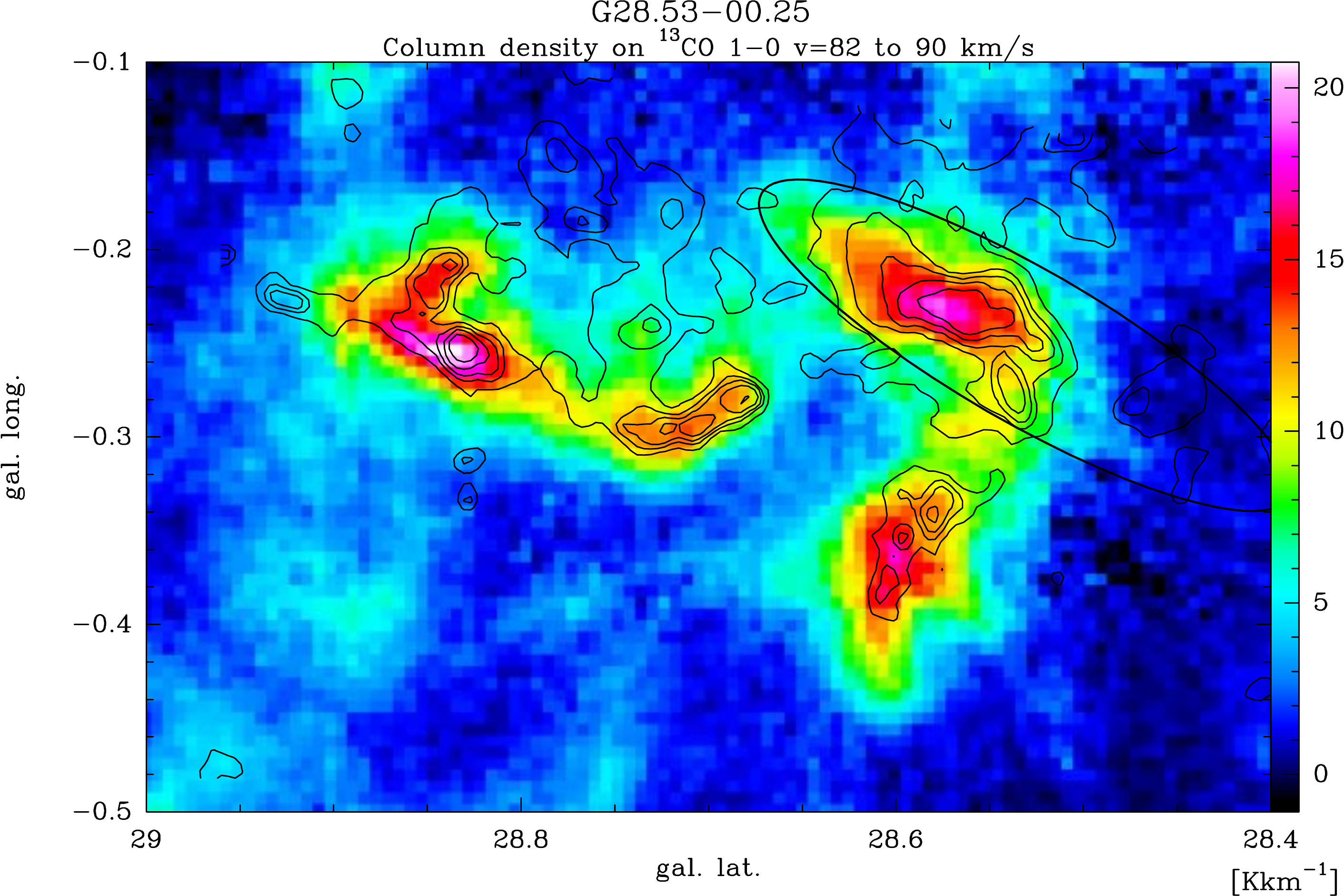}   
\caption[] {{\bf Left:} column density map of G28.53-0.25, obtained
  from SED fit to {\sl Herschel} 160, 250, 350, 500 $\mu$m data.
  Black contours indicate the levels $N_{dust}$=2.5, 3, 4,
  6$\times$10$^{22}$ cm$^{-2}$, white-dashed contour outlines the
  approximate completeness level, and ellipse is taken from Simon et
  al. (2006b), defining the IRDC. {\bf Right:} line integrated
  $^{13}$CO 1$\to$0 emission in colour scale (in [K km/s]) between
  v=82 and 90 km s$^{-1}$, the velocity range of the bulk emission of
  G28.53-0.25, and the associated GMC. The {\sl Herschel} H$_2$ column
  density is overlaid as black contours (levels 2, 2.5, 3, 4, 6
  10$^{22}$ cm$^{-2}$) and the IRDC is outlined by an ellipse (Simon
  et al. 2006b).}
\label{cloud-d-2}   
\end{figure*}   
 
\section{Probability distribution functions of column density for IRDCs} 
 
Probability distribution functions of column density, determined from
{\sl Herschel} data, for the infrared dark clouds G11.11-0.12
('snake'), G18.82-0.28 (Cloud A), and G28.53-0.25 (Cloud D) are shown.
The assumption of a constant line-of-sight temperature for each pixel
affects the accuracy of the column density maps in massive,
UV-illuminated GMCs with internal embedded (proto)stars, and can lead
to an underestimation of the column density map. We carefully checked
each SED fit for each pixel and always found very good fitting
results. In addition, the IRDC we study here are not strongly affected
by radiation; G28.53-0.25 is even not correlated with any protostellar
object, so we are confident in the validity of our column density
maps.

%%%%%%%%%%%%%%%%%%%%%%%%%%%%%%   
%%%%%%%%%%%%%%%%%%%%%%%%%%%%%%   
%%%  PDFs IRDCs %%%%%%%%%%%%%%%   
\begin{figure*}[!htpb]   
\begin{centering}   
\vspace{0cm}\includegraphics [width=6cm, angle={90}]{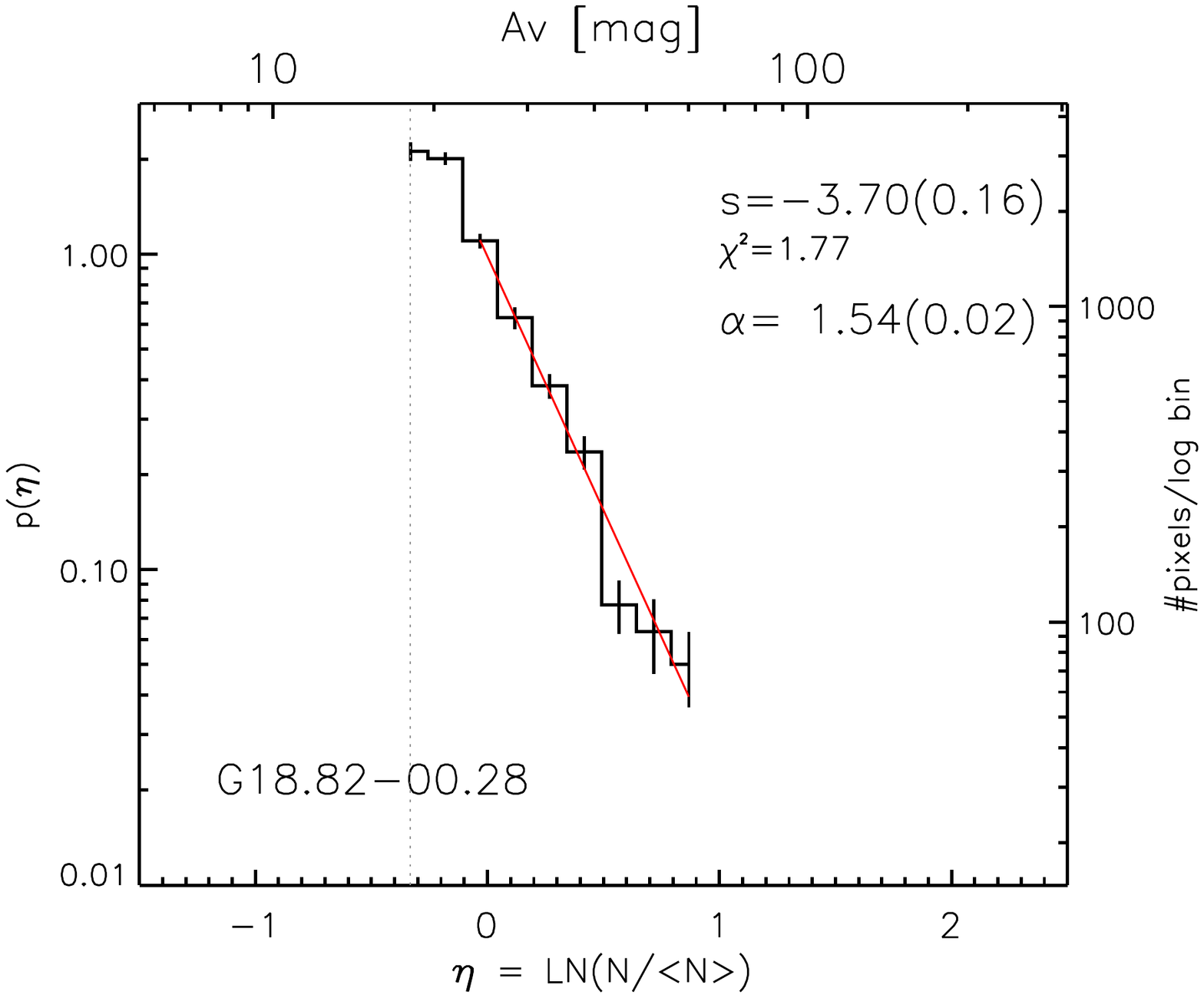}   
\hspace{-1.0cm}\includegraphics [width=6cm, angle={90}]{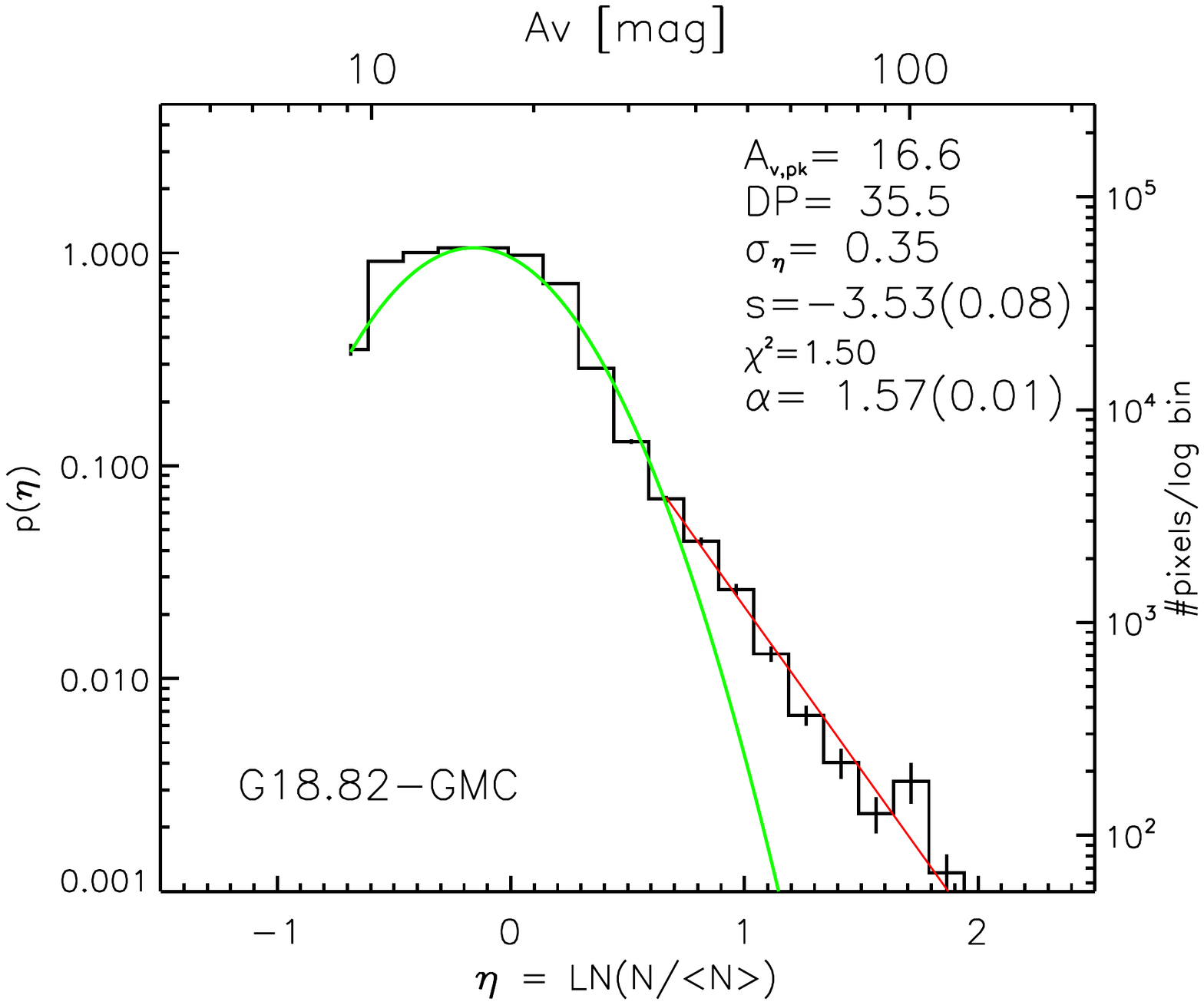}   
 
\vspace{-1.6cm}\includegraphics [width=6cm, angle={90}]{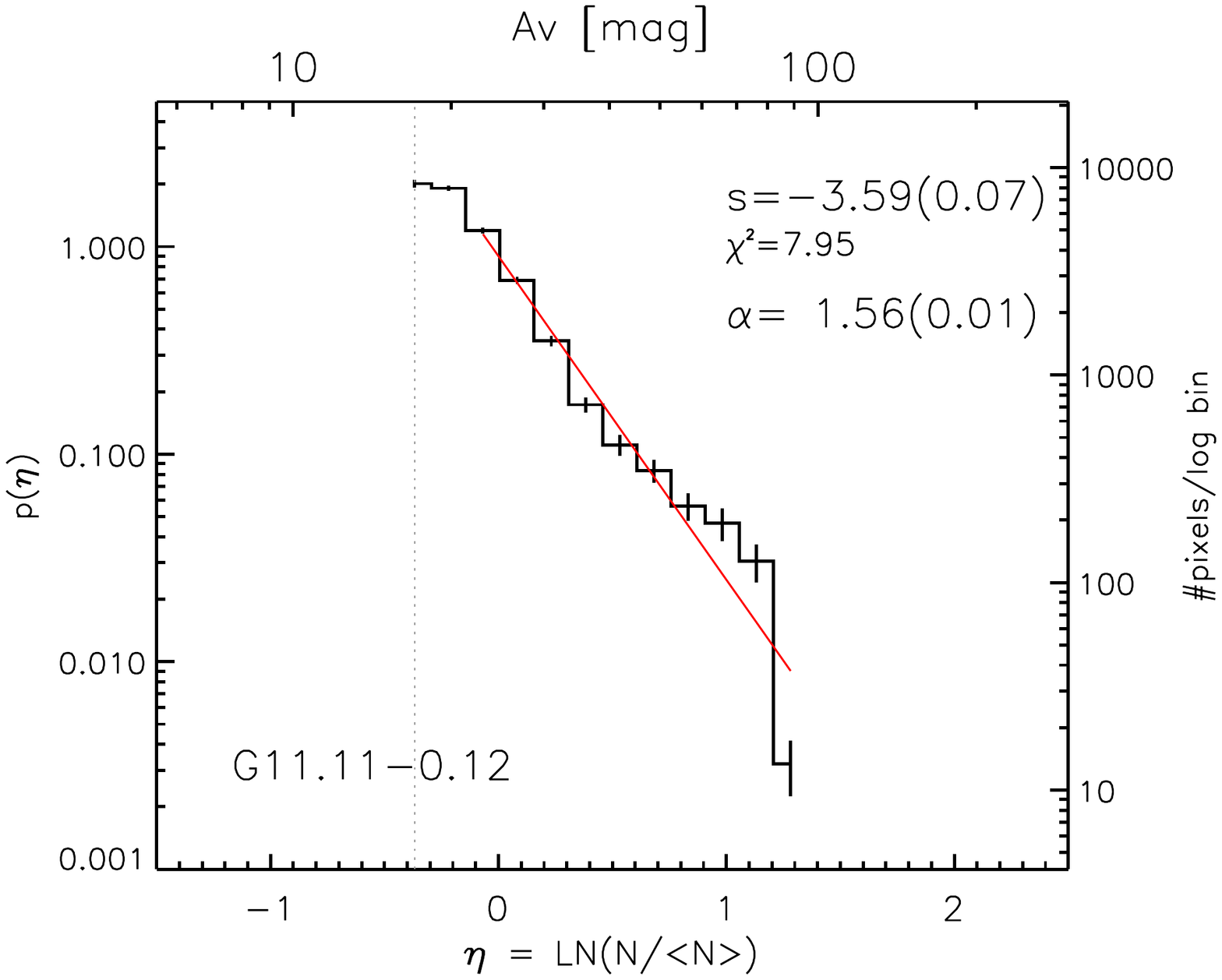}   
\hspace{-1.0cm}\includegraphics [width=6cm, angle={90}]{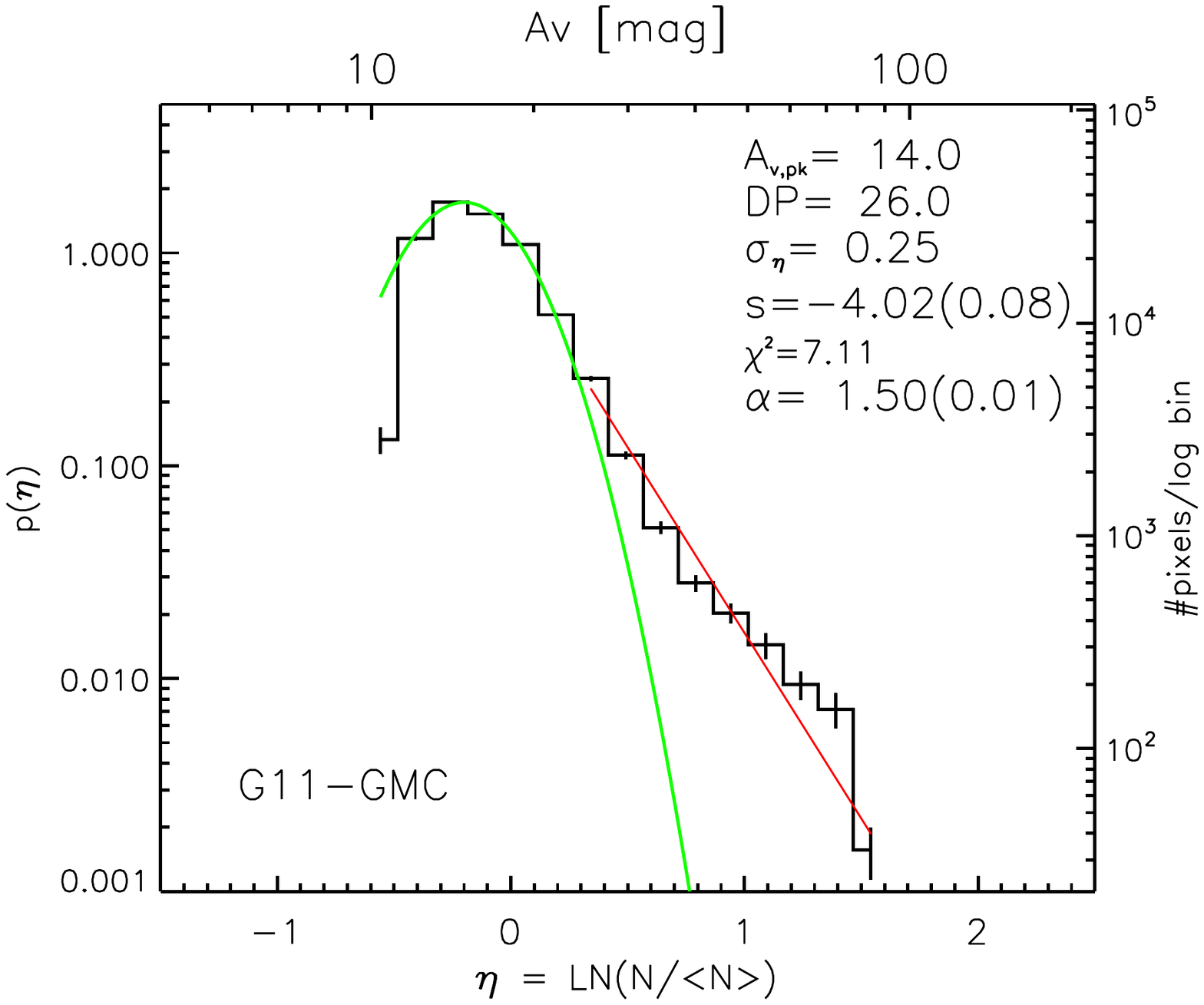}   
 
\vspace{-1.6cm}\includegraphics [width=6cm, angle={90}]{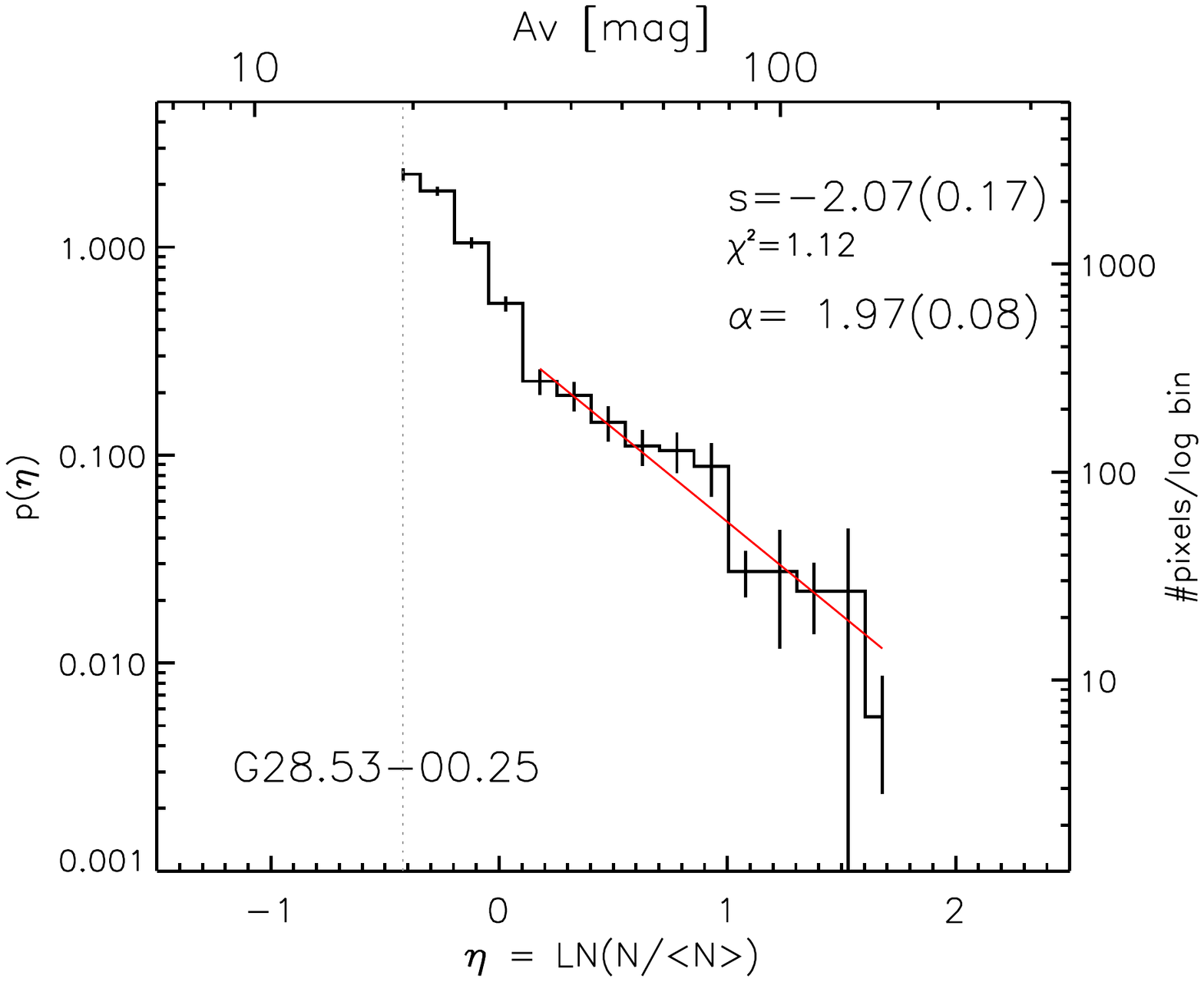}   
\hspace{-1.0cm}\includegraphics [width=6cm, angle={90}]{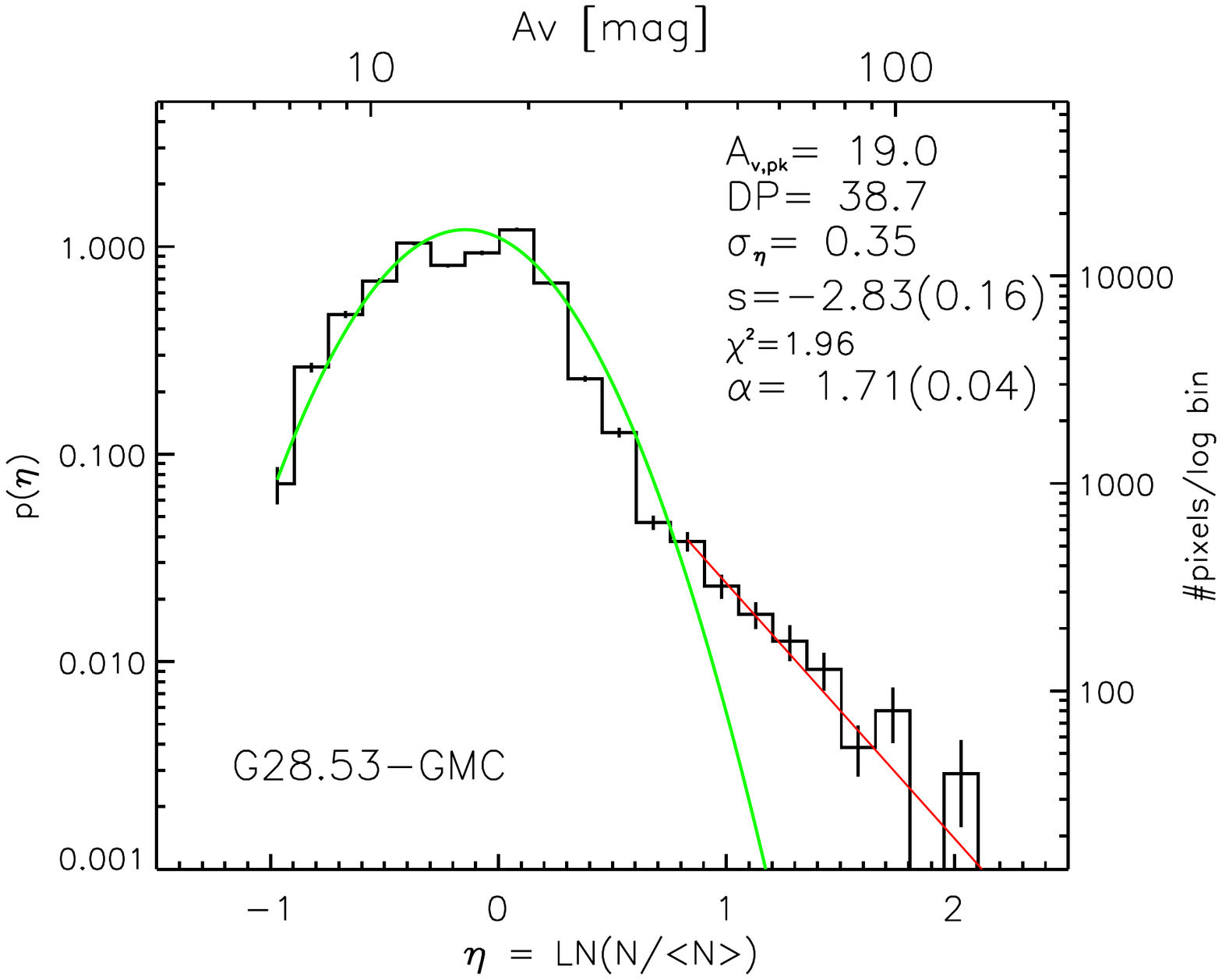}   
 
\caption[] {PDFs derived from {\sl Herschel} column density maps (SED 
  fit using only the {\sl Herschel} wavelengths 160, 250, 350, and 500 
  $\mu$m) at 36$''$ angular resolution. Left panel shows the PDFs 
  of pixels comprising only the IRDC i.e. G18.82-0.28 (Cloud A), 
  G28.53-0.25 (Cloud D), and G11.11-0.12. Right panel shows  
  PDFs from  associated GMC, including the IRDC.  The left y-axis 
  gives the normalised probability $p(\eta)$, the right y-axis the 
  number of pixels per log bin. The upper x-axis is the visual 
  extinction and the lower x-axis the logarithm of the normalised 
  column density.  The dashed vertical line indicates the completeness 
  level. The red line indicates a power-law fit to the high column 
  density tail with the slope $s$ together with its error and the 
  reduced $X^2$ goodness-of-fit. The exponent $\alpha$ of an 
  equivalent spherical density distribution $ \rho(r) \propto 
  r^{-\alpha}$ is also indicated in the panel.  The dashed line in the 
  upper panel indicates the completeness level; the PDF left of this 
  line is incomplete.} 
\label{pdf-irdcs}   
\end{centering}  
\end{figure*} 
   
\section{KS-test } 
 
%%%%%%%%%%%%%%%%%%%%%%%%%%%%%%   
%%% KS test %%%%%%%%%%%%%%%   
\begin{figure}[!htpb] \label{ks}   
\begin{centering}   
\includegraphics [width=7cm, angle={00}]{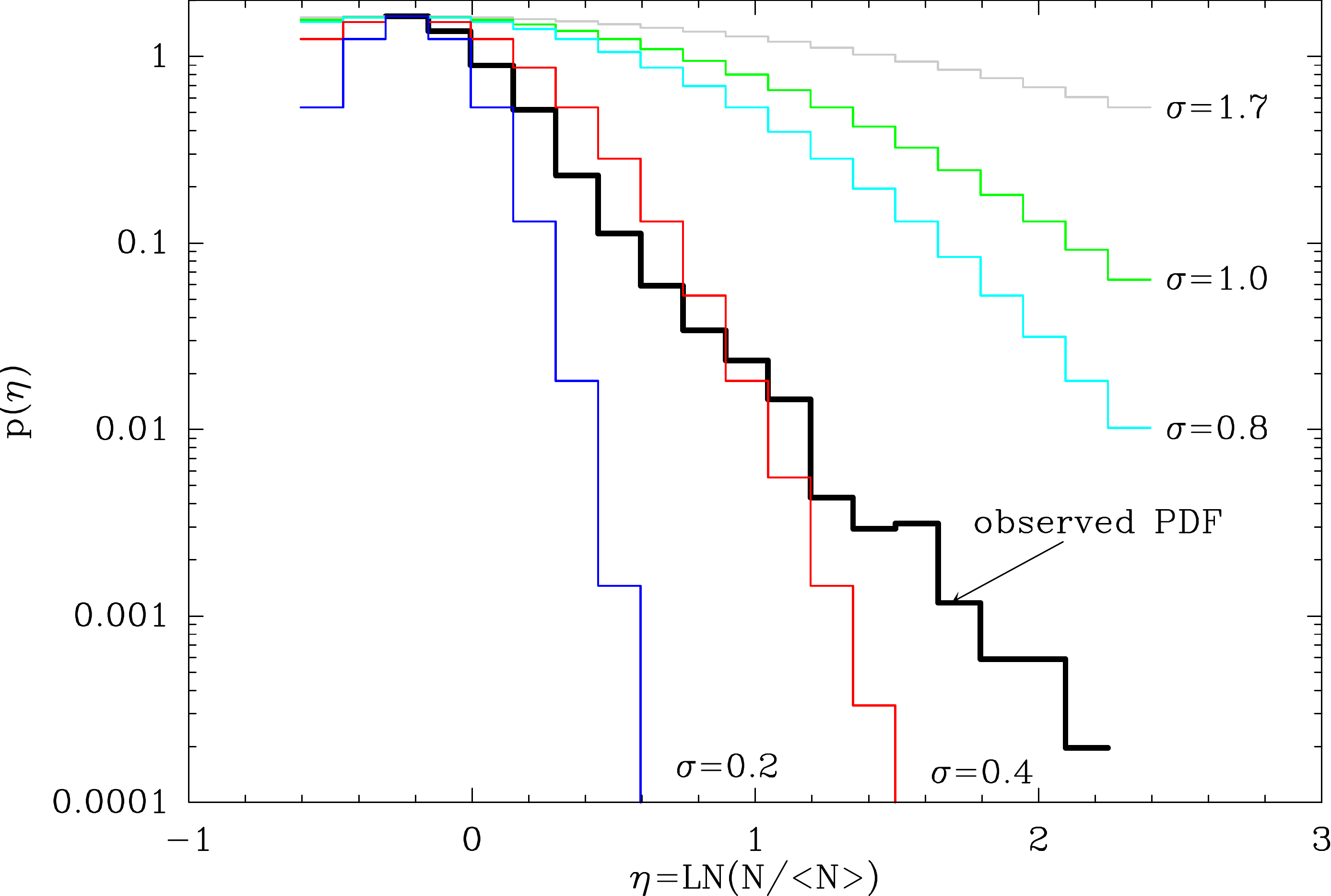}   
\includegraphics [width=7cm, angle={00}]{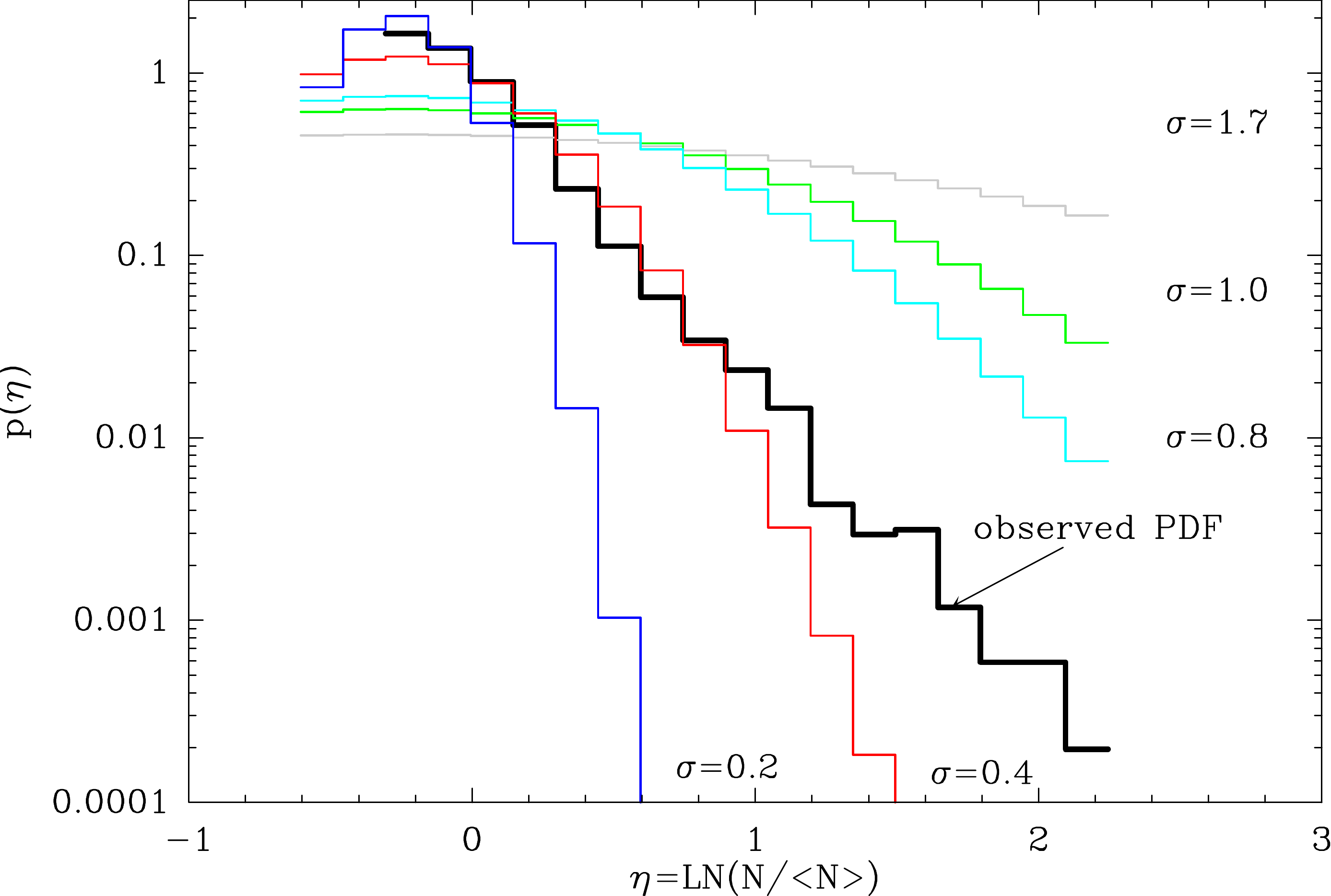}   
\caption[] {Synthetic distributions shown in various colours with 
  different widths (indicated in the panel). The observed PDF for 
  G28.37+0.07, constructed from {\sl Herschel} data is indicated in 
  bold black. Top panel shows distributions that are normalised to 
  the peak; the bottom panel shows distributions that are normalised to the 
  same number of pixels.} 
 \end{centering}  
\end{figure}   
 
\begin{table}  \label{ks-tab} 
  \caption{Results of KS-test. $D$ and $p$ are the parameters  
    of the KS-test for the two normalisations (on-peak value or number of pixels) used to generate the synthetic  
    distributions.}     
\begin{center}   
\begin{tabular}{lccccc}   
\hline  
\hline    
 &    &  \\ 
PDF width $\sigma$  & $D_{peak}$   &  $p_{peak}$ & $D_{integral}$ &  $p_{integral}$ \\  
\hline    
0.2 &  0.600  & 8.7843e-04 &  0.5790  & 1.7733e-03\\ 
0.4 &  0.301  & 2.8865e-01 &  0.3158  & 2.4667e-01 \\ 
0.8 &  0.382  & 8.7724e-02 &  0.3684  & 1.1612e-01 \\ 
1.0 &  0.579  & 1.4968e-03 &  0.4737  & 1.8105e-02 \\ 
1.7 &  0.789  & 2.2829e-06 & 0.63158  & 4.6629e-04 \\ 
\end{tabular} 
\end{center}   
\end{table}    
 
To verify if the column density distribution (PDF) we 
observed for G28.37+0.07 possibly arises from a log-normal 
distribution, we performed a two-sample KS-test. We first generated 
synthetic functions following a log-normal distribution 
    
\begin{equation}    
p_\eta\,{\rm d}\eta=\frac{1}{\sqrt{2\pi\sigma^2}}{\rm exp}\Big[ -\frac{(\eta-\mu)^2}{2\sigma^2} \Big]{\rm d}\eta    
,\end{equation}    
where $\sigma$ is the dispersion i.e. the width of the distribution, 
$\eta$ the mean logarithmic column density, and $\mu$ the peak (in 
units of $\eta$).  To have a larger parameter space to 
compare, we varied $\sigma$ between 0.2 and 1.7 (the latter is the 
width stated in KT). 
 
For the normalisation, we performed two approaches. First, we fixed
the peak of all synthetic distributions to the observed value.  The
x-axis value for this peak at \av\,$\sim$19 ($\mu$=-0.23) was
extracted from the PDF of the whole GMCs that revealed a log-normal +
power-law tail distribution (Fig.~\ref{allpdfs}).  Because these
distributions have the same peak value but not a normalisation to the
same number of pixels, we also produced a second set of functions that
have this kind of a normalisation (but different peaks).  Both sets of
resulting synthetic distributions together with our observed set are
shown in Fig.~C.1.
 
Eye inspection already shows that the observed PDF is not consistent
with any log-normal distribution, in particular not with one of a
large width ($\sigma$=1.7 was found in KT). To quantify this result,
we list in Table~C.1 the results of the two-sample KS-test, which
gives the statistic and associated probability that two data sets are
drawn from the same distribution.  We performed the test using the
IDL-routine 'ksto.pro' with the algorithm taken from procedure of the
same name in "Numerical Recipes" by Press et al.  (\cite{press1992}).
The parameter $D$ gives the maximum deviation between the cumulative
distribution of the observed PDF and the assumed underlying statistic
(the log-normal in our case), while $p$ shows the significance level
of the KS-statistic.  Small values of $p$ , which is the case for our
test, indicate that the pairwise distributions are significantly
different. The $ p$-values below 0.3 show that we can exclude the
scenario that our distribution can be fit with a log-normal
distribution by more than one sigma. Note that though the pixel
statstic is low and the binning is large, the general properties
(width, peak, slope of power-law tail) of the PDF do not depend on
binning in smaller or larger bins or varing the pixel size or
resolution (Schneider et al.  2015).
 
\section{The $^{12}$CO 3$\to$2 spectra of G11.11-0.12}

%%%%%%%%%%%%%%%%%%%%%%%%%%%%%%   
%%% Spectra of G11 %%%%%%%%%%%%%%%   
\begin{figure*}[!htpb]   
\begin{centering}   
\includegraphics [width=12cm, angle={00}]{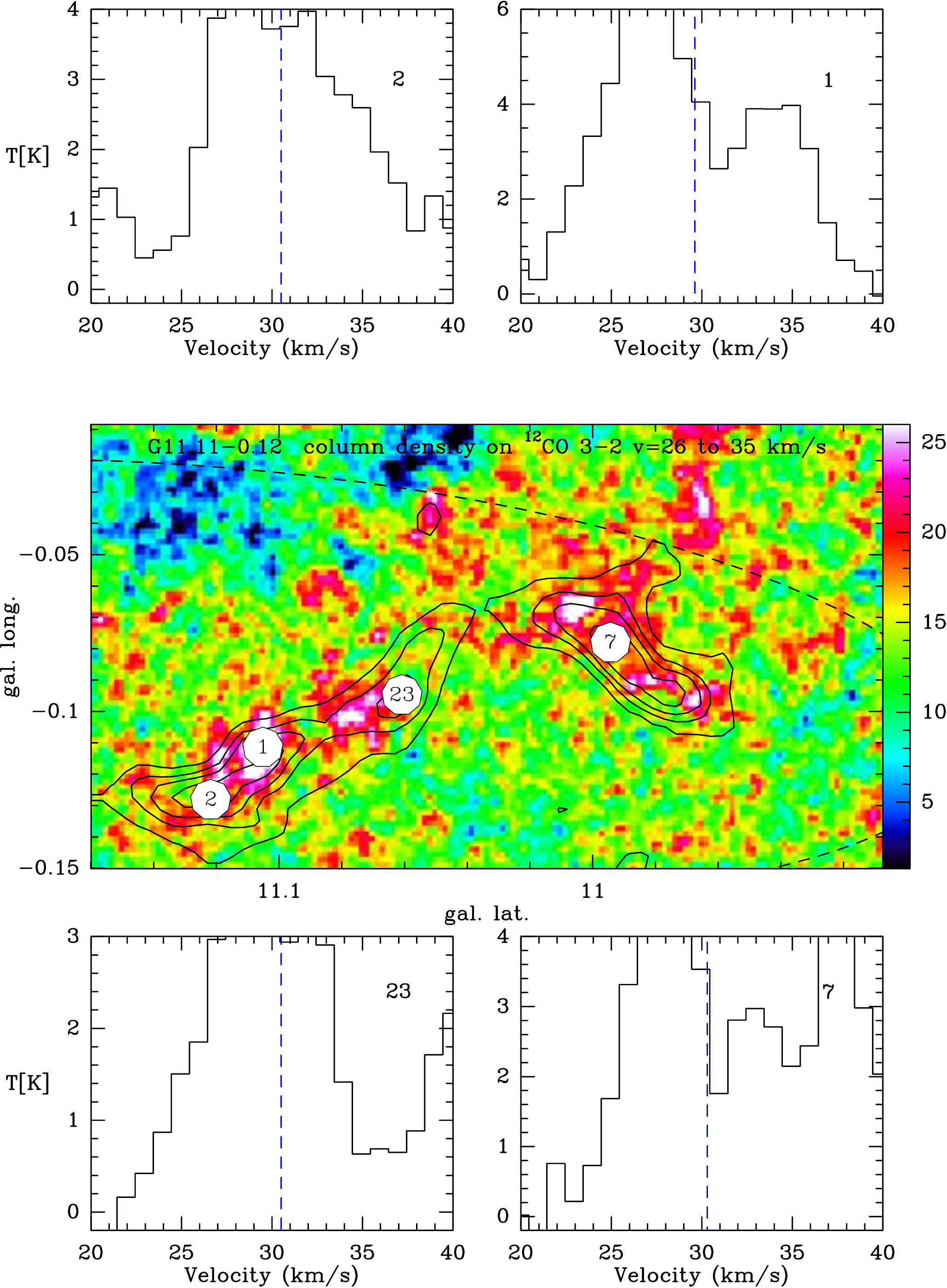}   
\caption[] {Middle panel shows in colour the line integrated $^{12}$CO
  3$\to$2 emission (between 26 and 35 km s$^{-1}$) of G11-0.12. The
  dust column density from {\sl Herschel} is overlaid as black
  contours (levels 3, 4, 5, 6 10$^{22}$ cm$^{-2}$).  The numbering
  from 1 to 4 indicates the position of submm-continuum sources
  detected with ATLASGAL and subsequently observed in N$_2$H$^+$
  (Tackenberg et al. 2014). The $^{12}$CO 3$\to$2 spectra at these
  positions are displayed in the panels . The blue-dashed line denotes
  the centre velocity of the N$_2$H$^+$ line. }
\label{spectra-g11}   
\end{centering}  
\end{figure*}   
 
Figure~\ref{spectra-g11} shows $^{12}$CO 3$\to$2 spectra at the
location of some of the continuum sources of the IRDC G11.11-0.12 (no
$^{13}$CO data are available) in the velocity range where the emission
is associated with the cloud. Similar as seen for G28.37+0.07, the
optically thick $^{12}$CO line shows a self-absorbed dip close to the
velocity of the optically thin N$_2$H$^+$ line (indicated by a
blue-dashed line). The velocity resolution of the CO data is only 1 km
s$^{-1}$ while that of N$_2$H$^+$ is 0.2 km s$^{-1}$ so that
resolution effects can play a role. The most straightforward
explanation for the observed profile is infall, i.e. a line shape
cause by the radial inwards motion of gas onto a central source.
 

\begin{thebibliography}{}    
  
 
\bibitem[2014]{catarina2014}     
Alves de Oliveira, C., Schneider, N., Merin, B., et al., 2014, A\&A568, 98 
 
\bibitem[2011]{ball2011}    
Ballesteros-Paredes, J., V{\'a}zquez-Semadeni, E., Gazol, A., 2011, MNRAS, 416, 1436   
 
\bibitem[2013]{beuther2013}    
Beuther, H., Linz, H., Tackenberg, J., et al., 2013, A\&A, 553, 81  
   
\bibitem[1978]{bohlin1978}    
Bohlin, R.C., Savage, B.D., Drake, J.F., 1978, ApJ 224, 132   
  
\bibitem[2014]{butler2014}   
Butler, M., Tan, J., Kainulainen, J., 2014,  ApJ, 782, L30 
   
\bibitem[1998]{carey1998}    
Carey, S.J., Clark, F.O., Egan, M.P., et al., 1998, ApJ, 508, 721  
 
\bibitem[2000]{carey2000}    
Carey, S.J., Feldman, P.A., Redman, R.O., et al., 2000, ApJ, 543, L157 
 
\bibitem[2013]{contreras2013}  
Contreras, Y., Schuller, F., Urquhart, J.S., et al., 2013, A\&A, 549, 45 
   
\bibitem[2011]{csengeri2011}    
Csengeri, T., Bontemps, S., Schneider, N., et al., 2011, A\&A, 527, 135   
 
\bibitem[2014]{csengeri2014}    
Csengeri, T., Urquhart, J.S., Schuller, F.,et al., 2014, A\&A, 565, 75   
 
\bibitem[2013]{dempsey2013}    
Dempsey, J.T., Thomas, H.S., Currie, M.J., ApJS, 209, 1, 13 
   
\bibitem[1998]{egan1998}    
Egan, M.P., Shipman, R.F., Price, S.D., et al., 1998, ApJ, 494, L199 
   
\bibitem[2008a]{fed2008a}    
Federrath, C., Klessen, R.S., Schmidt, W., 2008a, ApJ, 688, L79   
 
\bibitem[2008b]{fed2008b}    
Federrath, C., Glover, S., Klessen, R.S., 2008b, Physica Scripta, Volume 132, Issue , id. 014025 
  
\bibitem[2013]{fed2013}    
Federrath, C., Klessen, R. S., 2013, ApJ, 763, 51   
   
\bibitem[2010]{froebrich2010}    
Froebrich, D., Rowles, J., 2010, MNRAS, 406, 1350   
  
\bibitem[2014]{giri2014}    
Girichidis, P., Konstandin, L., Whitworth, A.P., et al., 2014, ApJ, 781, 91   
  
\bibitem[2010]{griffin2010} 
Griffin, M., Abergel, A., Abreau, A., et al., 2010, A\&A, 518, L3  
 
\bibitem[2010]{heiderman2010}    
Heiderman, A., Evans, N.J.E. II, Allen, L.E., et al., 2010, ApJ, 723, 1019 
    
\bibitem[2010]{henning2010}    
Henning, T., Linz, H., Krause, O., et al., 2010, 518, L95 
  
\bibitem[2012]{hennemann2012}    
Hennemann, M., Motte, F., Schneider, N., et al., 2012, A\&A, 543, L3   
   
\bibitem[2011]{hill2011}    
Hill, T., Motte F., Didelon P., et al., 2011, A\&A, 533, 94    
  
\bibitem[2006]{jackson2006}    
Jackson, J.M., Rathborne, J.M., Shah, R.Y., et al., 2006, ApJS, 163, 145 
   
\bibitem[2009]{kai2009}   
Kainulainen, J., Beuther, H., Henning, T., \& Plume, R., 2009, A\&A, 508, L35   
   
\bibitem[2011a]{kai2011a}   
Kainulainen, J., Alves, J., Beuther, H., et al., 2011a, A\&A, 536, 48   
 
\bibitem[2011b]{kai2011b}   
Kainulainen, J., Beuther, H., Banerjee, R., et al., 2011b, A\&A, 530, 64   
 
\bibitem[2013]{kai2013}   
Kainulainen, J., Tan, J.C., 2013, A\&A, 549, 53 
 
\bibitem[2000]{klessen2000}   
Klessen, R.~S., 2000, ApJ, 535, 869    
    
\bibitem[2013]{kirk2013}    
Kirk, H., Myers, P.C., Bourke, T.L., et al., 2013, ApJ, 766, 115 
    
\bibitem[2007]{kritsuk2007}    
Kritsuk, A.G., Norman, M.L., Padoan, P., Wagner, R., 2007, ApJ, 665, 416 
   
\bibitem[2011]{kritsuk2011}    
Kritsuk, A.G., Norman, M.L., Wagner, R., 2011, ApJ, 727, L20   
 
\bibitem[2010]{lada2010}    
Lada, C.J., Lombardi, M., Alves, J., 2010, ApJ, 724, 687   
   
\bibitem[2008]{lombardi2008}   
Lombardi, M., Lada, C., Alves, J., 2008, A\&A, 489, 143   
   
\bibitem[2004]{marston2004}   
Marston, A.P., Reach, W.T., Noriega-Crespo, A., et al., 2004, ApJS, 154, 333 
 
\bibitem[2010]{molinari2010}   
Molinari, S., Swinyard, B., Bally, J. et al., 2010,  A\&A, 518, L100   
 
\bibitem[1996]{myers1996}  
Myers, P.C., Maradones, D., Tafalla, M., et al., 1996, ApJ, 465, L133 
   
\bibitem[2011]{quang2011}   
Nguyen-Luong, Q., Motte, F., Hennemann, M., et al., 2011, A\&A, 535, 76  
 
\bibitem[2011]{ott2011}   
Ott,S., 2011, in Astronomical Society of the Pacific Conf. Ser., Vol. 442,  
ed. N. Evans, A. Accomazzi, D.J. Mink \& A.H. Rots, 347 
   
\bibitem[1997]{padoan1997}    
Padoan, P., Jones, J.T., Nordlund, A.A., 1997, ApJ, 474, 730    
   
\bibitem[2010]{peretto2010}    
Peretto, N., Fuller, G., 2010, ApJ, 723, 555 
 
\bibitem[2013]{peretto2013}    
Peretto, N.,  Fuller, G.A., Duarte-Cabral, A., et al., 2013, A\&A, 555, 112 
  
\bibitem[2010]{pilbratt2010} 
Pilbratt, G., Riedinger, J., Passvogel, T., et al., 2010, A\&A 518, L1    
 
\bibitem[2006]{pillai2006} 
Pillai, T, Wyrowski, F., Menten, K., et al., 2006, A\&A, 447, 929 
 
\bibitem[2010]{poglitsch2010} 
Poglitsch, A., Waelkens, C., Geis, N., et al., 2010, A\&A 518, L2  
 
\bibitem[1992]{press1992}   
Press, W.H., Teukolsky, S.A., Vetterling, W., Flannery, B.P., 1992, Cambridge University Press, 1992, 2nd ed.,  
Chap. 14 
   
\bibitem[2006]{rathborne2006}   
Rathborne, J.M., Jackson, J.M., Simon, R., 2006, ApJ, 641, 389  
 
\bibitem[2011]{rathborne2011}   
Rathborne, J.M., Garay, G., Jackson, J.M., et al., 2011, ApJ, 741, 120 
 
\bibitem[2014]{rathborne2014}   
Rathborne, J.M., Longmore, S., Jackson, J.M., et al., 2014, ApJ, 795, L25
 
\bibitem[2013]{roussel2013}   
Roussel, H., 2013, PASP, 125, 1126 
 
\bibitem[2013]{roy2013}  
Roy, A., Martin, P., Polychroni, D., et al., 2013, ApJ, 763, 55 
 
\bibitem[2013]{russeil2013}   
Russeil, D., Schneider, N., Anderson, L., et al., 2013, A\&A, 554, 42 
 
\bibitem[2013]{sakai2013}    
Sakai, T., Sakai, N., Foster, J.B., et al., 2013, ApJ, 775, 31 
    
\bibitem[2010]{schneider2010}    
Schneider, N.,  Csengeri T., Bontemps S., et al., 2010, A\&A, 520, 49  
  
\bibitem[2011]{schneider2011}    
Schneider, N., Bontemps, S., Simon, R., et al., 2011, A\&A,  529, 1   
   
\bibitem[2012]{schneider2012}    
Schneider, N., Csengeri, T., Hennemann, M., et al., 2012, A\&A, 540, L11   
  
\bibitem[2013]{schneider2013}    
Schneider, N., Andr\'e, Ph., K\"onyves, V., et al., 2013, ApJ, 766, L17 
 
\bibitem[2015]{schneider2015}    
Schneider, N., Ossenkopf, V., Csengeri, T., et al., 2015, A\&A, 575, 79
 
\bibitem[2009]{schuller2009}    
Schuller, F., Menten, K., Contreras, Y., et al., 2009, A\&A, 504, 415 
 
\bibitem[2014]{shipman2014}    
Shipman, R.F., van der Tak, F.F.S., Wyrowski, F., et al., 2014, A\&A, 570, 51
 
\bibitem[2012]{smith2012}  
Smith, R.J., Shetty, R., Stutz, A.M., Klessen, R.S., 2012, ApJ, 750, 64 
 
\bibitem[2013]{smith2013}  
Smith, R.J., Shetty, R., Beuther, H., Klessen, R.S., Bonnell, I.A., 2013, ApJ, 771, 24 
 
\bibitem[2001]{simon2001}  
Simon, R., Jackson, J.M., Clemens, D.P., Bania, T.M., 2001, ApJ, 551, 747 
 
\bibitem[2006a]{simon2006a}  
Simon, R., Rathborne, J.M., Shah, R., et al., 2006a, ApJ, 653, 1325 
 
\bibitem[2006b]{simon2006b}  
Simon, R., Jackson, J.M., Rathborne, J.M., et al., 2006b, ApJ, 639, 227 
 
\bibitem[2009]{siringo2009}    
Siringo, G., Kreysa, E., Kovacs, A., et al., 2009, A\&A, 497, 945 
 
\bibitem[1988]{strong1988}  
Strong, J., Bloemen, J., Dame, T., et al., 1988, A\&A, 207, 1 
 
\bibitem[2014]{tack2014}    
Tackenberg, J., Beuther, H., Henning, T., et al., 2014, A\&A, 565, 101 
 
\bibitem[2002]{tafalla2002}    
Tafalla, M., Myers, P.C., Caselli, P., et al., 2002, ApJ, 569, 815 
   
\bibitem[2002]{teyssier2002}    
Teyssier, D., Hennebelle, P., Perault,M., 2002, A\&A, 624, 638 
    
\bibitem[2011]{traficante2011}     
Traficante, A., Calzoletti, L., Veneziani, M., et al., 2011, MNRAS, 416, 2932 
 
\bibitem[2014]{tremblin2014}     
Tremblin, P., Schneider, N., Minier, V., et al., 2014, A\&A, 564, 106 
 
\bibitem[2008]{vaz2008}    
Vazquez-Semadeni, E., Gonzales, R.F., Ballesteros-Paredes, J., et al., 2008, MNRAS, 390, 769   
   
\bibitem[2014]{vikar2014}    
Vikar, Y., Clauset, A., 2014, The annals of applied statistics, Vol. 8, No. 1, 89 
 
\bibitem[1994]{walker1994}    
Walker, C.K., Narayanan, G., Boss, A.P., 1994, ApJ, 431, 767 
 
\bibitem[2014]{wang2014}   
Wang, K., Zhang, Q., Testi, L., et al. 2014, MNRAS, 439, 3275 
 
\bibitem[2008]{wang2008}   
Wang, Y., Zhang, Q., Pillai, T., et al., 2008, ApJ, 672, L33 
 
\bibitem[2001]{weiss2001}   
Weiss, A., Neininger, N., H\"uttemeister, S., Klein, U., 2001, A\&A, 365, 571 
 
\bibitem[2009]{zhang2009}    
Zhang, Q., Wang, Y., Pillai, T., \& Rathborne, J. 2009, ApJ, 696, 268 
 
\end{thebibliography}
\end{document}